\documentclass[twocolumn]{aastex63}

\usepackage{graphicx}
\usepackage{graphics}
\usepackage{amsmath}
\usepackage{amssymb}
\usepackage{color}
\usepackage{url}
\usepackage{hyperref}

\submitjournal{\apj}
\shorttitle{FOGGIE V: Modified Virial Temperature}
\shortauthors{Lochhaas et al.}

\newcommand{\rvir}{R_{200}}
\newcommand{\realrvir}{R_\mathrm{vir}}
\newcommand{\mvir}{M_{200}}
\newcommand{\tvir}{T_\mathrm{vir}}
\newcommand{\tmod}{T_\mathrm{mod}}
\newcommand{\KEth}{\mathrm{KE}_\mathrm{th}}
\newcommand{\KEnt}{\mathrm{KE}_\mathrm{nt}}

\begin{document}

\title{Figuring Out Gas \& Galaxies In Enzo (FOGGIE) V:\\ The Virial Temperature Does Not Describe Gas in a Virialized Galaxy Halo}
\correspondingauthor{Cassandra Lochhaas}
\email{clochhaas@stsci.edu}

\author[0000-0003-1785-8022]{Cassandra Lochhaas}
\affiliation{Space Telescope Science Institute, 3700 San Martin Dr., Baltimore, MD 21218}

\author[0000-0002-7982-412X]{Jason Tumlinson}
\affiliation{Space Telescope Science Institute, 3700 San Martin Dr., Baltimore, MD 21218}
\affiliation{Department of Physics \& Astronomy, Johns Hopkins University, 3400 N.\ Charles Street, Baltimore, MD 21218}

\author[0000-0002-2786-0348]{Brian W.\ O'Shea}
\affiliation{Department of Computational Mathematics, Science, and Engineering, 
Department of Physics and Astronomy, 
National Superconducting Cyclotron Laboratory,  
Michigan State University}

\author[0000-0003-1455-8788]{Molly S.\ Peeples}
\affiliation{Space Telescope Science Institute, 3700 San Martin Dr., Baltimore, MD 21218}
\affiliation{Department of Physics \& Astronomy, Johns Hopkins University, 3400 N.\ Charles Street, Baltimore, MD 21218}

\author[0000-0002-6804-630X]{Britton D.\ Smith}
\affiliation{Royal Observatory, University of Edinburgh, United Kingdom}

\author[0000-0002-0355-0134]{Jessica K.\ Werk}
\affiliation{Department of Astronomy, University of Washington, Seattle, WA 98195}

\author[0000-0001-7472-3824]{Ramona Augustin}
\affiliation{Space Telescope Science Institute, 3700 San Martin Dr., Baltimore, MD 21218}

\author[0000-0002-6386-7299]{Raymond C.\ Simons}
\affiliation{Space Telescope Science Institute, 3700 San Martin Dr., Baltimore, MD 21218}

\begin{abstract}
The classical definition of the virial temperature of a galaxy halo excludes a fundamental contribution to the energy partition of the halo: the kinetic energy of non-thermal gas motions. Using simulations of low-redshift, $\sim L^*$ galaxies from the FOGGIE project (Figuring Out Gas \& Galaxies In Enzo) that are optimized to resolve low-density gas, we show that the kinetic energy of non-thermal motions is roughly equal to the energy of thermal motions. The simulated FOGGIE halos have $\sim 2\times$ lower bulk temperatures than expected from a classical virial equilibrium, owing to significant non-thermal kinetic energy that is formally excluded from the definition of $\tvir$. We derive a modified virial temperature explicitly including non-thermal gas motions that provides a more accurate description of gas temperatures for simulated halos in virial equilibrium. Strong bursts of stellar feedback drive the simulated FOGGIE halos out of virial equilibrium, but the halo gas cannot be accurately described by the standard virial temperature even when in virial equilibrium.  Compared to the standard virial temperature, the cooler modified virial temperature implies other effects on halo gas: (i) the thermal gas pressure is lower, (ii) radiative cooling is more efficient, (iii) \ion{O}{6} absorbing gas that traces the virial temperature may be prevalent in halos of a higher mass than expected, (iv) gas mass estimates from X-ray surface brightness profiles may be incorrect, and (v) turbulent motions make an important contribution to the energy balance of a galaxy halo.
\end{abstract}


\section{Introduction}
\label{sec:intro}

The current paradigm of hierarchical structure formation has been in place for decades \citep{Rees1977,Silk1977,White1978,White1991}. The general mechanism by which galaxies form starts with random density fluctuations in dark matter that collapse over cosmic time into massive dark matter halos. Baryonic matter is similarly gravitationally attracted to dark matter halos, where it collects and eventually forms galaxies \citep[for a review, see][]{Benson2010}. The standard paradigm of galaxy formation supposes that gas falling onto a massive halo, $M_{\rm halo} \gtrsim {\rm few}\times10^{11} M_{\odot}$ \citep{Birnboim2003}, shock-heats to the virial temperature before later cooling at the halo center to form stars. More recent models \citep{Keres2005,Dekel2006,Dekel2009,Nelson2013} show that infalling gas need not shock-heat to high temperatures, but may instead be accreted to the central galaxy along filaments while remaining cold ($T \lesssim 10^{4-5}$ K). Modern simulations show that both hot halos and cold filaments can exist surrounding a galaxy simultaneously, but the presence of a hot, shock-heated gaseous halo is still expected surrounding massive galaxies, even if it is not the primary mode of gas accretion \citep{Bennett2020,Fielding2017,Stern2020a}.

In order for gas to form stars at the center of halos, it must be cold (with $T \ll 10^4$ K). The standard paradigm assumes gas either cools radiatively before accreting onto the galaxy or flows onto the galaxy from the intergalactic medium (IGM) without being heated. However, the exact processes by which cold gas forms from, or interacts with, the expected hot halo are not fully understood. The conditions for the formation and survival of the cold gas are strongly dependent on the properties of the hot, diffuse, volume-filling phase of the circumgalactic medium (CGM). 

In models that attempt to describe hot gas observations or cool gas formation and survival within the hot halo, the hot gas is usually assumed to be static, in hydrostatic equilibrium near the virial temperature, $\tvir$, of the halo \citep{Maller2004,Anderson2010,Miller2013,Faerman2017,Mathews2017,McQuinn2018,Qu2018,Stern2019,Voit2019a,Faerman2020}. `Idealized' simulations commonly adopt a hydrostatic hot halo at $\tvir$ as part of their initial conditions \citep{Armillotta2017,Fielding2017,Li2020a}. Small, cold gas clouds may then condense from the hot medium, seeded by thermal instabilities \citep{McCourt2012,Voit2015}. Cold gas may also be seeded by galactic winds, where the hot flow can entrain, precipitate, or carry cold clumps into the CGM \citep{Thompson2016,Schneider2018,Lochhaas2018} to re-accrete later. If cold CGM gas is instead in the form of extended filamentary structures, these structures may pierce through the expected virial shock and hot halo \citep{Keres2005,Dekel2006,Keres2009,Dekel2009,Bennett2020}, interacting with the hot diffuse gas and creating hydrodynamical instabilities at the hot-cold interface \citep{Mandelker2016,Mandelker2020}. Alternatively, cold accreting gas may take the form of a cooling flow, where the hot halo undergoes bulk cooling as it is compressed on its journey to the central galaxy \citep{Mathews1978,Fabian1984,Malagoli1987,Li2012,Stern2019,Stern2020a}.

Observations of the CGM typically find both hot and cold gas traced by high- and low-ionization state absorption observed in the UV and optical  \citep{Wakker2009,Rudie2012,Werk2013,Stocke2013,Bordoloi2014,Lehner2015,Borthakur2016,Heckman2017,Keeney2017,Chen2018,Berg2018,Berg2019,Rudie2019,Chen2020,Lehner2020} or emission in the X-ray \citep{Anderson2010}. The densities and temperatures are derived from ionization modeling, where generally it is assumed that high-ion absorbers (\ion{O}{6} or \ion{O}{7}) trace a warmer, collisionally-ionized gas phase than low-ionization state absorbers (e.g., \ion{Mg}{2}, \ion{Si}{3}), which trace a cooler, photoionized gas phase \citep[e.g.,][]{Tumlinson2017}. Studies that find a significant mass of cold gas in the CGM of $L^*$ galaxies, $\sim10^{10}M_\odot$, have raised  questions about how so much cold gas could be supported in the CGM \citep[e.g.,][]{Werk2014,Keeney2017}. Fitting small, thermal-pressure-supported cold clouds into the standard paradigm of a hydrostatic hot halo is difficult while also matching the cold and hot gas densities inferred from photoionized modeling \citep[but see \citeauthor{Haislmaier2021} \citeyear{Haislmaier2021}]{Werk2014,McQuinn2018}. All such modeling is laden with assumptions about the thermal balance of the CGM that could prove to be mistaken. 

At larger scales, galaxy cluster and intra-cluster medium (ICM) analytic, simulation, and observational studies have shown that the ICM is not in perfect hydrostatic equilibrium because non-thermal kinetic gas motions are crucial to the overall energy balance of the halo. Bulk non-thermal gas motions, such as turbulence, contribute a significant fraction of pressure support to the cluster gas \citep{Shi2015,Shi2018,Simionescu2019}. This fraction is significant enough to produce a ``hydrostatic mass bias", i.e. the cluster mass derived without including non-thermal pressure support differs from the ``true" cluster mass by $\sim15\%$ on average \citep{Lau2013,Shi2016}.

At $\sim L^*$ galaxy halo scales, only recently have simulations begun to show that the standard picture of a hot gaseous halo in hydrostatic equilibrium may not be accurate. \citet{Lochhaas2020} showed that even in idealized $L^*$ CGM simulations initiated with hydrostatic hot halos, galactic feedback creates bulk flows that induce significant turbulence and rapidly drive the halo out of hydrostatic equilibrium. Instead, the halo evolves toward a dynamical equilibrium in which non-thermal turbulent and ram pressure combine with the usual thermal pressure to hold the CGM up against gravity. The simulations of \citet{Oppenheimer2018} also showed the importance non-thermal pressure support of the CGM. \citet{Salem2016}, \citet{Ji2020} and \citet{Butsky2020} found that cosmic rays are also an important non-thermal supporting pressure in the CGM of simulated galaxies. Clearly, the structure of the hot phase, which is so important to models of observed and simulated cold CGM gas at the galaxy halo scale, warrants further investigation beyond a simple assumption of hydrostatic equilibrium at $\tvir$.

In this paper, we apply a virial analysis to simulated galaxy halos from the Figuring Out Gas \& Galaxies In Enzo (FOGGIE) cosmological zoom-in simulations to quantify when and where $\sim L^*$ galaxy halos are in virial equilibrium. We find that dynamic gas motions drive the temperature of the diffuse hot halo below the classical $\tvir$ by a factor of $\sim2$, even when the halo is in or close to virial equilibrium. We derive a ``modified" virial temperature, which adds explicit treatment of bulk gas motions to the classical definition of $\tvir$. This modified virial temperature more accurately describes the temperature of gas in the outskirts of the $\sim L^\star$ FOGGIE galaxy halos. A cooler than expected ``hot" halo has significant implications on the thermal pressure and cooling rates of the gas as well as on inferences made from UV absorption line and X-ray emission CGM observations.

Section~\ref{sec:derivation} provides the derivation of the modified virial temperature and explains how it differs from the standard virial temperature. Section~\ref{sec:FOGGIE} describes the FOGGIE simulations and the basic analysis we use throughout the paper. Section~\ref{sec:virial} presents how we assess the virial equilibrium of the FOGGIE halos (\S\ref{subsec:assessing}) and the results of this assessment (\S\ref{subsec:virial_results}). Section~\ref{sec:Tmod} describes how we calculate the modified virial temperature in the FOGGIE simulations (\S\ref{subsec:calc_Tmod}) and shows that the modified virial temperature accurately describes the simulated halo gas when the halo is in virial equilibrium (\S\ref{subsec:T_results}). Section~\ref{sec:xcorr} explores the impact of strong bursts of star formation feedback on the energy and temperature of the CGM. In Section~\ref{sec:implications}, we describe the implications of a lower temperature for the thermal pressure of the CGM (\S\ref{subsec:th_pres}), the CGM cooling time (\S\ref{subsec:tcool}), CGM mass estimates from X-ray observations (\S\ref{subsec:xray}), the origin of the \ion{O}{6} ion (\S\ref{subsec:OVI}), and the importance of turbulence to the CGM (\S\ref{subsec:turb}). We conclude in Section~\ref{sec:summary}. Appendix~\ref{sec:Rvir_def} shows our results do not depend on the precise definition of the virial radius or the virial theorem.

\section{Deriving Virial Temperature}
\label{sec:derivation}

In the standard paradigm of galaxy formation, the gaseous halo bound to a galaxy is virialized within the potential well of the dark matter halo such that
\begin{equation}
    \mathrm{PE} - \Sigma = -2\mathrm{KE}, \label{eq:virial}
\end{equation}
where PE is the potential energy of the galaxy and its dark matter halo, KE is the kinetic energy of the halo gas, and $\Sigma$ is a boundary pressure working with gravity to confine the halo gas from the outside. Gas falling into the halo is heated by passing through a virial shock at roughly the virial radius, so it is assumed that the kinetic energy of gas infall is completely thermalized into a thermal energy, $\KEth$, which is given by
\begin{equation}
    \KEth = \frac{3}{2}\frac{k_B T}{\mu m_p}, \label{eq:TE}
\end{equation}
where $T$ is the temperature, $\mu=0.6$ is the molecular weight per particle for fully ionized gas of primarily primordial composition, $k_B$ is the Boltzmann constant and $m_p$ is the mass of the proton. Note that Equation~\ref{eq:TE} gives the specific thermal energy of the gas, which is the energy per unit gas mass. Through the virial equation (Eq.~\ref{eq:virial}), the (specific) potential energy of the gas is thus directly related to the temperature of the gas, and this temperature is defined as virial temperature $\tvir$ \citep[e.g.,][]{Mo2010}:
\begin{equation}
    \tvir=\frac{1}{2}\frac{\mu m_p}{k_B}\frac{G\mvir}{\rvir}, \label{eq:Tvir}
\end{equation}
where $\mvir$ and $\rvir$ are the halo virial mass and radius, respectively, and $G$ is the gravitational constant. This definition of the virial temperature assumes the halo gas can be adequately described by a singular isothermal sphere density profile. In general, this may not be applicable for real galactic halos, but we show in Appendix~\ref{sec:SIS} that this approximation holds reasonably well in the outskirts of our simulated halos, near $\rvir$.

Throughout this paper, we define $\rvir$ as the radius enclosing an overdensity 200 times the critical density of the universe, which evolves with redshift (although we always use an overdensity factor of 200, regardless of redshift). We show in Appendix~\ref{sec:Rvir_def} that our results are insensitive to the exact choice of overdensity in the definition of virial radius, and so robust against inconsistent practice for this quantity in the existing literature. Equation~(\ref{eq:Tvir}) assumes the specific potential energy of gas in a dark matter halo at the virial radius is described by
\begin{equation}
    \mathrm{PE} = -\frac{G\mvir}{\rvir} \label{eq:PE}
\end{equation}
and the boundary pressure term is described by
\begin{equation}
    \Sigma=\frac{1}{2}\frac{G\mvir}{\rvir}. \label{eq:Sigma}
\end{equation}
Again, both equations~\ref{eq:PE} and~\ref{eq:Sigma} assume a singular isothermal sphere halo gas density profile. See Appendix~\ref{sec:SIS} for explicit calculation of these terms without assuming a particular gas density profile.

The definition of virial temperature (\ref{eq:Tvir}) makes a strong, deeply embedded assumption about the energy partition in gas-filled dark-matter halos: that all the potential energy of gas flowing into the halo is fully thermalized into internal thermal energy and that gas turbulence and bulk flows contribute nothing, by definition, to the overall energy of the halo gas when the system is in virial equilibrium. To explore the consequences of explicitly including non-thermal motions (such as turbulence and bulk flows) in the energy partition of the halo, we rewrite the virial equation (Eq.~\ref{eq:virial}) to explicitly include kinetic energy from both thermal and non-thermal motions:
\begin{equation}
    \mathrm{PE}-\Sigma=-2(\KEth+\KEnt)
\end{equation}
where PE is still given by Equation~\ref{eq:PE}, the thermal kinetic energy $\KEth$ is given by Equation~\ref{eq:TE}, and the kinetic energy due to non-thermal gas motions is $\KEnt$. Plugging these in and rewriting, we find a modification to the virial temperature, $\tmod$, that explicitly includes non-thermal gas motions, given by: 
\begin{equation}
    \tmod=\frac{1}{2}\frac{\mu m_p}{k_B}\frac{G\mvir}{\rvir} - \frac{2}{3}\frac{\mu m_p}{k_B}\KEnt \label{eq:Tmod}
\end{equation}
or
\begin{equation}
    \tmod=\tvir - \frac{2}{3}\frac{\mu m_p}{k_B}\KEnt. \label{eq:Tvirmod}
\end{equation}

Both $\tvir$ and $\tmod$ assume the virial equation (Eq.~\ref{eq:virial}) holds. A halo in perfect virial equilibrium does not contain any sources or sinks of energy --- the gas can only transfer energy between its potential and kinetic energies. Star formation, feedback, and radiative cooling provide sources and sinks of energy in the halo that can drive a departure from virial equilibrium. Therefore, we expect the virial temperature (either the standard $\tvir$ or the modified $\tmod$) to be a good descriptor of halo gas only far from the galaxy where these sources and sinks operate, at galactocentric radii $\gtrsim0.5\rvir$. By comparing the thermal energy of the gas in the FOGGIE halos (see \S\ref{sec:FOGGIE} below) to its radiative cooling rate, we confirm that the cooling time in the outskirts of the halos we study here is longer than the Hubble time and radiative energy losses are small compared to thermal and kinetic energy of the gas, and thus we neglect energy sinks in the outskirts of the halos. However, we note that the radiative cooling is not entirely negligible, and neglecting it is a caveat to the virial equilibrium argument at a $\sim10\%$ level (see Appendix~\ref{sec:cooling}). In detail, there may be events in a galaxy's history that lead to temporary departures from virial equilibrium even near the virial radius: mergers may lead to especially strong bursts of star formation feedback that may unbind a portion of the halo's gas. Likewise, there may be spatially distinct parts of the halo that do not participate in the overall balance of virial equilibrium, such as cosmological filaments that can pierce inward through the virial shock without being heated \citep[e.g.][]{Bennett2020} or strong outflows faster than the escape velocity of the halo (see \S\ref{subsec:calc_Tmod}). \citet{Fielding2020b} analyzed the properties of the CGM across many idealized and cosmological simulations, finding that the properties of the outer CGM ($r\gtrsim0.5\rvir$) tend to be set by cosmological structure formation whereas the properties of the inner CGM ($r\lesssim0.5\rvir$) tend to be set by feedback processes in the central galaxy \citep[a result also corroborated by][]{Stern2020b}, validating our choice to focus on the outer CGM.

We note that neither $\tvir$ nor $\tmod$ will capture all relevant physics that sets the temperature of the warm gas in a galaxy halo. Both are built on the assumption of a singular isothermal sphere profile, an assumption which only holds in the outskirts of the halo (see Appendix~\ref{sec:SIS}). If the warm gas is in hydrostatic equilibrium, its temperature must increase toward the center of the gravitational potential well (see Figure~\ref{fig:temp_radius} and \S\ref{subsec:T_results}), a property that is not captured by converting the potential energy near $\rvir$ into a single temperature through the virial theorem. In this paper, we work within the limitations of the concept of virial temperature to describe the warm halo gas temperature in the outskirts of simulated galaxy halos. Our goal is to formulate a more accurate estimate of the widely-used virial temperature through the concept of energy budget accounting within the halo gas, and ensure our findings are still as intuitive as the concept behind the standard $\tvir$.

\begin{figure}
    \centering
    \includegraphics[width=\linewidth]{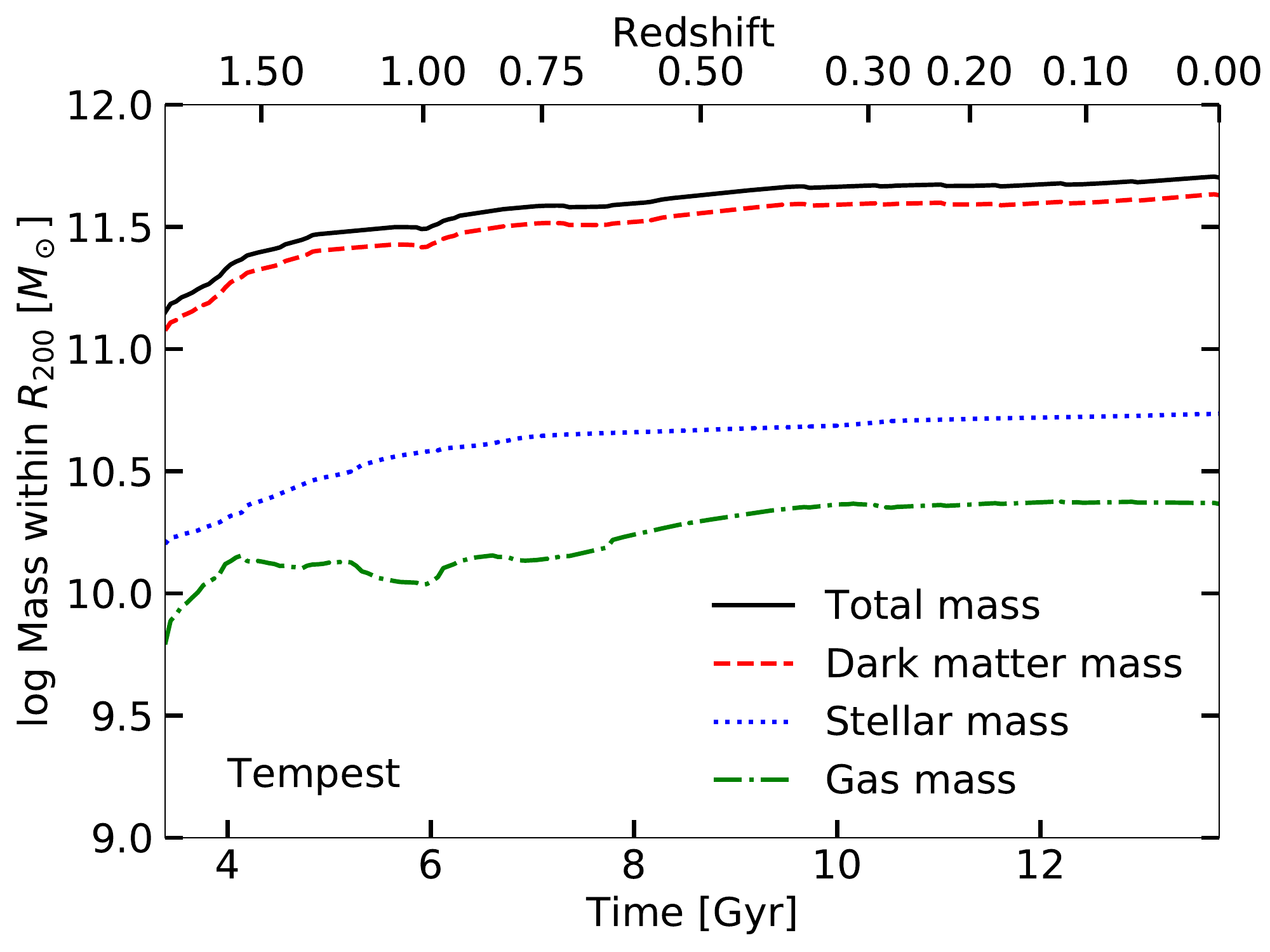}
    \includegraphics[width=\linewidth]{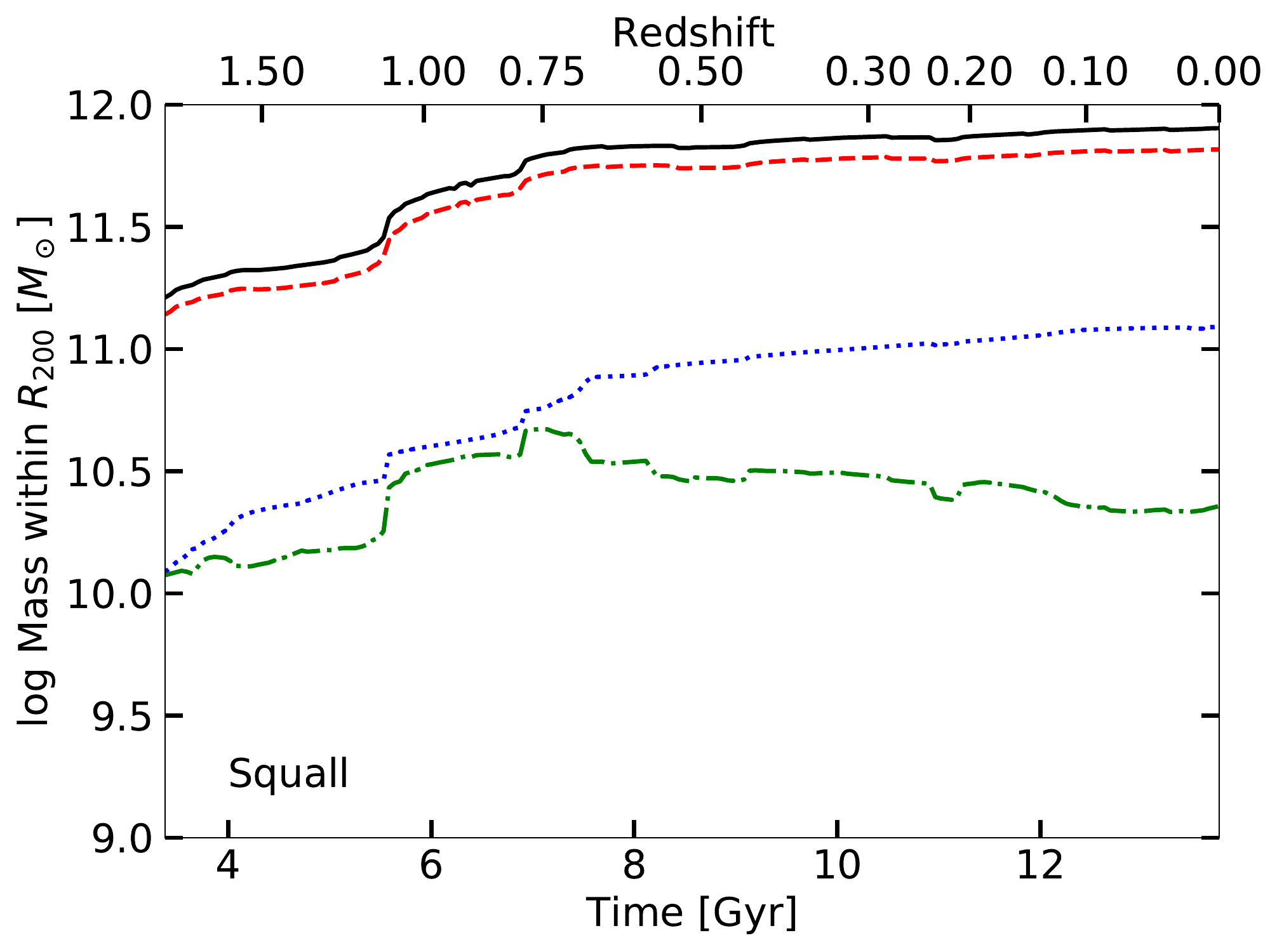}
    \includegraphics[width=\linewidth]{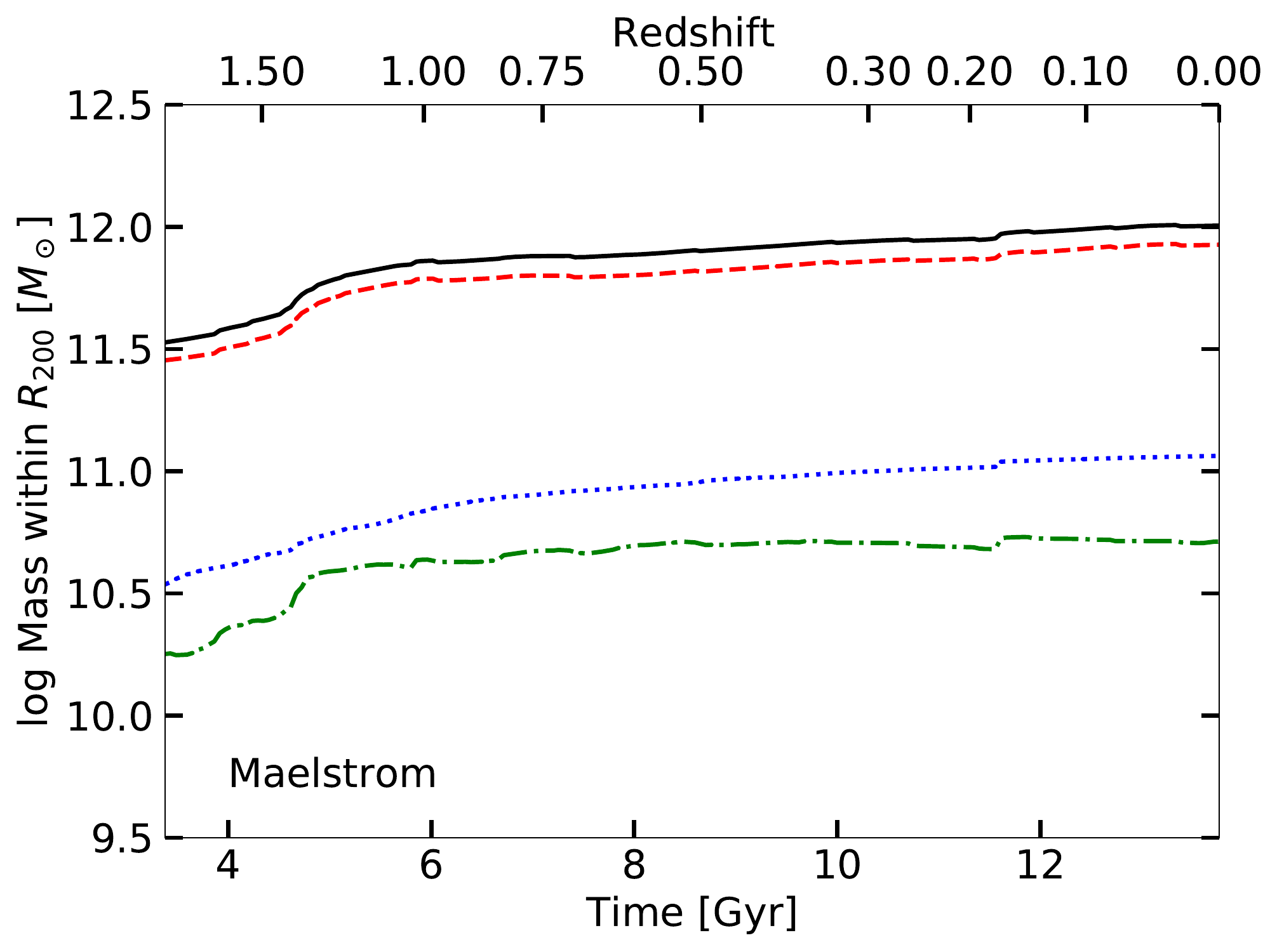}
    \caption{The total (black solid), dark matter (red dashed), stellar (blue dotted) and gaseous (green dash-dotted) masses within $\rvir$ for the three galaxies from the FOGGIE suite considered in this work, arranged from least massive to most massive at $z=0$ top to bottom.}
    \label{fig:masses}
\end{figure}

\section{The Simulations: Figuring Out Gas \& Galaxies In Enzo}
\label{sec:FOGGIE}

To explore energy partition in realistic halo simulations and assess the viabilty of the modified virial temperature for characterizing the bulk properties of the CGM, we use simulations from the Figuring Out Gas \& Galaxies In Enzo (FOGGIE) project. These simulations are described fully in the previous papers FOGGIE I -- IV \citep{Peeples2019,Corlies2020,Zheng2020,Simons2020}, but we briefly describe the relevant parts here for convenience.

\begin{table}
    \centering
    \begin{tabular}{l r r r}
        Property & Tempest & Squall & Maelstrom \\ \hline
        $^\mathrm{a}R_{200}$ [kpc] & 168.3\phantom{0} & 195.93 & 211.87 \\
        $^\mathrm{b}M_{200}$ [$10^{11}M_\odot$] & 5.04 & 8.02 & 10.12 \\
        $^\mathrm{c}M_\mathrm{DM,200}$ [$10^{11}M_\odot$] & 4.26 & 6.56 & 8.45 \\
        $^\mathrm{d}M_{\star,200}$ [$10^{10}M_\odot$] & 5.44 & 12.34 & 11.55 \\
        $^\mathrm{e}M_\mathrm{gas,200}$ [$10^{10}M_\odot$] & 2.33 & 2.28 & 5.15 \\
    \end{tabular}
    \caption{Properties of the three FOGGIE halos studied in this paper at $z=0$.\\$^\mathrm{a}$ Radius enclosing an average density of $200\times$ the critical density of the universe at $z=0$.\\$^\mathrm{b}$ Total mass enclosed within $\rvir$.\\$^\mathrm{c}$ Dark matter mass enclosed within $\rvir$.\\$^\mathrm{d}$ Stellar mass enclosed within $\rvir$. Includes satellites.\\$^\mathrm{e}$ Gas mass enclosed within $\rvir$. Includes ISM of central and satellites.}
    \label{tab:halo_props}
\end{table}

FOGGIE is run using the adaptive mesh refinement (AMR) code Enzo \citep{Bryan2014,BrummelSmith2019}\footnote{https://enzo-project.org}. As introduced in \citet{Simons2020}, six halos with roughly the Milky Way's present day total mass were selected from a cosmological volume 143.88 comoving Mpc on a side to be re-simulated in ``zoom-in" regions, where additional spatial refinement of at least 1.10 comoving kpc is forced in a box 287.77 comoving kpc on a side centered on the galaxy as it moves through the cosmological domain. Within this ``forced-refinement'' box, the resolution is refined further up to 274.44 comoving pc using an adaptive ``cooling refinement'' criterion in which one cell is replaced with 8 cells when the product of gas cooling time and sound speed is smaller the original cell size. The cooling refinement scheme places high resolution elements where they are needed most, in the high density and/or rapidly cooling cells, saving computational resources with less refinement in the hot and/or lowest density phases. However, the forced refinement region keeps the warm, diffuse gas resolved to a high level even in the absence of short cooling times, allowing detailed kinematics to be resolved and reducing the degree of artificial mixing frequently present in simulations with standard refinement schemes. In the outskirts of the halo that we focus on in this paper, the typical spatial resolution is set at the fixed, minimum refinement of the forced-refinement tracking box, where cells are 1.10 comoving kpc on a side, because there is not much gas there with very short cooling times.

The galaxies chosen to be simulated at high resolution have their last significant merger ($<$ 10:1 mass ratio) prior to $z = 2$, to be similar to the Milky Way merger history. This generally means they do not have strong bursts of star formation or feedback driving their halo gas significantly away from equilibrium at low redshifts (see \S\ref{subsec:virial_results}), making them an excellent choice for this study that depends on the halo gas being in or near virial equilibrium. At the time of this writing, three of the six FOGGIE galaxies have been run to $z = 0$, Tempest, Squall, and Maelstrom, so we focus on just these three in this paper (see \citeauthor{Simons2020} \citeyear{Simons2020} for more information on these halos). However, we expect our results to be generally applicable and not specific to the properties of these galaxies and their halos.

Figure~\ref{fig:masses} shows the build-up of gaseous, stellar, dark matter, and total masses within $\rvir$ in these three halos over the redshift range considered here, $z\sim2$ to $z=0$. Table~\ref{tab:halo_props} shows the final properties of each halo at $z=0$. By $z\sim 1$, each galaxy's total mass is above the threshold where a virial shock is expected to  form and remain stable, $M_h\sim {\rm few} \times 10^{11}M_\odot$ \citep{Birnboim2003}. Maelstrom, being the most massive of the three galaxies, surpasses this threshold by $z\sim1.5$. In general, the build-up of all types of mass within $\rvir$ becomes smooth and slowly increasing at late times. Squall is an exception because it continues to undergo gas-rich minor mergers at late times that drive the star formation rate up and lead to bursty changes in the stellar or gas masses within $\rvir$.

We select $\sim190$ snapshots in time between $z=2$ and $z=0$, separated by $\sim50$ Myr, for each of the halos. The FOGGIE runs are set up to output a snapshot every $\sim5$ Myr but for the bulk of our analysis (Sections~\ref{sec:virial} and~\ref{sec:Tmod}), we use only every tenth snapshot in time and perform all analysis on each of these snapshots. We divide up the outer CGM between $0.3\rvir$ and $1.3\rvir$ into 100 radial bins (of width $0.01\rvir$) to compute the properties of the CGM gas as functions of radius. In what follows, we take the radial bin $0.99\rvir < r <\rvir$ as the bin representing the gas near $\rvir$.\footnote{This shell of width $0.01\rvir$ contains $\sim200,000-300,000$ cells. We recalculated all results with a shell of width $0.1\rvir$ and found no qualitative difference in using shells of different widths except for smoother radial profiles, so we continue with the thinner shell.}

At low redshift, $\rvir$ for the FOGGIE halos examined here falls partially inside and partially outside of the forced refinement region. We test the impact of combining high and low resolution cells within a single spherical shell by re-calculating all results using only the high-resolution cells in the shell near $\rvir$ and find it has a minor quantitative and no qualitative effect on our results. The higher-resolution cells can better resolve the gas kinematics and thus contain somewhat more non-thermal kinetic energy overall than the low-resolution cells, which serves to somewhat \emph{strengthen} the difference between $\tvir$ and $\tmod$ and strengthen our qualitative conclusions. We proceed with using all cells within the shell at $\rvir$ rather than only the high-resolution cells, but note that perhaps a cosmological simulation with higher forced resolution than FOGGIE would find an even stronger difference between $\tvir$ and $\tmod$ than we report here. This result shows the importance of high resolution within the diffuse CGM gas.

\begin{figure*}
\centering
    \includegraphics[width=0.49\linewidth]{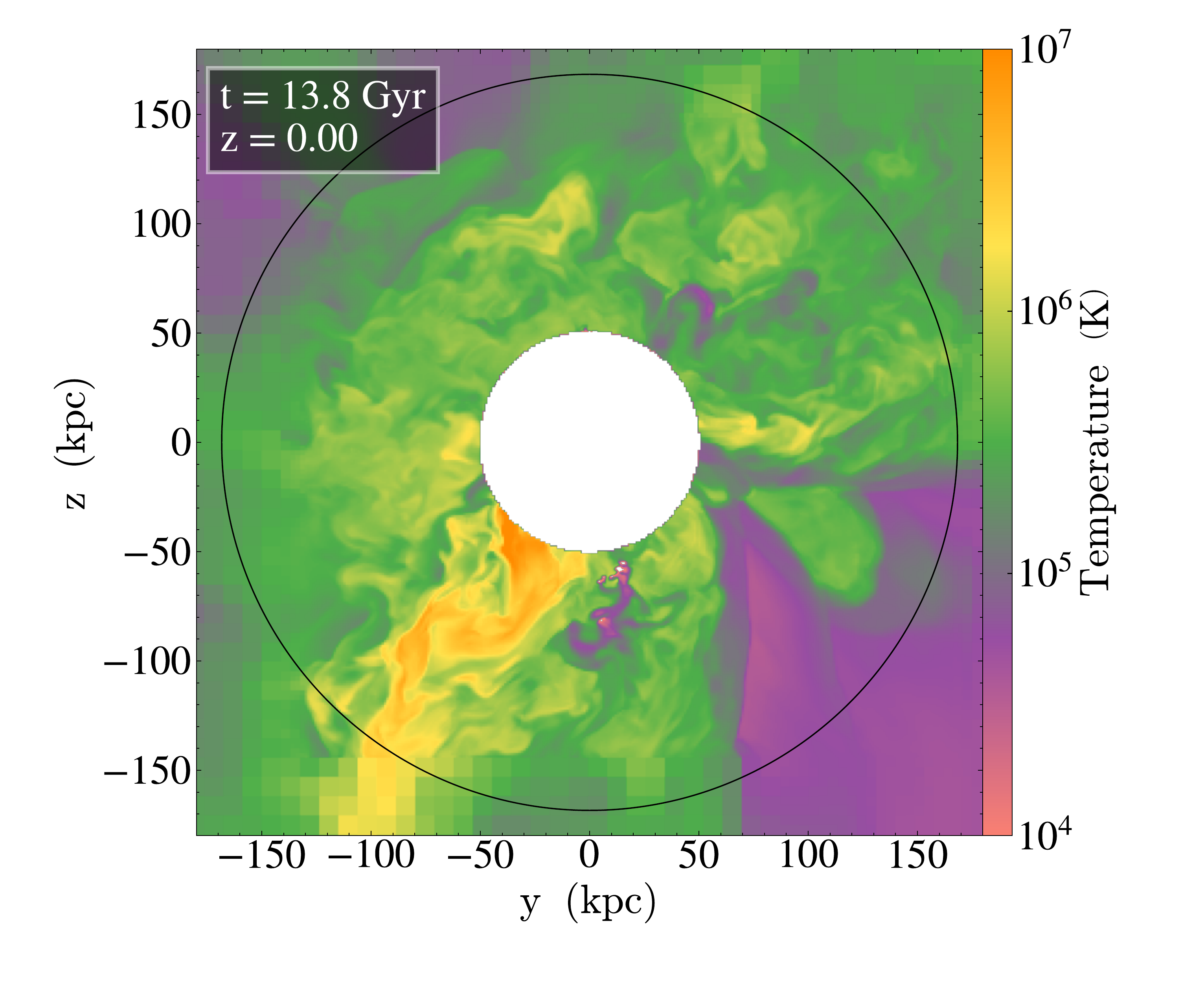}
    \includegraphics[width=0.49\linewidth]{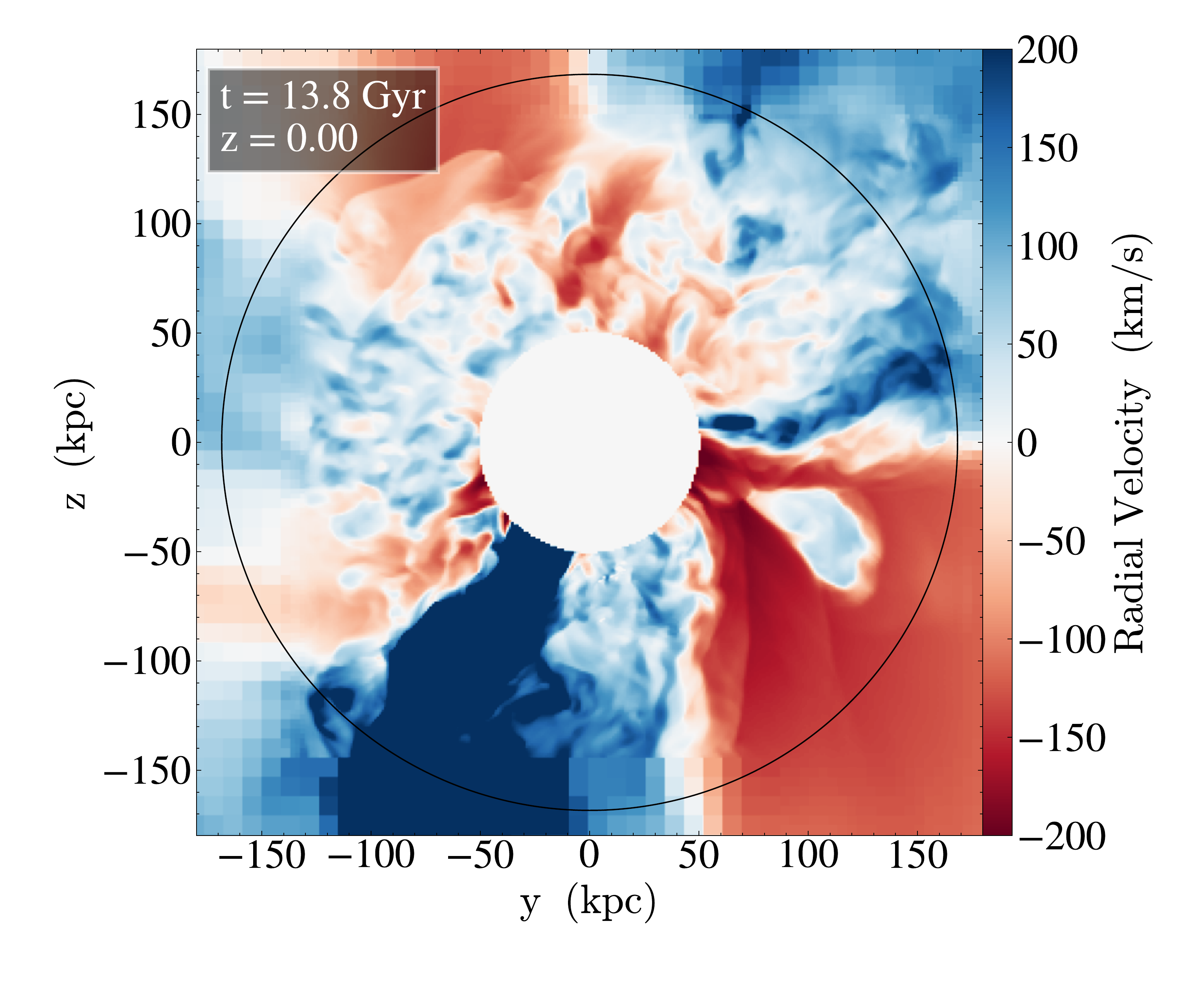}
    \includegraphics[width=0.49\linewidth]{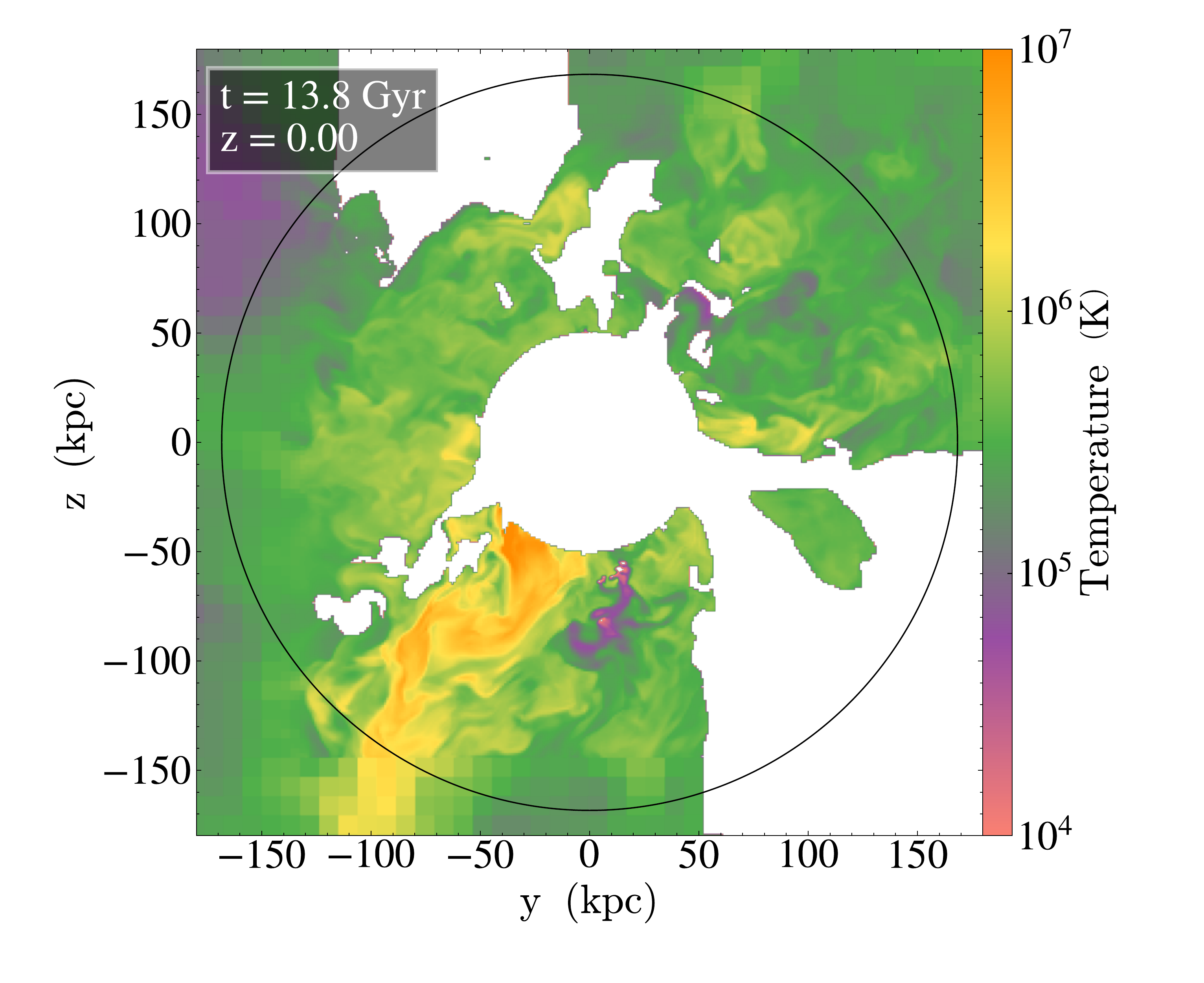}
    \includegraphics[width=0.49\linewidth]{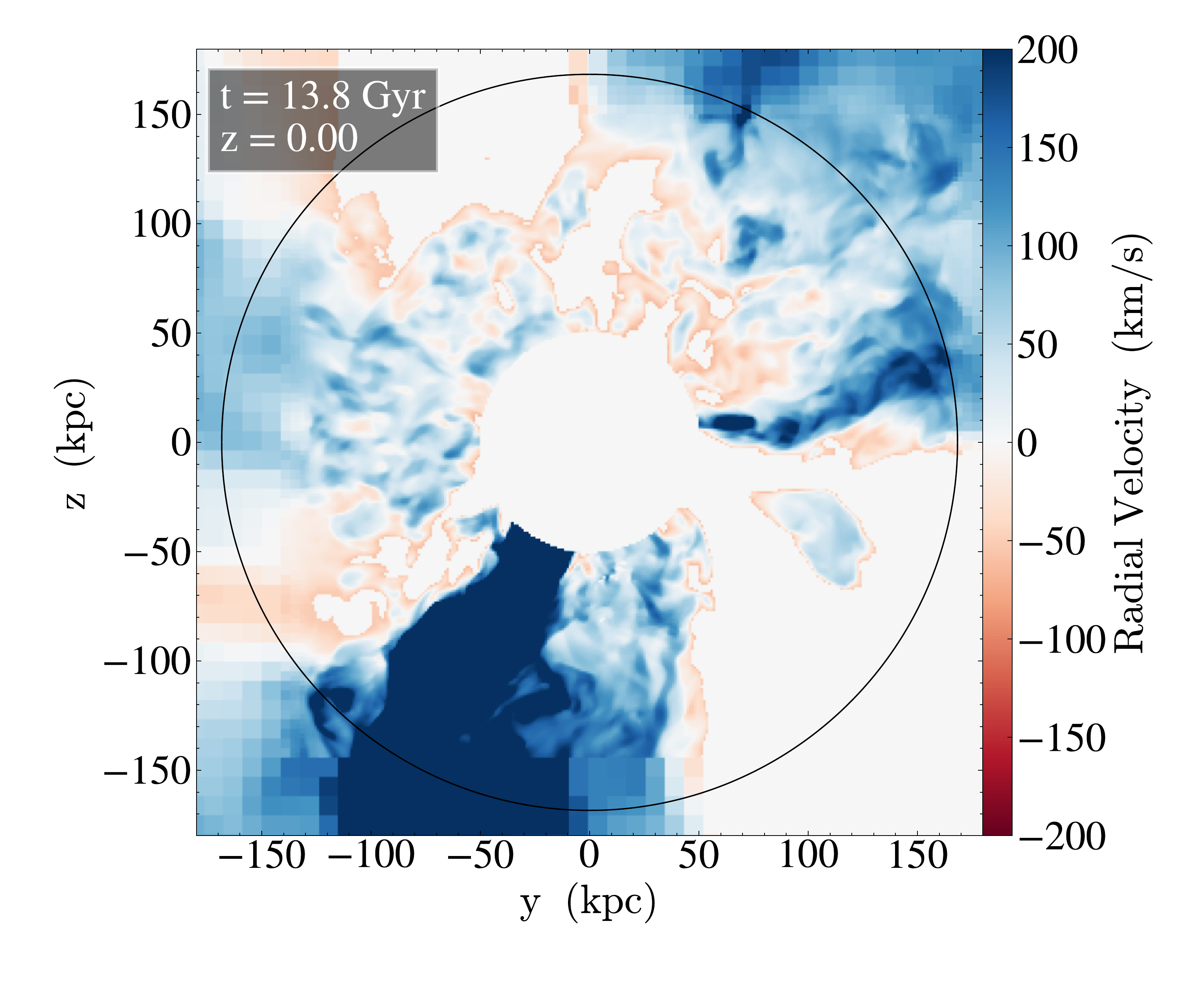}
    \caption{Slices of gas temperature (left) and radial velocity (right) in the Tempest halo at $z=0$ before and after cutting satellite ISM and filament gas (top and bottom panels, respectively). The black circle shows the location of $\rvir$. The satellite cut removes a negligible amount of mass and volume from the domain for this halo at this redshift, but the filament cut removes $\sim37\%$ of the gas mass and $\sim22\%$ of the volume between $0.3\rvir$ and $\rvir$.}
    \label{fig:slices}
\end{figure*}

\section{Virial Energy of the FOGGIE Halos}
\label{sec:virial}

Because both the standard virial temperature and the modified virial temperature are built on the assumption of virial equilibrium, we start by assessing when and where the FOGGIE halos are in virial equilibrium. Throughout the paper, when we say a halo is ``virialized", what we mean is that {\it the warm halo gas} has kinetic and potential energies that nearly or exactly satisfy Equation~\ref{eq:virial}. We focus exclusively on the warm halo gas because it is this component of the CGM for which we are interested in deriving a temperature. In principle, all components of the galaxy system --- including the dark matter, the stars in the disk, and the interstellar medium gas --- contribute their energies to the overall energy balance and virialization of the system. The standard definition of virial temperature (Equation~\ref{eq:Tvir}) assumes the warm halo gas is {\it virialized into} the gravitational potential well set by the dark matter halo, and thus separates out just this part of the system. We note that an energy budget accounting of the entire galactic halo system would nearly always produce a virialized system, even if the energy accounting of the warm halo gas alone does not strictly satisfy equation~\ref{eq:virial}. In this way, we use the term ``virialized" to refer to an energy budget accounting for the warm halo gas alone, rather than as a statement about the equilibrium for the entire system.

Note, as well, the difference between ``virialized" and ``thermalized." If the warm halo gas satisfies Equation~\ref{eq:virial}, it is {\it virialized}. However, that does not imply that all kinetic energy of the gas is in the form of thermal kinetic energy, that is, it does not imply the gas is fully {\it thermalized}. Indeed, we will show below that, in our simulations, only half of the halo gas's kinetic energy is thermal.

\subsection{Assessing Virial Equilibrium}
\label{subsec:assessing}

First, we remove the parts of the CGM that do not contribute to the warm halo gas that we focus on here: satellites and filaments. Satellites are excised from the domain by removing all gas cells with a density $>2\times10^{-26}$ g cm$^{-3}$ and temperature $<1.5\times10^4$ K. In some cases, this method does not perfectly remove all gas associated with a satellite, but it does eliminate confusion of satellite ISM gas with host halo CGM gas. To remove filaments, we excise all gas with an inward radial velocity faster than half of the local free-fall velocity, $v_\mathrm{ff}=r/\sqrt{3\pi/(32G\rho)}$, where $r$ is the galactocentric radius of the gas parcel, $G$ is the gravitational constant, $\rho=\frac{3}{4\pi}\frac{M_\mathrm{enc}(<r)}{r^3}$ is the average mass density within $r$, and $M_\mathrm{enc}(<r)$ is the total mass (dark matter, gas, and stars) contained within radius $r$. We chose to remove gas with inward radial velocities greater than $\frac{1}{2}v_\mathrm{ff}$ because some inflow filaments have enough tangential motion that their radial velocities are not exactly $v_\mathrm{ff}$, and we found that a fraction of $\frac{1}{2}$ removes most filament contamination. Again, this cut does not perfectly remove all filament gas, but it removes enough that the filament contamination to virialized CGM gas is small in most cases.

Outflows, in the form of galactic winds launched from the central galaxy and any of its satellites, have a different source than the warm, volume-filling halo gas and are unlikely to have an energy budget that follows the virial theorem. However, outflows have a range of velocities, making them difficult to select and remove cleanly, and they may also mix into the ambient CGM to become part of the virialized halo gas. Rather than attempt to remove coherent structures of outflows like we do with satellites and filaments, we do not consider any gas with an outward radial velocity greater than the escape velocity at its location in the dark matter halo to be contributing to the virial balance of the halo gas. Near the virial radius, very little gas is moving fast enough to escape the halo at times when the halo gas is in virial equilibrium (see Figure~\ref{fig:vel_dist} and discussion in \S\ref{subsec:calc_Tmod}).

Figure~\ref{fig:slices} shows temperature and radial velocity slices through the center of the Tempest halo at $z=0$, before and after the cuts to remove satellite ISM and filaments. In both cases, we remove the central $0.3\rvir$ to eliminate the galaxy and extended disk. This particular snapshot of this halo does not have any satellites in the plane of this slice, but the filament cut removes an extended wedge-shaped filament in the bottom right of the panel, primarily gas with $T\lesssim10^5$ K and $v_r<-100$ km s$^{-1}$. For Tempest at $z=0$, the satellite cut removes $0.4\%$ of all gas mass between $0.3\rvir$ and $\rvir$ while the filament cut removes $37.4\%$ of all gas mass in the same region. The total gas mass between $0.3\rvir$ and $\rvir$ is $5.5\times10^9M_\odot$ before any cuts and is $3.45\times10^9M_\odot$ after removing satellites and filaments. By volume, the satellite cut removes $0.01\%$ of the volume between $0.3\rvir$ and $\rvir$ and the filament cut removes $21.6\%$ of the volume.

Both the standard and the modified virial temperatures assume the halo gas near $\rvir$ is in virial equilibrium, i.e. that Equation~(\ref{eq:virial}) is satisfied. We generally expect this to be true unless the halo is experiencing galaxy mergers or a strong burst of energy input in the form of feedback \citep{Fielding2020b,Stern2020b}. Rather than assuming the virial equation holds, we explicitly measure it within the FOGGIE halos.

We measure directly whether the gas at $\rvir$ is in virial equilibrium by summing the potential, kinetic, and thermal energies of the gas in a thin spherical shell\footnote{Formally, the virial theorem applies to all gas in the system, not just within a thin shell. However, the assumption of a singular isothermal sphere as the density profile implies that the energy of gas located within a shell will be in equilibrium if the entire system is in equilibrium. We use the shell definition of the virial energy throughout the paper as it allows us to focus on just the outskirts of the halo, but see Appendix~\ref{sec:SIS} for a comparison of shell virial energies to whole-system virial energies.}. We focus on the outer CGM, where we expect virial equilibrium to hold. We define the ``virial energy", VE, to be this sum:
\begin{equation}
    \mathrm{VE} = \mathrm{PE} - \Sigma + \sum2\left(\KEnt + \KEth\right) \label{eq:VE}
\end{equation}
where PE is given by Equation~(\ref{eq:PE}) multiplied by the gas mass in the shell, the boundary pressure term $\Sigma$ is given by Equation~(\ref{eq:Sigma}), and the thermal and non-thermal kinetic energies are obtained by direct sum over all cells in the spherical shell. We use the total energies, not the specific energies as we did in Eqs.~(\ref{eq:virial}) through Eq.~(\ref{eq:Tvirmod}), making Equation~(\ref{eq:VE}) a true measurement of the total energies of the gas contained within the shell. If the gas satisfies virial equilibrium, then VE $=0$.

\subsection{Halos are in Virial Equilibrium Only When Non-Thermal Kinetic Energy is Included}
\label{subsec:virial_results}

\begin{figure}
    \centering
    \includegraphics[width=\linewidth]{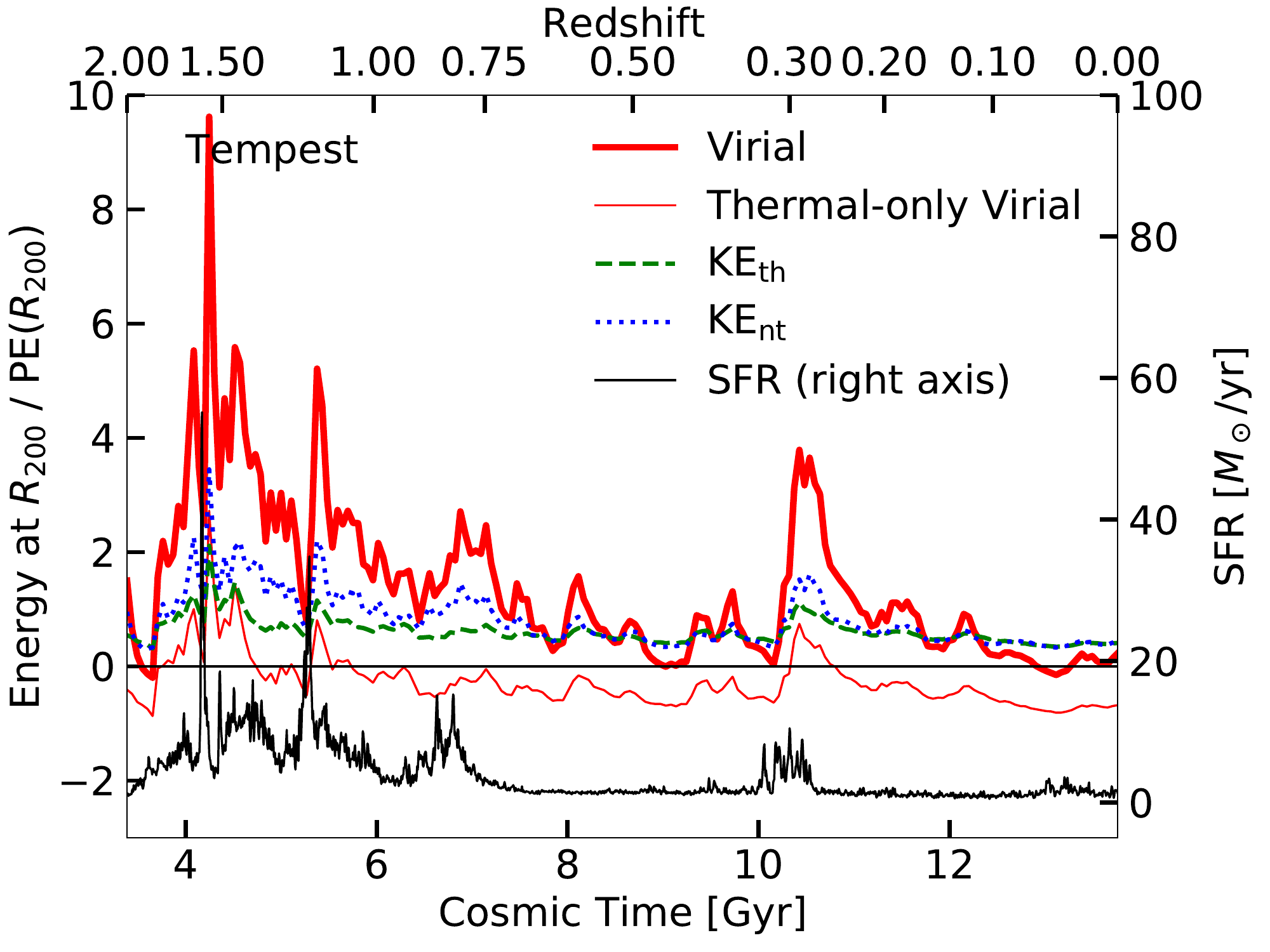}
    \includegraphics[width=\linewidth]{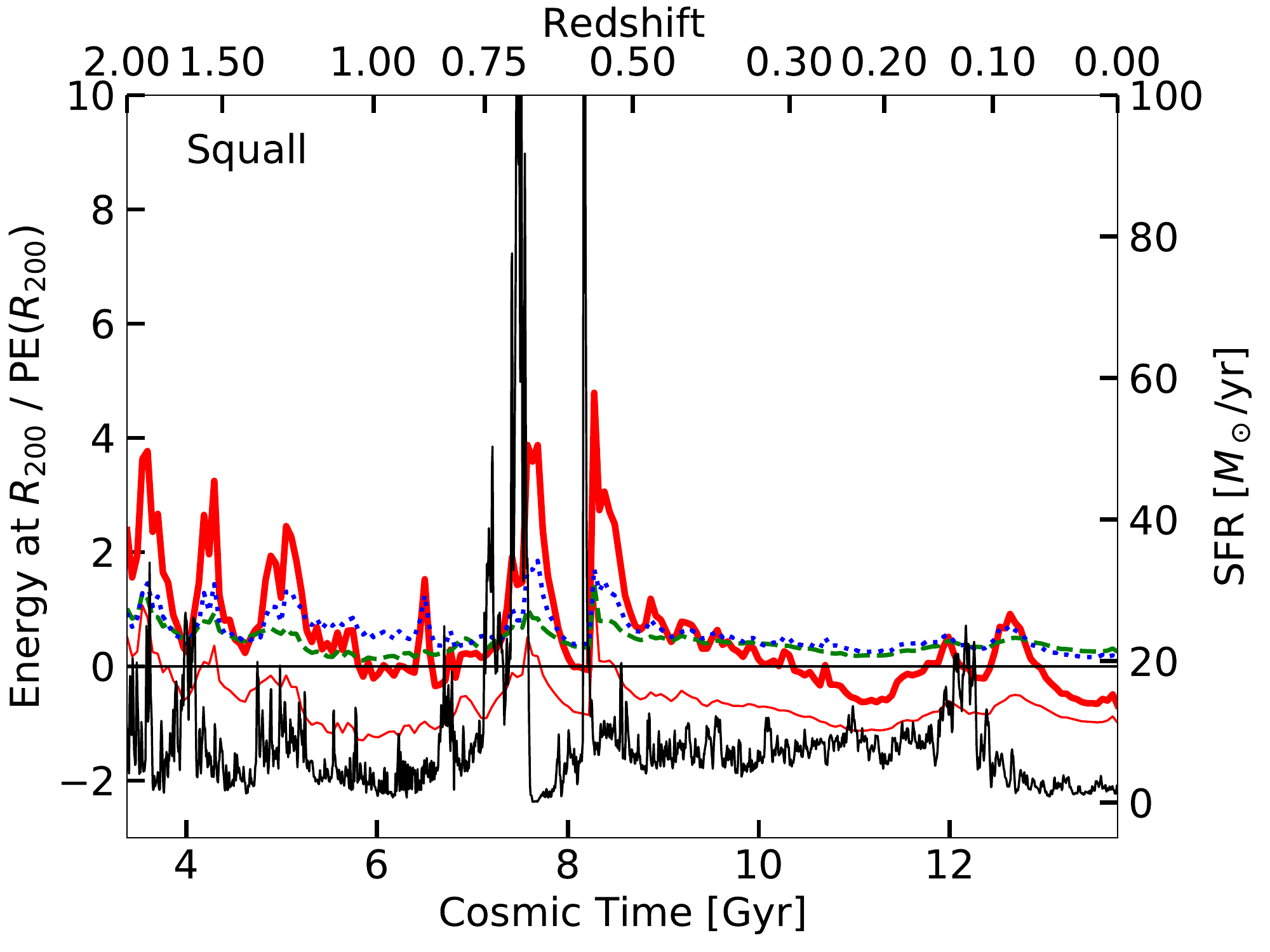}
    \includegraphics[width=\linewidth]{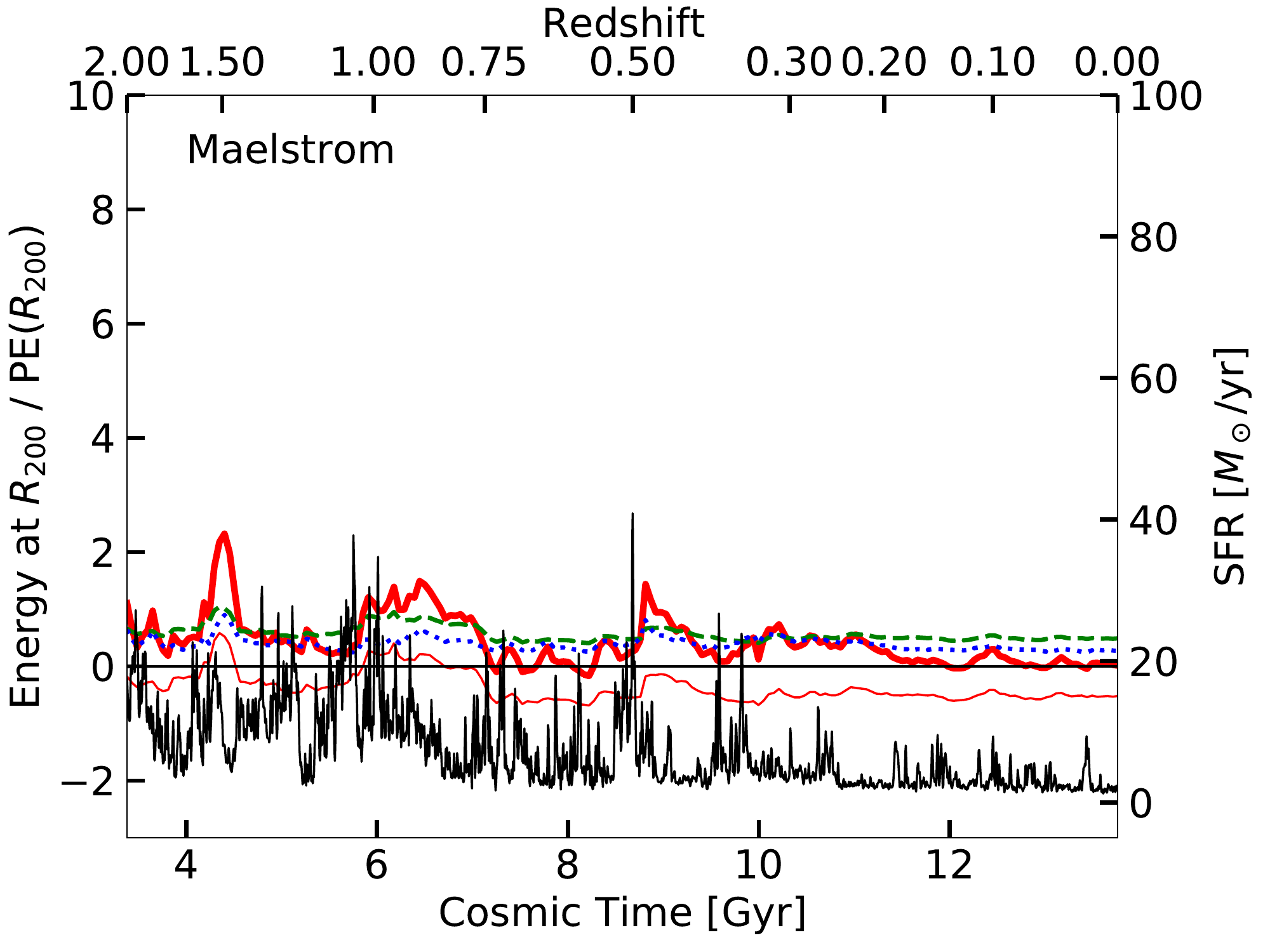}
    \caption{Energies of the gas within $0.99\rvir<r<\rvir$ as a function of cosmic time (bottom axis) and redshift (top axis). All energies are normalized by $\frac{G\mvir}{\rvir}$. The virial energy (Eq.~\ref{eq:VE}) is plotted as the thick, red line. The thermal kinetic energy of the gas is plotted as the green dashed line and the non-thermal kinetic energy is plotted as the blue dotted line. A thin solid red line indicates the virial energy of the gas near $\rvir$ if $\KEnt$ is neglected in the virial equation. The SFR of the central 20 kpc of the halo is shown as the thin black line, with values indicated on the right axis.}
    \label{fig:virial_time}
\end{figure}

Figure~\ref{fig:virial_time} shows the virial energy VE (thick red line) given by Equation~(\ref{eq:VE}) in a radial shell $0.99\rvir<r<1.0\rvir$ over cosmic time as the halos evolve from $z=2$ to $z=0$. None of the halos have gas near $\rvir$ in perfect virial equilibrium for extended periods of time; instead their VE oscillates as the halos evolve, approaching values near zero only at low redshift or during periods of low star formation rate (SFR). The SFR is plotted as the thin black line, with values marked on the right axis, and is calculated as all new stars formed since the previous time step within 20 kpc of the center of the halo. It is clear that there is a correlation between star formation bursts and when the gas near $\rvir$ is out of virial equilibrium, for example at $z\sim0.3$ or $z\sim1.5$ in Tempest (top panel). At lower redshift, the gas near $\rvir$ stabilizes and approaches virial equilibrium, but bursts of stellar feedback still drive the gas near $\rvir$ away from equilibrium temporarily (this direct cause-and-effect relationship will be discussed further below, see \S\ref{sec:xcorr}).

Tempest, being the lowest-mass halo, has its gas near $\rvir$ significantly perturbed away from equilibrium by relatively small bursts of star formation and does not approach steady virial equilibrium until rather late in its evolution, $z\lesssim0.2$. Squall is more massive (see Figure~\ref{fig:masses}), but has very strong bursts of star formation that drive its gas significantly out of equilibrium. Nonetheless, its gas approaches virial equilibrium somewhat earlier than Tempest, $z\lesssim0.4$. Maelstrom, the most massive halo of the three, has gas roughly in virial equilibrium (thick red line close to zero) throughout much of its evolution $z\lesssim0.75$, despite a significant number of star formation bursts. Maelstrom's SFR peaks to higher values, and more frequently, than Tempest's (at $z\lesssim1$), and yet the bursts do not drive its gas as far out of virial equilibrium as Tempest's bursts do. It seems the ability of strong feedback events to perturb gas near $\rvir$ is dependent on the mass of the halo.

\begin{figure}
    \centering
    \includegraphics[width=\linewidth]{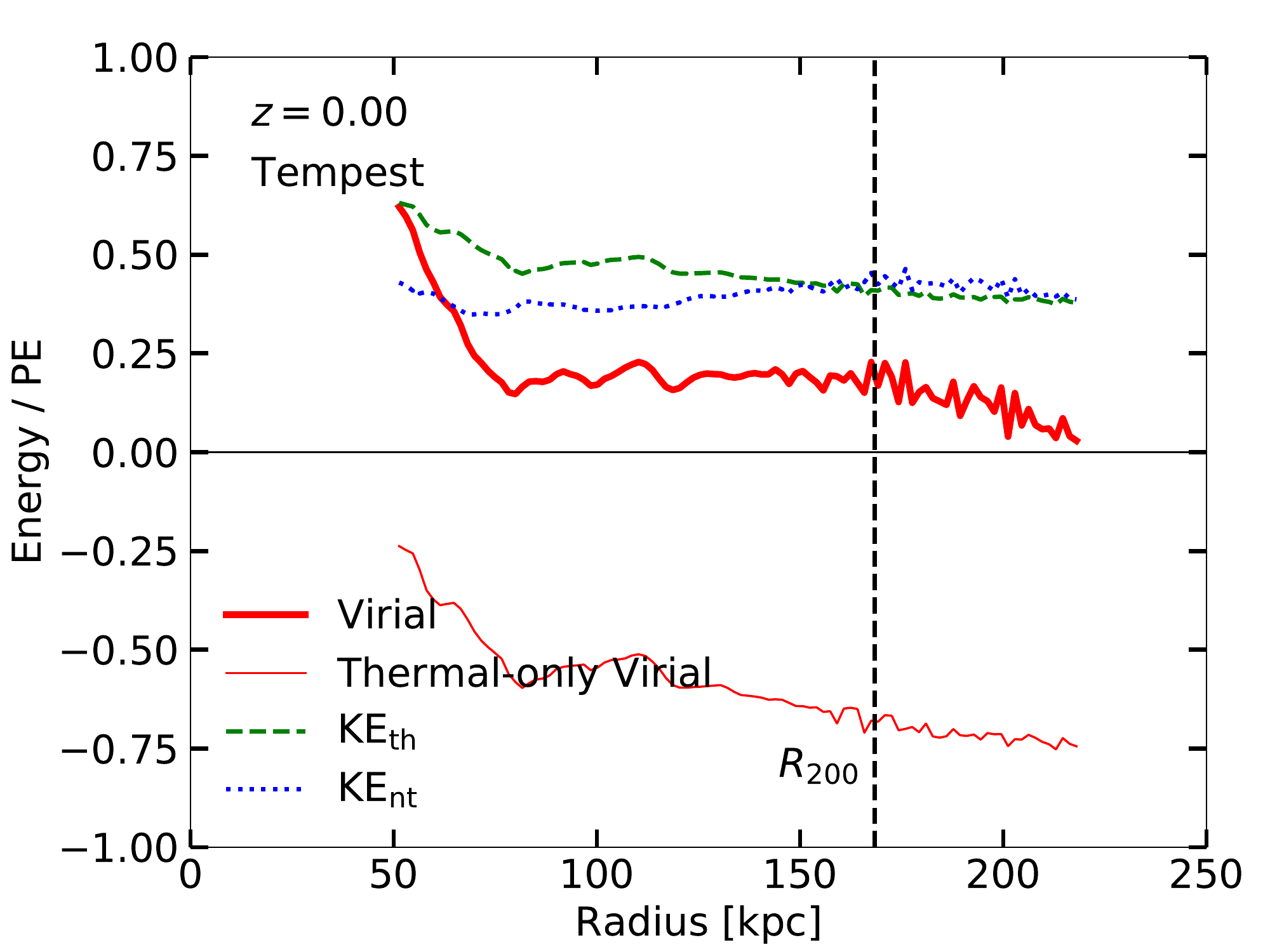}
    \includegraphics[width=\linewidth]{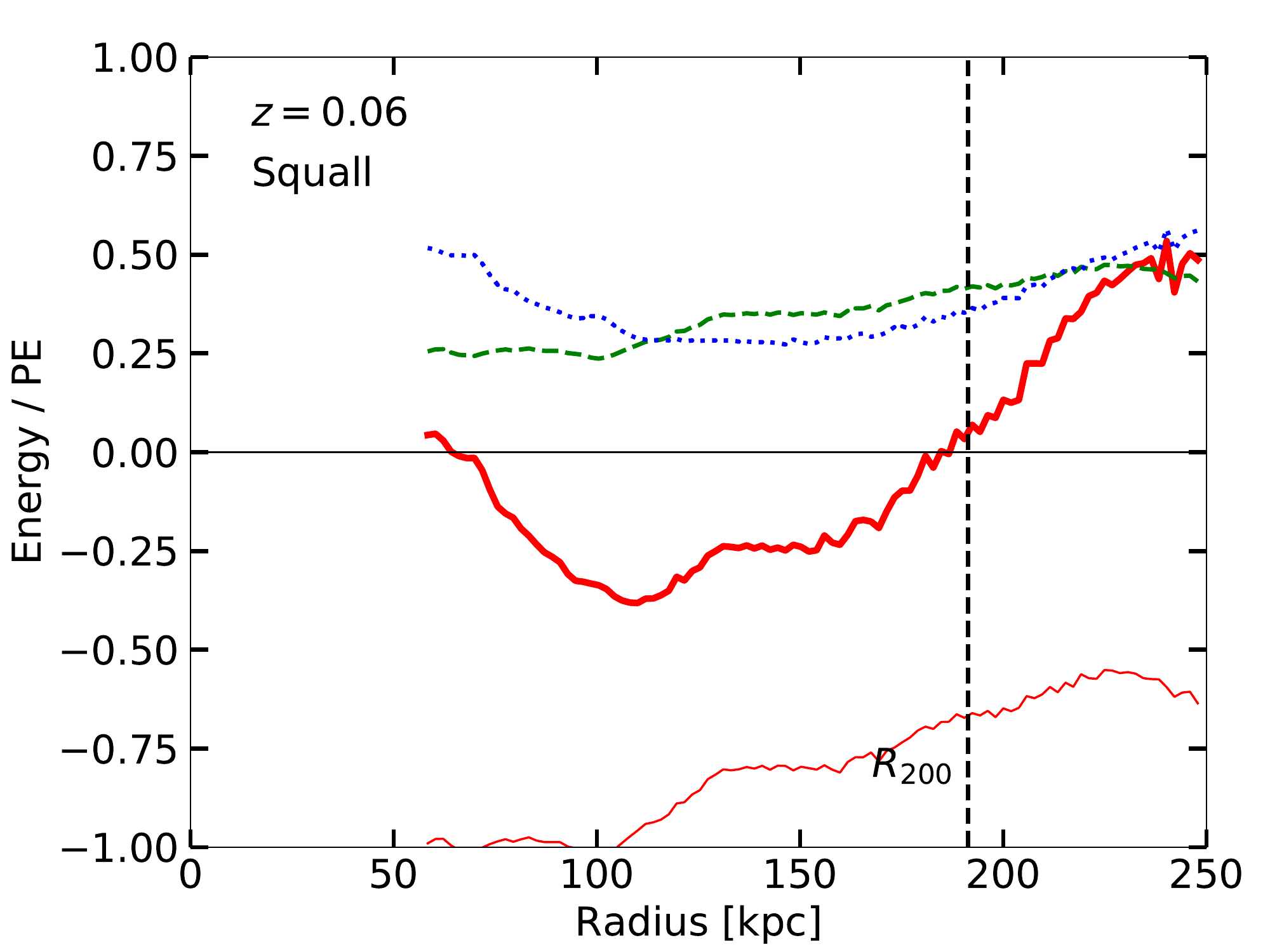}
    \includegraphics[width=\linewidth]{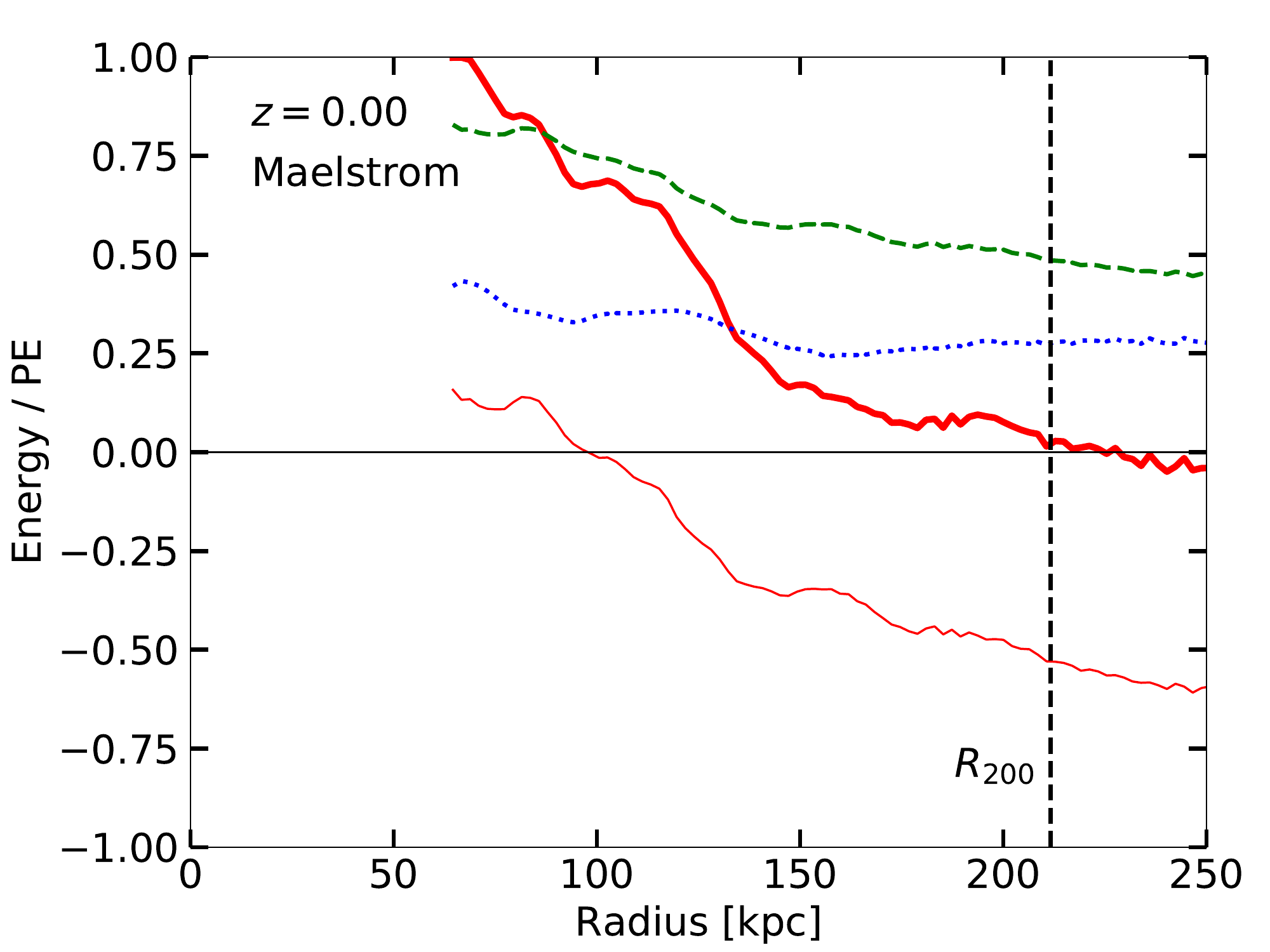}
    \caption{Energies of the gas within the halos, as in Figure~\ref{fig:virial_time}, as a function of distance from the center of the halo and normalized by $\frac{GM_\mathrm{enc}(r)}{r}$. The snapshots shown here are chosen as the latest times when the gas near $R_{200}$ is close to virial equilibrium (VE $\approx0$), marked in the panels for each halo. Squall is far from equilibrium at $z=0$ so we show the next closest time it is roughly in equilibrium, $z=0.06$. The other halos are roughly in equilibrium at $z=0$.}
    \label{fig:virial_radius}
\end{figure}

We expect the gas near $\rvir$ to be at the virial temperature (either $\tvir$ or $\tmod$) only when the gas near $\rvir$ satisfies VE $\sim0$. However, the gas near $\rvir$ is \emph{not} in perfect virial equilibrium throughout much of the halos' evolution --- instead, the gas's virial energy oscillates near zero and is perturbed by feedback events, especially at higher redshift.

The thin red line in Figure~\ref{fig:virial_time} shows the virial energy of the gas near $\rvir$ if the non-thermal kinetic energy of bulk flows is neglected, like in the standard definition of virial temperature. This curve falls below the VE $=0$ at all times other than following strong bursts of star formation for all halos. At late times, when the halos are massive enough to maintain virialized halos, neglecting the non-thermal kinetic energy in the energy balance of the halo would lead to the conclusion that the halo gas is under-virialized and under-supported and should be collapsing. It is only when the non-thermal kinetic energy is included that the halo gas can be said to be close to virialized (even if perfect virial equilibrium is not achieved long-term).

Figure~\ref{fig:virial_time} also shows the thermal and non-thermal kinetic energies of the gas near $\rvir$ as the green dashed and blue dotted curves, respectively. This figure illustrates our basic finding that the gas near $\rvir$ has roughly equal amounts of thermal and non-thermal kinetic energy at nearly all times. Shortly after strong bursts of feedback, both the thermal energy and the non-thermal kinetic energy increase, as feedback both heats and accelerates gas.

Figure~\ref{fig:virial_radius} shows the same energy components as in Figure~\ref{fig:virial_time}, but as a function of radius at a given snapshot in time, over the radius range $0.3\rvir$ to $1.3\rvir$ (note that the vertical scale differs from Figure~\ref{fig:virial_time}). Each halo's snapshot was chosen to reflect a time when the gas near $\rvir$ in each halo is roughly in virial equilibrium, which is $z=0$ for Tempest, $z=0.06$ for Squall, and $z=0$ for Maelstrom. The halo gas is closest to virial equilibrium for $r\gtrsim0.5\rvir$, and again we see that neglecting the non-thermal kinetic energy in the virial equation leads to a configuration that is far out of equilibrium. Feedback drives the gas away from equilibrium in the inner regions of each halo, and some residual feedback-driven outflows that traveled to the outer halo can push it out of virial equilibrium near $\rvir$ as well, as in the case of Squall (middle panel). Figure~\ref{fig:virial_radius} shows that at $z=0$ we expect any temperature derived from the virial equation to be a poor description of the gas within $r\lesssim0.5\rvir$, where the virial energy deviates strongly from zero. However, the virial energy is not exactly zero for most of the volume and time so we do not expect either $\tvir$ or $\tmod$ to be a perfect descriptor of the gas temperature.

In summary, we find that while the gas in the outskirts of the FOGGIE halos are rarely in perfect virial equilibrium (VE $=0$), their virial energies (Equation~\ref{eq:VE}) are close to zero for much of the later stages of their evolution, when they are massive enough to be expected to host a hot halo \citep[see discussion surrounding Figure~\ref{fig:masses} in \S\ref{sec:derivation} and][]{Birnboim2003}. Their normal state is to be close to VE near $\rvir$, except when strong bursts of star formation feedback temporarily drive the halo out of equilibrium, after which it settles back into an equilibrium state. Neglecting the non-thermal kinetic energy contribution to the overall energy balance of the halo would suggest the gas in these halos are further out of virial equilibrium than they really are, suggesting that the halo gas is in fact \emph{virialized} without being fully \emph{thermalized}. We have shown that the non-thermal kinetic energy is an important component of energy partition in halos and should not be neglected in analyses that rely on accurate characterization of their main properties.

\section{Modified Virial Temperature in FOGGIE Halos}
\label{sec:Tmod}

With an understanding of when (periods of low SFR) and where ($r\gtrsim0.5\rvir$) the CGM gas is in virial equilibrium and thus when and where we expect $\tmod$ to be a good descriptor of the gas temperature, we move on to calculating the modified virial temperature for the FOGGIE halos. We measure the distribution of gas temperatures at each radius in the simulations and compare to both $\tmod$ and $\tvir$ to determine if the modified virial temperature is a better descriptor of the mass-weighted peak gas temperature in the CGM than the standard virial temperature.

\subsection{Calculating Modified Virial Temperature}
\label{subsec:calc_Tmod}

\begin{figure*}
    \centering
    \includegraphics[width=\linewidth]{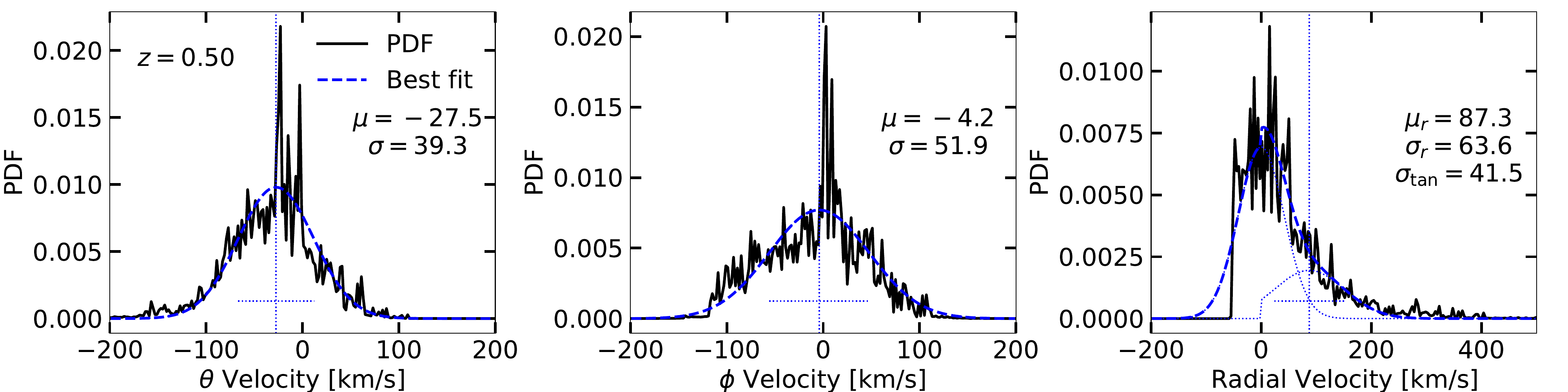}
    \includegraphics[width=\linewidth]{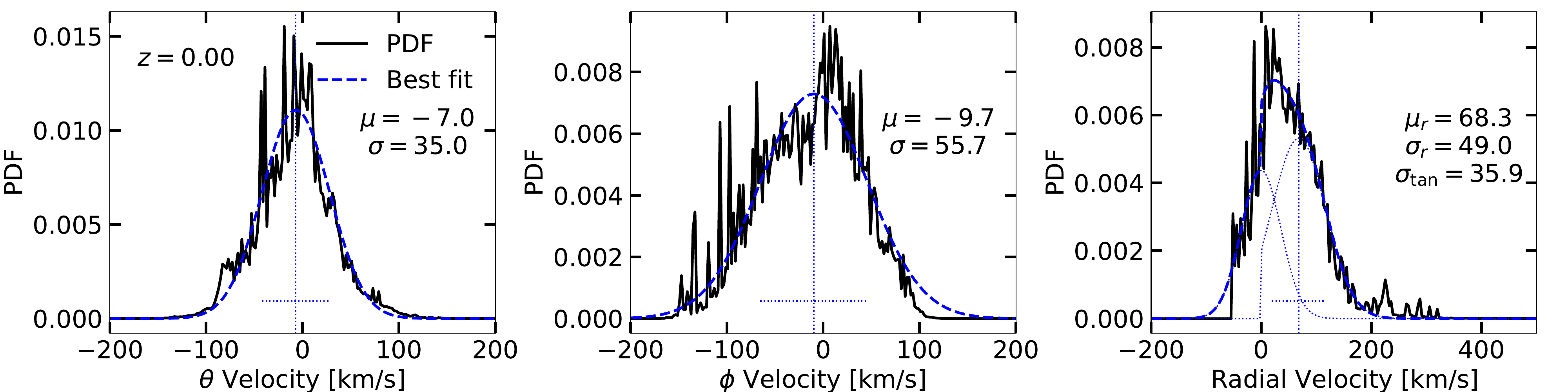}
    \caption{Mass-weighted distributions of velocities of gas within $0.99\rvir<r<\rvir$ are shown as the black solid lines, split into the three spherical directions: $\theta$ (left), $\phi$ (middle), and radial (right). The top row shows gas near $\rvir$ at $z=0.5$ and the bottom row shows gas near $\rvir$ at $z=0$, both in the Tempest halo. The best-fit Gaussian model (to the two tangential velocity directions) and the best-fit double-Gaussian model (to the radial velocity direction) are shown as the blue dashed lines. For the double-Gaussian model, the two Gaussians that make it up are shown with thin blue dashed lines. Dotted vertical lines show the location of the best-fit mean of the Gaussian and dotted  horizontal lines show $\pm1$ best-fit standard deviation. The best-fit values are printed in each panel.}
    \label{fig:vel_dist}
\end{figure*}

To compute $\tmod$ we need a measurement of the non-thermal kinetic energy $\KEnt$, and thus both the turbulent velocity and the bulk flow velocity of the gas. As gas velocities are tracked explicitly, cell-by-cell, at runtime, we could use the simulated velocity fields to obtain a measure of the non-thermal kinetic energy as in \S\ref{sec:virial}. However, rather than integrate all the cell-level data directly, we will use statistical descriptions of their distributions in velocity space within radial shells, as gas velocities are typically more accessible in CGM observations than the unknown sum of all kinetic energies. In doing so, we must be careful to consider coherent bulk flows, such as filaments and fast outflows, apart from localized turbulence or convective motions. 

First, we decompose the CGM velocity into spherical components: radial velocity $v_r$ and velocities $v_\theta$ and $v_\phi$ tangential to the radial direction (Figure~\ref{fig:vel_dist}). The tangential velocities are defined arbitrarily, \emph{not} relative to the disk of the galaxy. Near the virial radius, we find that any rotation in the filament-removed CGM gas is negligible at times when the halo is near virial equilibrium (i.e. at $z=0$ in the Tempest halo, Figure~\ref{fig:vel_dist}), so we do not include bulk CGM rotation in our accounting of the halo's non-thermal gas motions.\footnote{Observations and simulations alike have found the CGM to be rotating within $\sim0.5\rvir$ \citep{HodgesKluck2016,Ho2019,Martin2019,Ho2020}, although \citet{Oppenheimer2018} shows rotation is sub-dominant to other forms of non-thermal gas motions at the virial radius.} The spread of the tangential velocity distributions will thus be a good tracer for turbulent or convective non-thermal motions. We perform a least-squares fit of Gaussian distributions of the form
\begin{equation}
    f_\mathrm{tan}(v)=A\exp{\frac{-(v-\mu)^2}{2\sigma^2}}
\end{equation}
to the two tangential velocity distributions to obtain the peak velocity, $\mu$, and velocity dispersion, $\sigma$, of the two tangential velocity distributions (the amplitude $A$ is a free parameter that has no physical meaning in this case, as all of the velocity distributions being fit are normalized). This also allows us to confirm that the peak of these distributions are close to zero, indicating a small net rotation. The left and center panels of Figure~\ref{fig:vel_dist} show the two tangential velocity distributions with their best-fit Gaussian distributions for the gas near $\rvir$ in the Tempest halo at $z=0.5$ and $z=0$, as an example. At $z=0.5$, the gas does show some bulk rotation in the $\theta$ direction as indicated by a non-zero $\mu=-27$ km s$^{-1}$, but the rotation is not long-lived or coherent, generally appearing and disappearing from snapshot to snapshot as the halo evolves. By $z=0$, there is no significant rotation of the Tempest halo's gas near $\rvir$ in either the $\theta$ or $\phi$ directions, as indicated by $\mu\approx0$ for both tangential velocity directions.

The radial velocity distribution will have a contribution from turbulent motions, assuming turbulence in the CGM is isotropic, but it will also have contributions from galactic outflows that reach large distances. Thus, we do not expect the radial velocity distribution to be symmetric, nor necessarily peaked close to zero. To fit the radial velocity distribution, we perform a least-squares fit to the sum of two Gaussian distributions of the form
\begin{equation}
    f_r(v)=A_1\exp{\frac{-v^2}{2\sigma_\mathrm{tan}^2}} + A_2\exp{\frac{-(v-\mu_r)^2}{2\sigma_r^2}}
\end{equation}
where the first term in the sum represents the contribution due to turbulence and the second term represents the contribution due to bulk outflows. The turbulence Gaussian has its mean fixed to 0 km s$^{-1}$ and its standard deviation fixed to $\sigma_\mathrm{tan}$, where $\sigma_\mathrm{tan}$ is the best-fit standard deviation of a Gaussian fit to the tangential velocity distribution (not shown), given by $v_\mathrm{tan}=\sqrt{v_\theta^2+v_\phi^2}$. The outflow Gaussian's mean and standard deviation are unconstrained, but it is defined to be zero for $v<0$ km s$^{-1}$ so that it measures strictly the bulk outflow component of the radial velocity. Both Gaussians' amplitudes are unconstrained. The right panels in Figure~\ref{fig:vel_dist} show the radial velocity distribution along with the best-fit sum of Gaussians to the distribution. The cut to remove filaments removes all gas with $v_r\lesssim50$ km s$^{-1}$ (this value changes slightly with redshift), and galactic outflows can be seen as the Gaussian component shifted toward large positive velocities.

With the velocity distributions characterized, we can compute the non-thermal kinetic energy from either turbulence only or turbulence and bulk flows. Because we do not expect significant rotation of the CGM near the virial radius and the means of the best-fit Gaussians to the tangential velocity distributions are close to zero, we use only the standard deviation of the best-fit Gaussians to the tangential velocity distributions as a measure of turbulent velocity. Turbulence will always contribute to the virialization and non-thermal kinetic energy of the gas, but it is unclear how much, if at all, bulk outflows with velocities $<v_\mathrm{esc}$ in the radial velocity distribution contribute to the virialization of the halo. It could be that outflows produce a perturbation from virial equilibrium for the halo (like filaments and satellite galaxies) and thus should not be included in the derivation of the modified virial temperature, or it could be that outflows provide a necessary supporting force for the halo and thus should be included. 

Instead of attempting to characterize how much outflows contribute to the virialized, non-thermal kinetic energy, we define two ways of calculating the modified virial temperature: one with only turbulence, and one with both turbulence and bulk outflows. The total kinetic energy per mass (specific kinetic energy) of a velocity distribution is $\int \frac{1}{2}v^2f(v)\ \mathrm{d}v$ where $f(v)$ is the probability of a parcel of gas having a velocity between $v$ and $v+\mathrm{d}v$, normalized such that $\int f(v)\ \mathrm{d}v = 1$. For a Gaussian velocity distribution, like we find for the tangential velocity distributions, the specific kinetic energy is simply $\frac{1}{2}\sigma^2$ where $\sigma$ is the standard deviation of the Gaussian, the velocity dispersion. For the radial velocity distribution, we model $f(v)$ as the sum of two Gaussians, one of which cuts off for $v_r<0$ km s$^{-1}$, so we compute $\int \frac{1}{2}v_r^2f(v_r)\ \mathrm{d}v_r$ directly from the best-fit function, which includes both the radial-direction turbulent velocity distribution and the outflow velocity distribution. The specific kinetic energy of a velocity distribution that can be described as a three-dimensional Gaussian is given by $\mathrm{KE}=\frac{1}{2}(\sigma_\theta^2+\sigma_\phi^2+\sigma_r^2)$, where each $\sigma$ is the velocity dispersion in one of the three directions. In our case, we have measured velocity dispersions for the two tangential dimensions and assume that the velocity dispersion in the third, radial dimension can be described as the average of the tangential dispersions, $\sigma_r^2=\frac{1}{2}(\sigma_\theta^2+\sigma_\phi^2)$. The non-thermal kinetic energy due to  turbulence alone is then
\begin{equation}
    \KEnt^{\mathrm{turb}}=\frac{1}{2}\left(\frac{3}{2}\sigma_\theta^2+\frac{3}{2}\sigma_\phi^2\right) \label{eq:KEnt_turb}
\end{equation}
where $\sigma_\theta$ and $\sigma_\phi$ are the standard deviations of the best-fit Gaussians to the $\theta$ and $\phi$ velocity distributions, respectively. The non-thermal specific kinetic energy of both turbulence and bulk outflows is given by
\begin{equation}
    \KEnt^{\mathrm{turb+out}}=\frac{1}{2}\left(\sigma_\theta^2+\sigma_\phi^2\right) + \frac{1}{2}\int_{-0.5v_{\rm ff}}^{v_\mathrm{esc}} v_r^2 f(v_r)\ \mathrm{d}v_r \label{eq:KEnt_turb_out}
\end{equation}
where the lower bound of the integral reflects the cut made to remove the inflow filaments (see \S\ref{subsec:assessing}), and the upper bound at the escape velocity $v_\mathrm{esc}$ assumes that outflows that are fast enough to escape the halo do not contribute to the virialization of the halo. Note that the contribution of turbulence in the radial direction is included in $f(v_r)$ within the integral, so the factor of $3/2$ has been dropped from the first term in the sum.

The standard virial temperature $\tvir$ is a single temperature for all locations in the halo, by the singular isothermal sphere definition. However, for $\tmod$ we can derive a radius-dependent form by making some simple substitutions. In Equation~(\ref{eq:Tmod}), we replace $\mvir$ with $M_\mathrm{enc}(<r)$, the enclosed mass within a radius $r$, and replace $\rvir$ with $r$ (this is equivalent to replacing $\mvir$ and $\rvir$ in the potential energy term given by equation~\ref{eq:PE} with $M_\mathrm{enc}(<r)$ and $r$, respectively)\footnote{We allow the temperature $\tmod$ to vary although we still use the singular isothermal sphere assumption for the form of PE (equation~\ref{eq:PE}). In this case, the singular isothermal sphere assumption governs the gas density and halo gravitational potential profiles, rather than the temperature.}. We also measure $\sigma_\mathrm{\theta}$, $\sigma_\mathrm{\phi}$, and $\int \frac{1}{2}v_r^2f(v_r)\ \mathrm{d}v_r$ within radial bins. This gives
\begin{equation}
    \tmod^{\mathrm{turb}}=\frac{1}{2}\frac{\mu m_p}{k_B}\left[\frac{GM_\mathrm{enc}(<r)}{r} - \left(\sigma_\theta(r)^2+\sigma_\phi(r)^2\right)\right]. \label{eq:Tvirmod_r_turb}
\end{equation}
and
\begin{align}
    &\tmod^{\mathrm{turb+out}}=\frac{1}{2}\frac{\mu m_p}{k_B}\left[\frac{GM_\mathrm{enc}(<r)}{r} - \right. \nonumber \\ &\left.\frac{2}{3}\left(\sigma_\theta(r)^2+\sigma_\phi(r)^2 + \int_{-0.5v_{\rm ff}(r)}^{v_\mathrm{esc}} v_r^2 f(v_r,r)\ \mathrm{d}v_r \right)\right]. \label{eq:Tvirmod_r_turb_out}
\end{align}
In equations~(\ref{eq:Tvirmod_r_turb}) and~(\ref{eq:Tvirmod_r_turb_out}), we find $\sigma_{\theta}(r)$, $\sigma_{\phi}(r)$, and $f(v_r,r)$ by fitting these functional forms to the velocity distributions in bins of radius, from $0.3\rvir$ to $1.3\rvir$, of radial width $0.01\rvir$. We now have all the analytic tools we need to compare $\tvir$ and $\tmod$ as descriptions of simulated halo gas. 

\subsection{$\tmod$ Better Describes the CGM Temperature than $\tvir$}
\label{subsec:T_results}

\begin{figure}
    \centering
    \includegraphics[width=\linewidth]{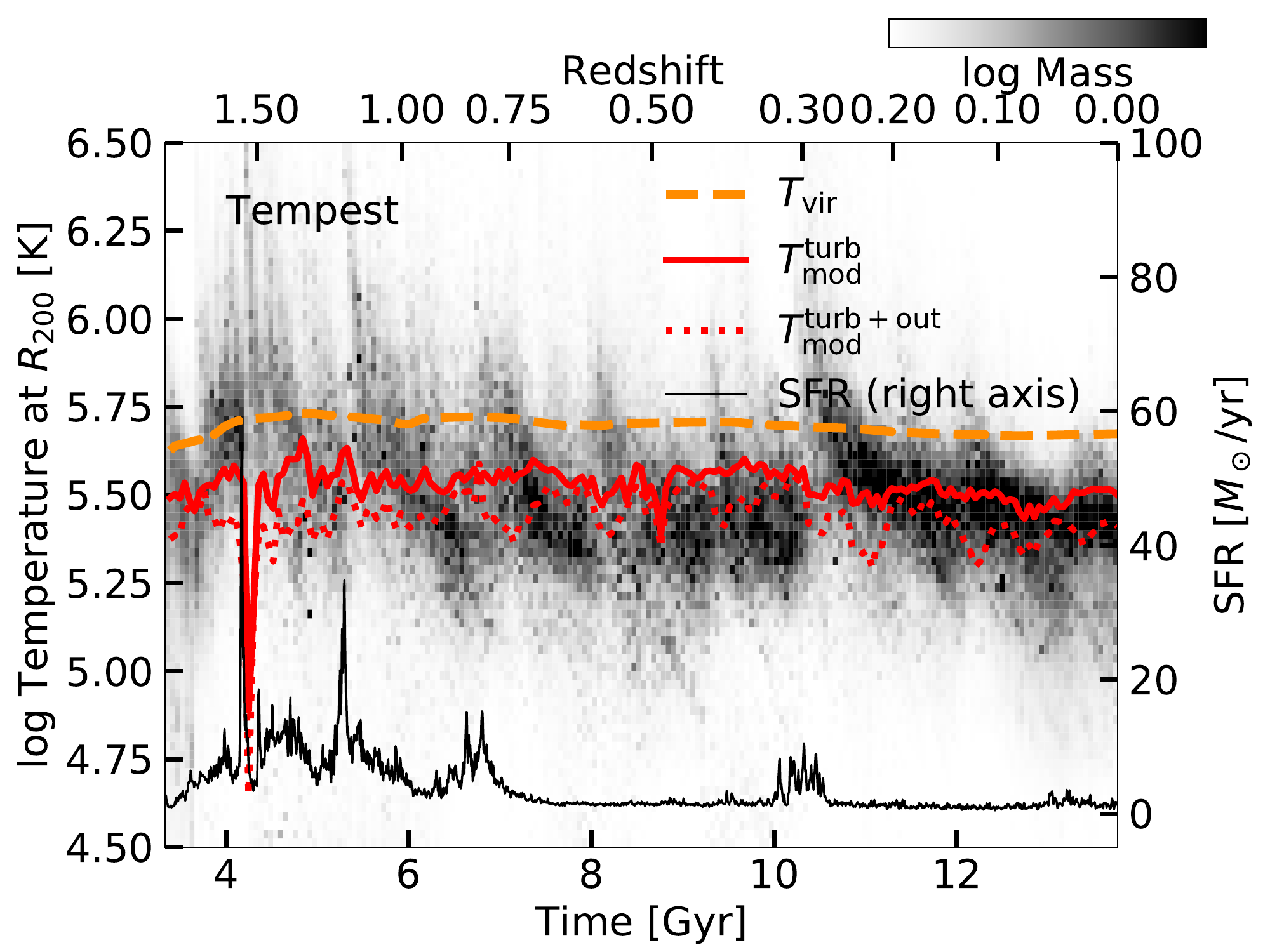}
    \includegraphics[width=\linewidth]{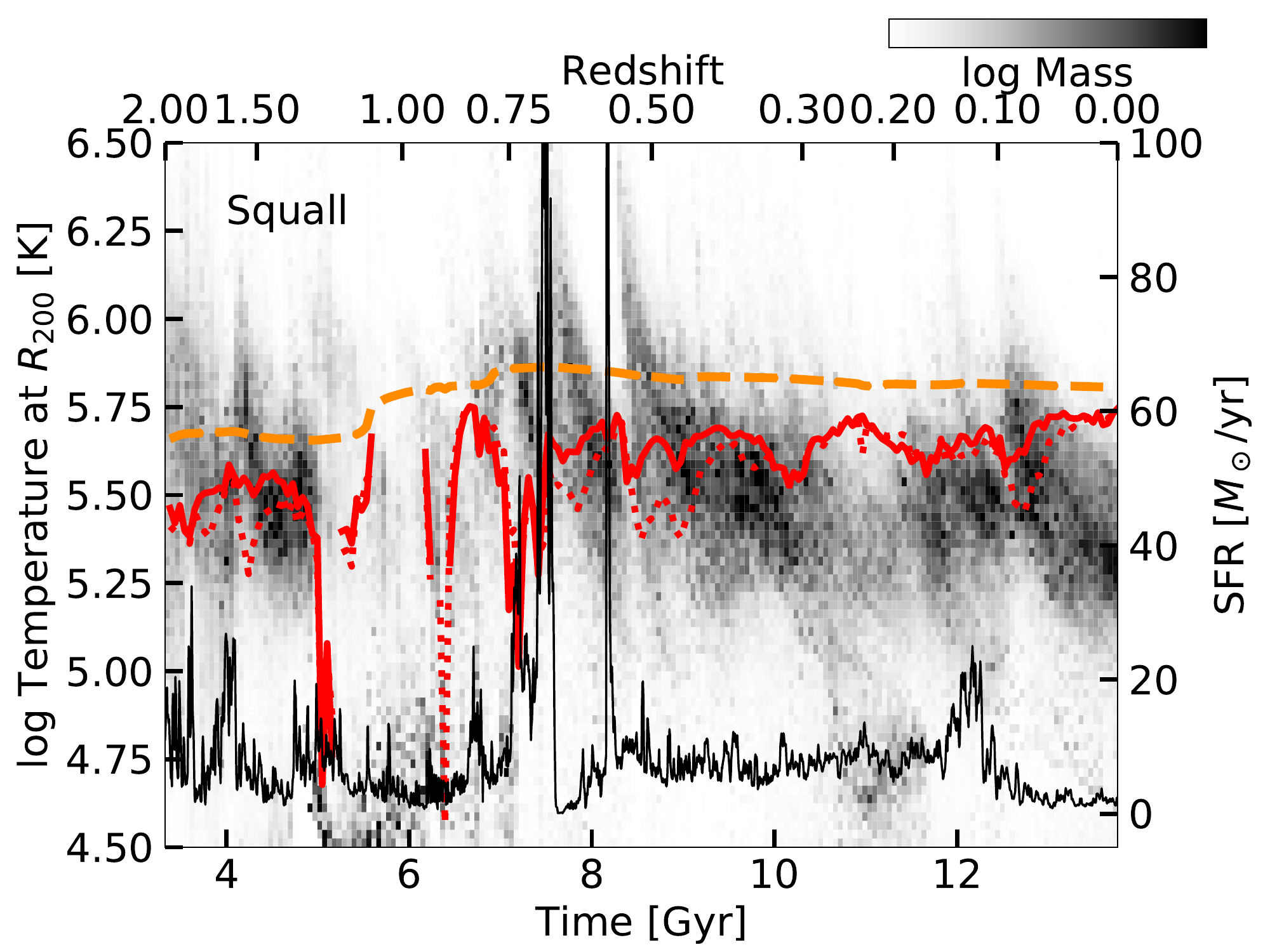}
    \includegraphics[width=\linewidth]{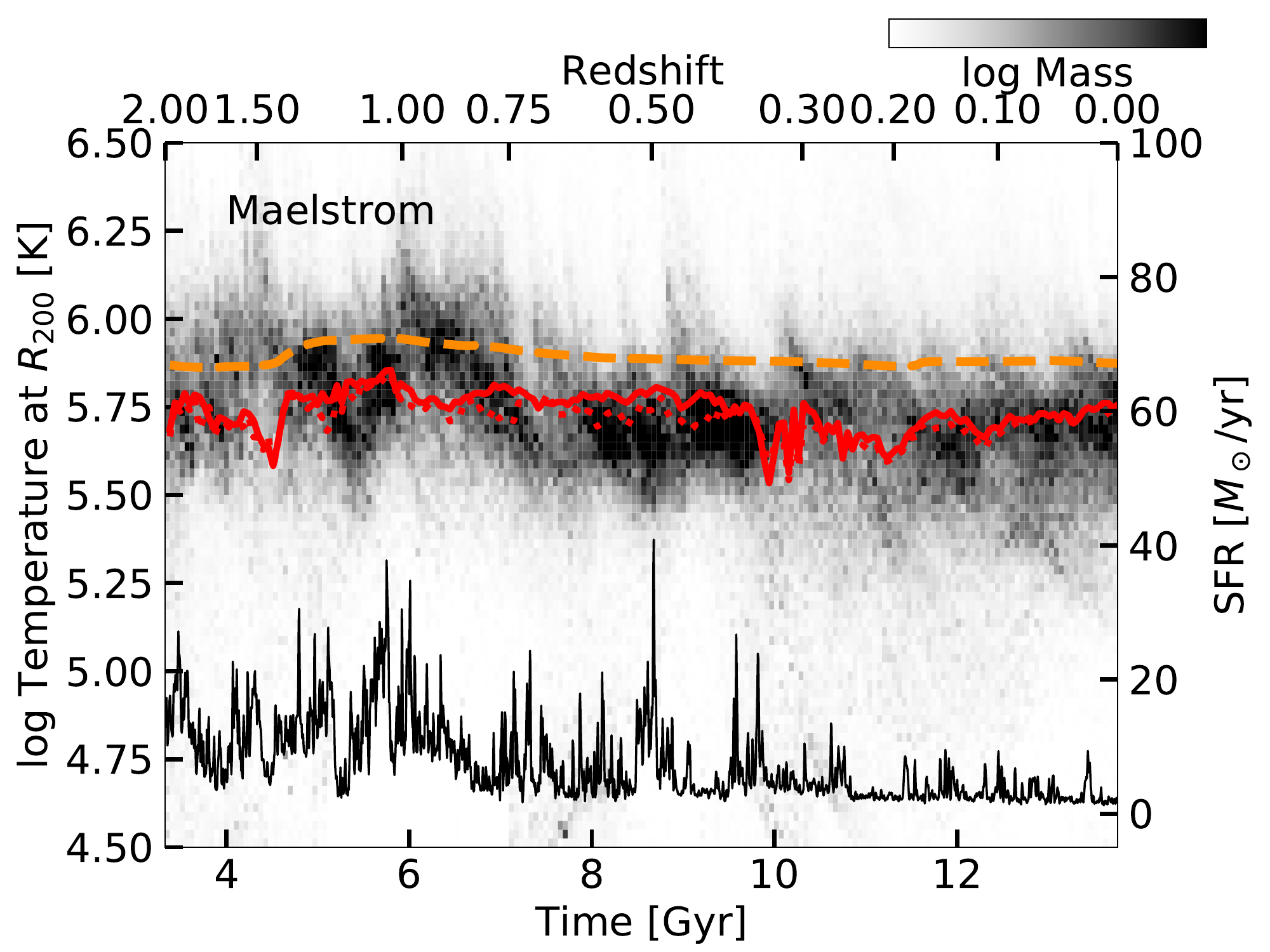}
    \caption{The temperature of gas with $0.99\rvir<r<\rvir$ is shown as a mass-weighted distribution as a function of cosmic time (bottom axis) and redshift (top axis), with dark colors indicating the peak of the mass-weighted temperature distribution. The orange dashed line shows the standard virial temperature (Equation~\ref{eq:Tvir}) and red solid and red dashed lines show the modified virial temperature, calculated using the kinetic energy due to turbulence only (Equation~\ref{eq:Tvirmod_r_turb}) or both turbulence and outflows (Equation~\ref{eq:Tvirmod_r_turb_out}), respectively. The thin black line shows the star formation rate, with values on the right axis.}
    \label{fig:temp_time}
\end{figure}

Figure~\ref{fig:temp_time} shows two-dimensional mass-weighted histograms of the gas within $0.99\rvir<r<\rvir$ in temperature-time space. Darker shading indicates the temperature of a larger fraction of the gas mass at $\rvir$. The standard virial temperature (Equation~\ref{eq:Tvir}) is shown as the dashed orange line and the modified virial temperature, calculated either with turbulence alone or with turbulence and bulk outflows (equations~\ref{eq:Tvirmod_r_turb} or~\ref{eq:Tvirmod_r_turb_out}, respectively) is shown as the dotted red and solid red lines, respectively. When there is a strong burst of star formation, the temperature histogram of gas near $\rvir$ shifts upward to higher temperatures as feedback heats the CGM. At low redshift and during quiescent periods, when the gas near $\rvir$ is closest to virial equilibrium (see Figure~\ref{fig:virial_time}), The darkest parts of the histograms indicate the temperature range of the majority of the gas mass, and $\tmod$ passes closer to this temperature range than the standard $\tvir$, which over-estimates the temperature of the gas near $\rvir$ by roughly a factor of two at nearly all times unless feedback is coincidentally heating the CGM gas.

\begin{figure}
    \centering
    \includegraphics[width=\linewidth]{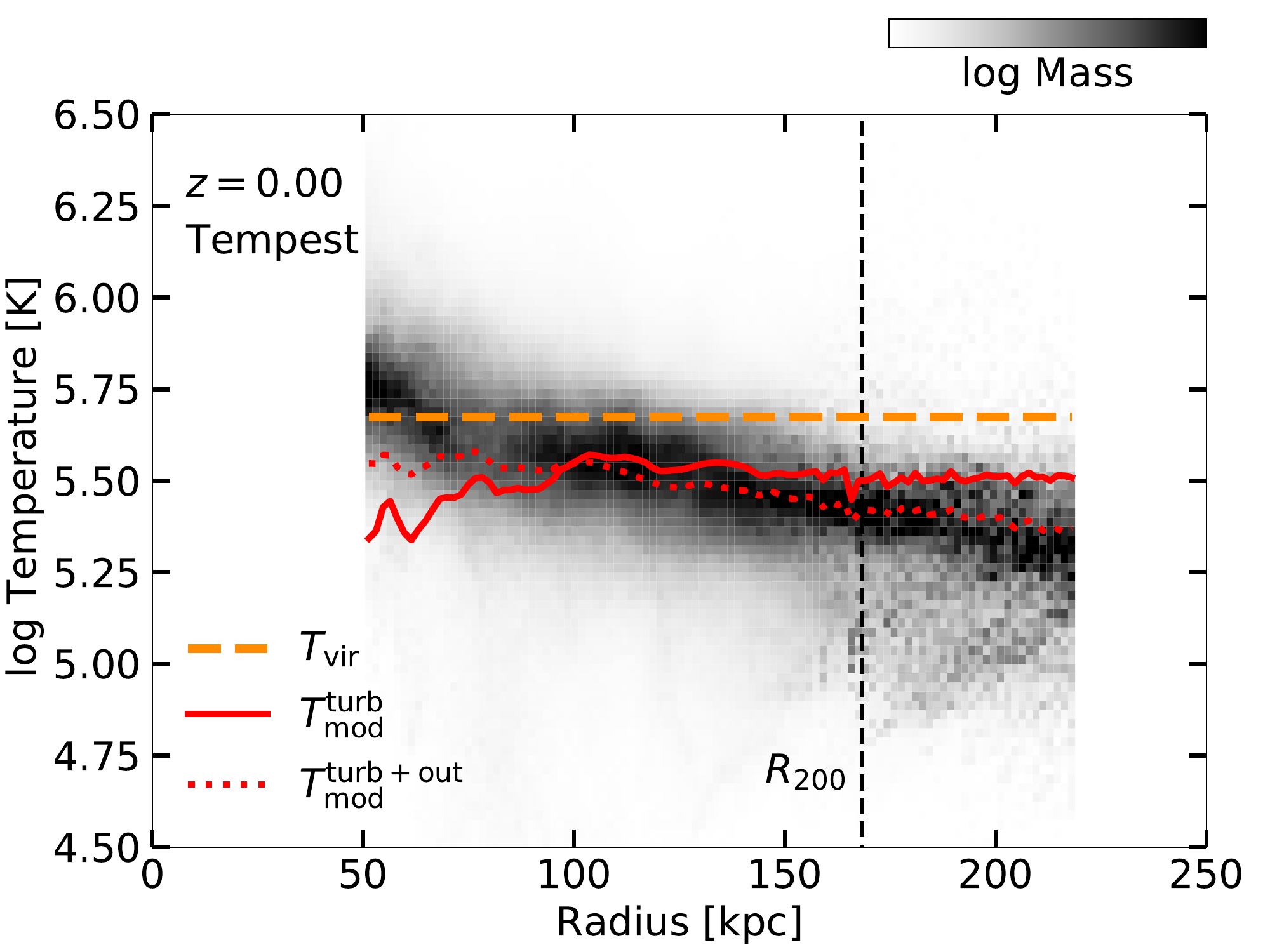}
    \includegraphics[width=\linewidth]{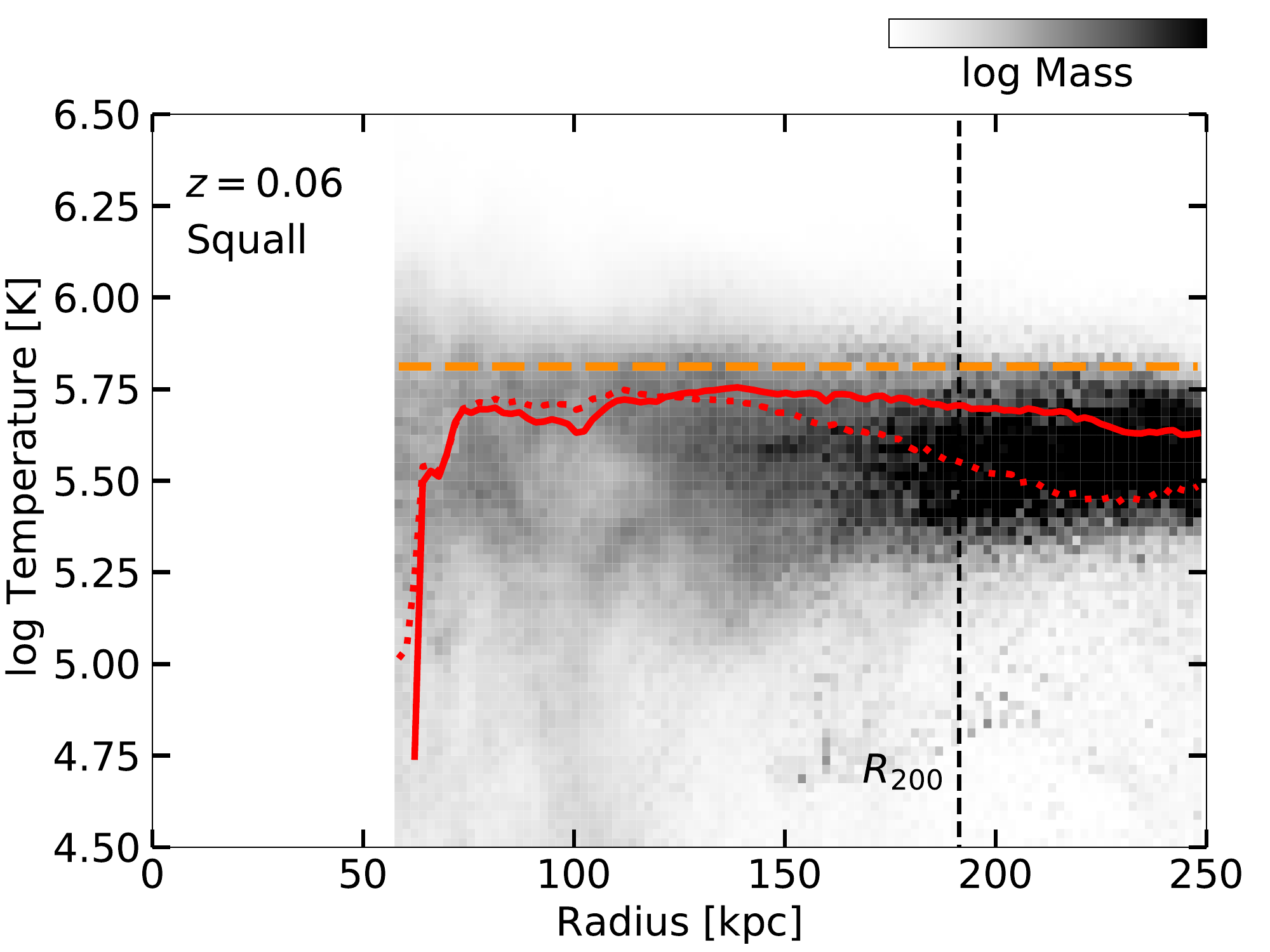}
    \includegraphics[width=\linewidth]{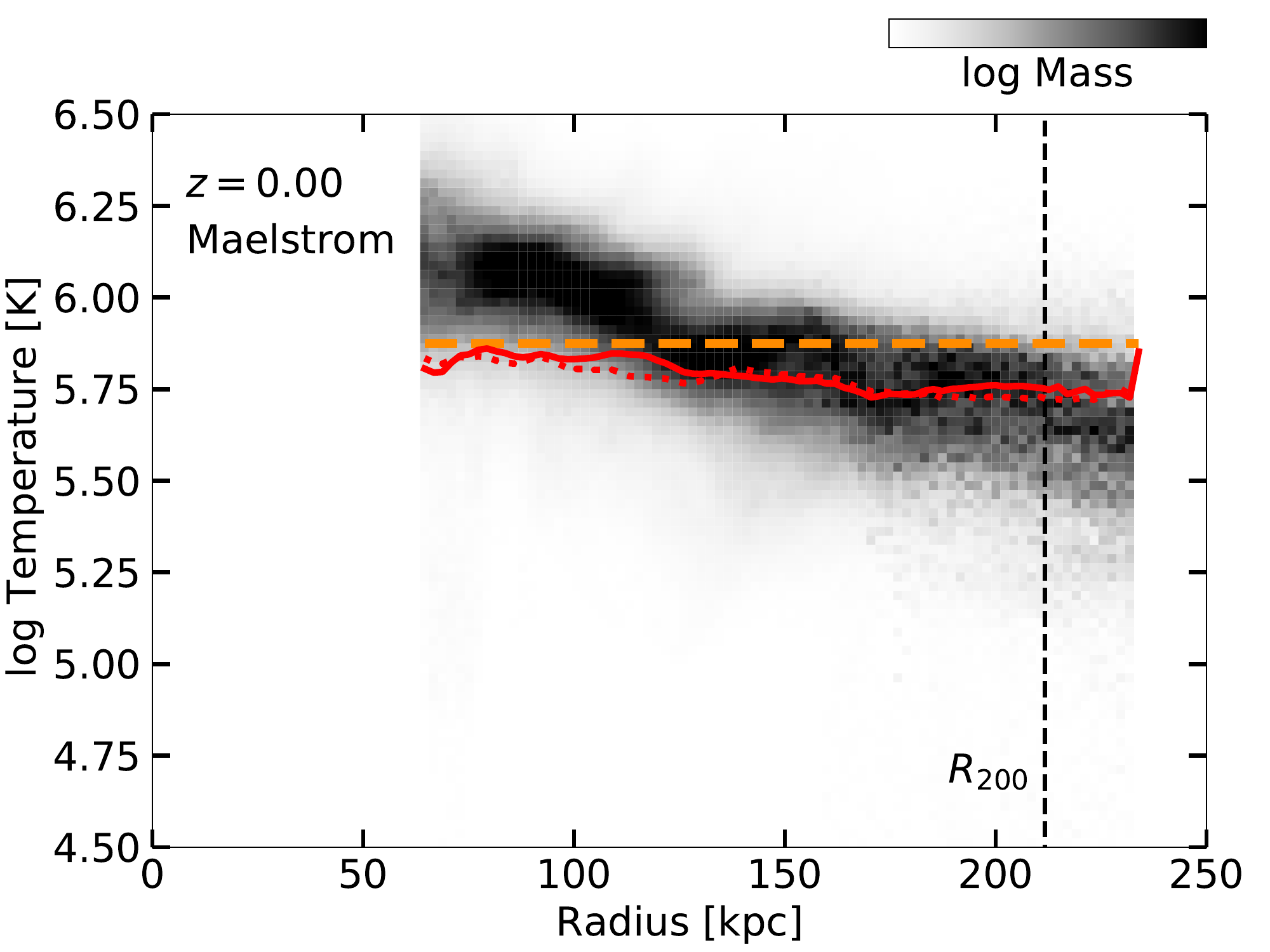}
    \caption{The mass-weighted temperature of the filament- and satellite-removed CGM gas at times when the region near $\rvir$ is in virial equilibrium (as indicated in each panel) is shown as a two-dimensional histogram in temperature-radius space, with dark colors indicating the peak of the mass-weighted temperature distribution at a given radius. The standard virial temperature (Equation~\ref{eq:Tvir}) is radius-independent and shown as the orange dashed line. The two ways of calculating the modified virial temperature, with and without bulk outflows (Equations~\ref{eq:Tvirmod_r_turb_out} and~\ref{eq:Tvirmod_r_turb}) are shown as the dotted and solid red lines, respectively.}
    \label{fig:temp_radius}
\end{figure}

Figure~\ref{fig:temp_radius} shows a mass-weighted two-dimensional histogram of temperature of CGM gas as a function of distance from the halo center compared to the standard virial temperature (orange dashed line) and the two calculations for the modified virial temperature (solid red for turbulence-only, Equation~\ref{eq:Tvirmod_r_turb}, dotted red for turbulence and outflows, Equation~\ref{eq:Tvirmod_r_turb_out}), at the same time snapshots as in Figure~\ref{fig:virial_radius} for each halo. The distribution is smoother at smaller radii where more of the cells in a given radial bin are refined to a higher resolution (see \S\ref{sec:FOGGIE}). In the outskirts of the halos, where the gas is closest to virial equilibrium (see Figure~\ref{fig:virial_radius}), $\tmod$ passes closer to the darkest regions of the mass-weighted temperature distribution than the standard $\tvir$ does. Both $\tvir$ and $\tmod$ fail to describe the temperature of the gas in the inner regions of the halo, where the gas temperature rises. In these inner regions, the halo gas is not in virial equilibrium (see Figure~\ref{fig:virial_radius}), so it is not expected that any temperature derived from the virial theorem will accurate describe this gas. In the inner regions of halos, there are several physical processes occurring that are not captured by an energy budget accounting through the lens of the virial theorem: there are sources (feedback) and sinks (radiative cooling) of energy, the singular isothermal sphere assumption used in both $\tvir$ and $\tmod$ does not describe the gravitational potential as well as it does in the outskirts of the halo (Appendix~\ref{sec:SIS}), and an increased gas temperature where the gravitational potential is stronger is expected from hydrostatic equilibrium but not captured by virial equilibrium arguments. For these reasons, we focus on the outskirts of the halo when using the virial theorem to estimate the halo gas temperature (although note that even at large radii, radiative cooling plays a minor role in the gas energy balance; see Appendix~\ref{sec:cooling}).

The standard $\tvir$ overestimates the temperature of the majority of the gas (by mass) by a factor of $\sim2$ at the virial radius, where we would expect the $\tvir$ to describe the gas temperature best. The modified virial temperature calculated including both turbulence and bulk outflows, $\tmod^\mathrm{turb+out}$, performs somewhat better in describing the peak of the temperature distribution (darkest shading) near $\rvir$ than $\tmod^\mathrm{turb}$ calculated from turbulence alone, without bulk outflows, in Tempest and Squall, whereas it makes no difference in Maelstrom, perhaps indicating that the kinetic energy due to outflows is more important to include in the overall energy balance of lower-mass halos.

When the kinetic energy due to bulk outflows is included in the energy balance of gas near $\rvir$, the corresponding $\tmod$ is smaller because the $\KEnt$ term in Equation~(\ref{eq:Tvirmod}) is larger, driving a larger deviation below the standard $\tvir$. Because it is unclear how much, if at all, outflows with $v<v_\mathrm{esc}$ contribute to the energy budget of the CGM gas, we report both $\tmod^\mathrm{turb}$ and $\tmod^\mathrm{turb+out}$ and do not pick one or the other as a better description of the temperature of the majority by mass of the CGM gas. However, Figure~\ref{fig:temp_time} indicates there are some times when it appears $\tmod^\mathrm{turb+out}$ passes closer to the peak of the temperature distribution near $\rvir$ than $\tmod^\mathrm{turb}$, which indicates there are some times in the halo's history when outflows are an important contributor to the energy budget of the halo gas, and some times when they are not (although note that outflows faster than the escape velocity are never included in $\tmod^\mathrm{turb+out}$). In addition, at different radii within the halo one or the other form of $\tmod$ may be a closer description of the peak of the gas temperature distribution, indicating that outflows may only be important to the energy budget at certain radii. We also see from Figure~\ref{fig:temp_radius} that neither form of $\tmod$ is an appropriate descriptor of the gas temperature at small radii where the assumption of virial equilibrium breaks down, as expected (see \S\ref{subsec:virial_results}).

\section{Feedback Drives Deviations from Equilibrium}
\label{sec:xcorr}

While $\tmod$ is clearly a better descriptor of the gas temperature in the outskirts of the FOGGIE halos than $\tvir$ for most each halos' evolution, there are times following strong bursts of star formation when outflows push the halo gas out of  virial equilibrium and away from $\tmod$ or $\tvir$ (Figure~\ref{fig:temp_time}). Here, we quantify this cross-correlation between the SFR and energy or temperature of the gas near $\rvir$ explicitly.

\begin{figure}
    \centering
    \includegraphics[width=\linewidth]{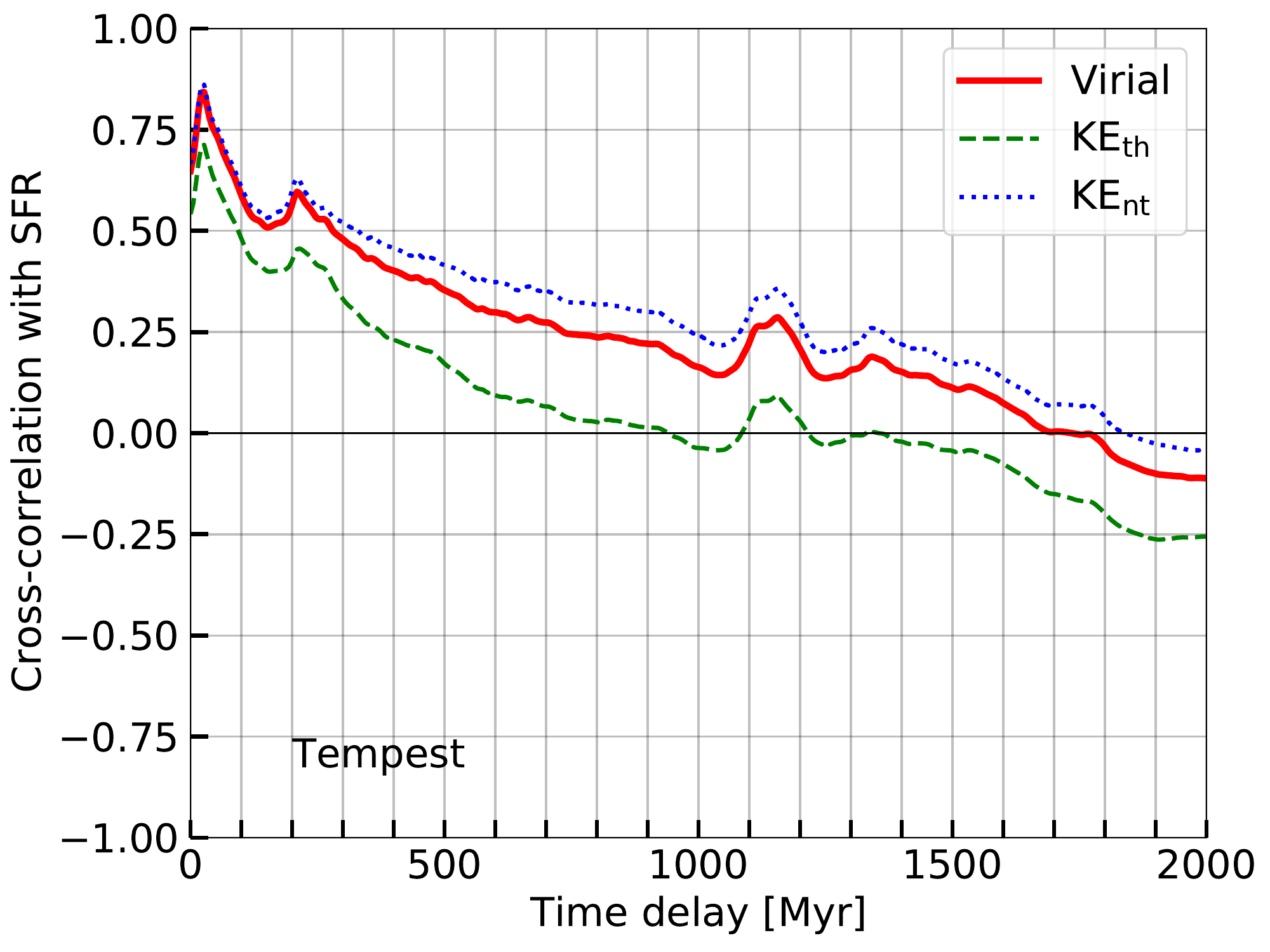}
    \includegraphics[width=\linewidth]{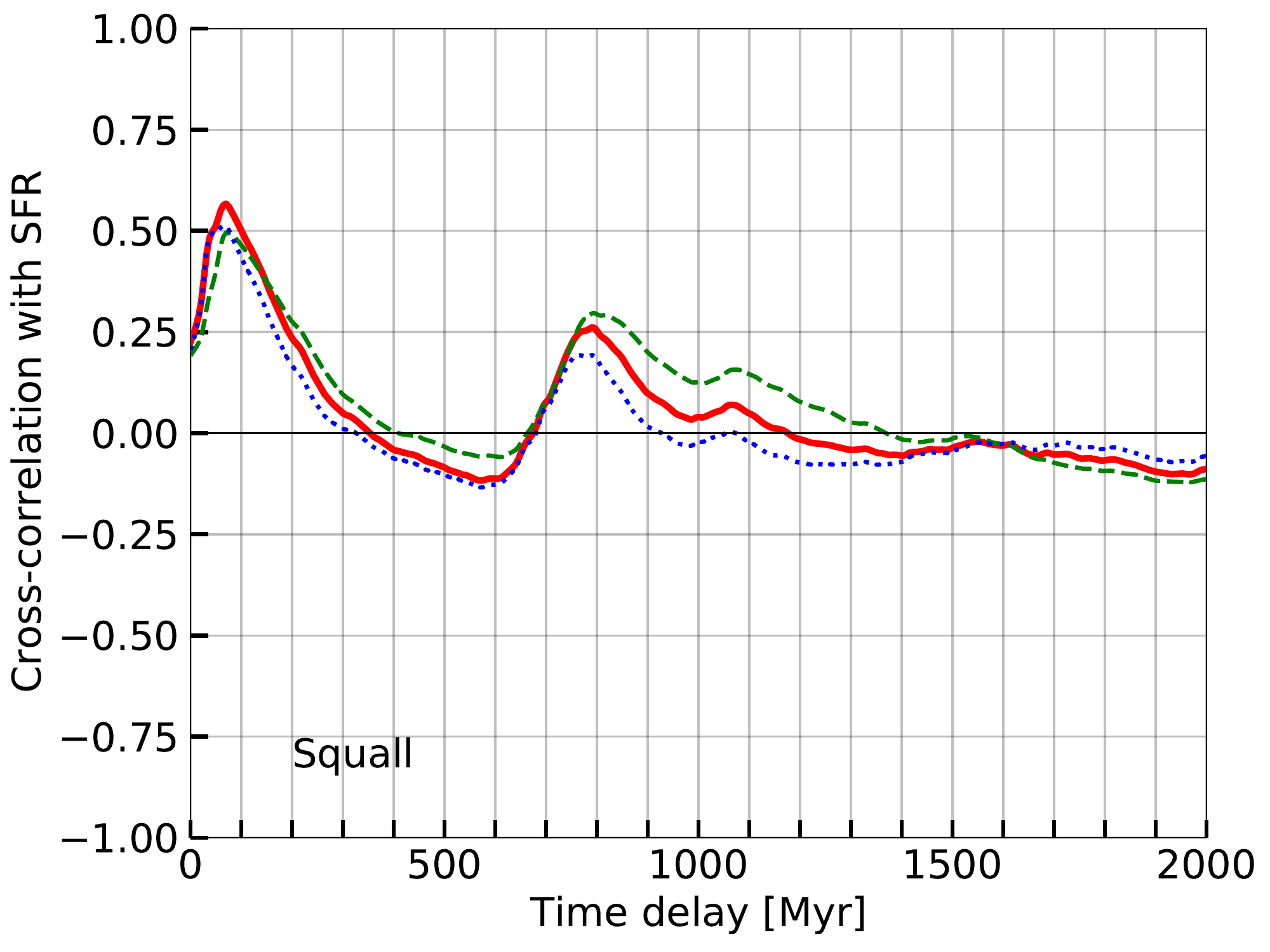}
    \includegraphics[width=\linewidth]{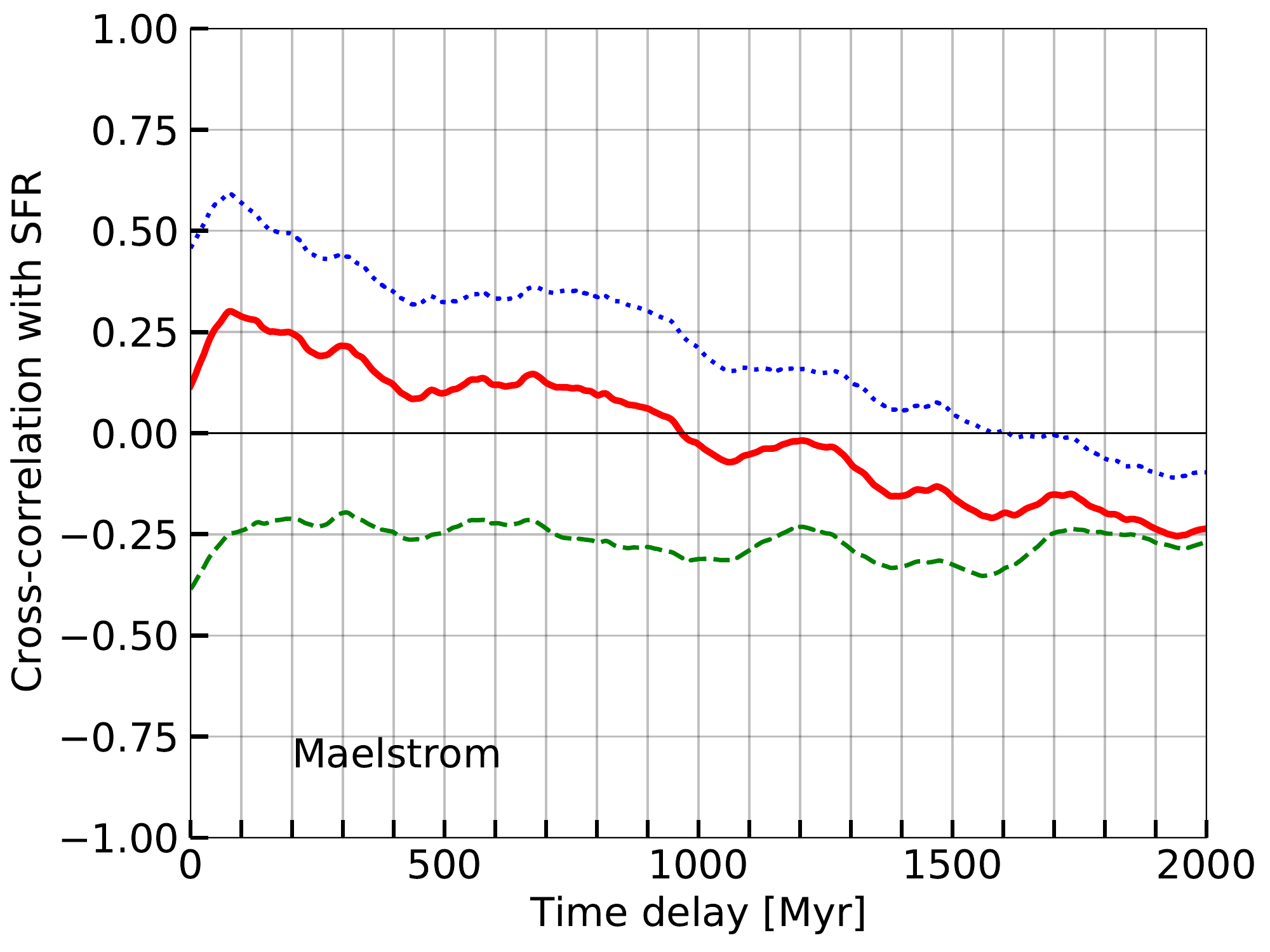}
    \caption{The cross-correlation between SFR and various energies of the gas near $\rvir$ as a function of time delay (Equation~\ref{eq:xcorr}). The correlation is positive for $\tau\lesssim1-1.5$ Gyr and peaks at $\tau\sim25-75$ Myr, quantifying what is easy to see by eye in Figure~\ref{fig:virial_time}, that the halo is driven out of virial equilibrium by strong bursts of feedback.}
    \label{fig:energy_xcorr}
\end{figure}

Figure~\ref{fig:energy_xcorr} shows a time-delay cross-correlation between the central galaxy's SFR and the energy of the gas at $\rvir$. The cross-correlation is computed as
\begin{equation}
    \xi(\tau)=\frac{1}{N}\sum_{i}^{N} \frac{[\mathrm{SFR}(t_i) - \overline{\mathrm{SFR}}][\mathrm{E}(t_i-\tau) - \overline{\mathrm{E}}]}{\sigma_\mathrm{SFR}\sigma_\mathrm{E}} \label{eq:xcorr}
\end{equation}
where SFR$(t_i)$ is the star formation rate and E$(t_i)$ is the energy, which can be VE (Equation~\ref{eq:VE}), $\KEth$, or $\KEnt$, at time snapshot $t_i$. $\overline{\mathrm{SFR}}$ and $\overline{\mathrm{E}}$ are the averages of SFR and energy over the time from $z=2$ to $z=0$, $\sigma_\mathrm{SFR}$ and $\sigma_\mathrm{E}$ are the standard deviations of SFR and energy over time, $\tau$ is a time-delay shift of one function relative to the other, and the sum is taken over all $N$ time snapshots. This function is normalized such that $\xi(\tau)=1$ would indicate a perfect correlation between SFR and energy at a time delay of $\tau$, $\xi(\tau)=-1$ would indicate a perfect anti-correlation at $\tau$, and $\xi(\tau)=0$ indicates no correlation at $\tau$. We perform this calculation over every snapshot output by the FOGGIE runs, separated by $\sim5$ Myr, in order to capture short-time variations in the energy and temperature of the halo. This is a factor of ten increase in time resolution over the analysis in Sections~\ref{sec:virial} and~\ref{sec:Tmod}.

Figure~\ref{fig:energy_xcorr} shows that at time delays $\tau\lesssim1-1.5$ Gyr, SFR and VE, $\KEnt$, and $\KEth$ (except for Maelstrom, see below) are all positively correlated, peaking at $\tau\sim25$ Myr for Tempest and $\tau\sim75$ Myr for Squall and Maelstrom. This means that stronger star formation is correlated with a higher energy of gas near $\rvir$ $\sim25-75$ Myr later\footnote{Note that the value of $\tau$ where the cross-correlation peaks is not strictly a travel time of outflows from the galaxy to $\rvir$ because the duration of the star formation burst also helps set $\tau$.}, and it takes $\sim1-1.5$ Gyr for the energy of the gas near $\rvir$ to fully ``relax" to what it was before the period of higher star formation, as can also be seen in Figure~\ref{fig:virial_time}. The strong correlation between SFR and gas energy at $\rvir$ confirms what was suspected from Figure~\ref{fig:virial_time}: strong bursts of feedback, driven by large SFRs, is what drives the gas near $\rvir$ away from the halo's ``natural state" of virial equilibrium. This further emphasizes why both $\tvir$ and $\tmod$ fail at describing the halo gas following a burst of star formation or in the inner regions of the halo most strongly affected by feedback: the virial theorem provides a tool for accounting for different forms of energies in the halo gas, and when the virial energy of the gas defined by equation~(\ref{eq:VE}) is not close to zero, the prerequisite for $\tvir$ or $\tmod$ to be descriptive is not met.

Interestingly, Squall shows two prominent peaks in the cross-correlation, but the second peak at $\sim800$ Myr is likely driven by the two extremely strong bursts of star formation at $\sim7.5$ Gyr and $\sim8.2$ Gyr (see middle panel of Fig.~\ref{fig:virial_time}) rather than predicting a cause-and-effect behavior. We verify this is the case by capping the SFR at $20 M_\odot$ yr$^{-1}$ and re-calculating the cross-correlation, which greatly diminishes the strength of the second peak without affecting the primary peak at 75 Myr (not shown). For large $\tau$, the cross-correlation samples fewer points and so it can be disproportionately driven by a handful of extreme events.

Maelstrom shows the weakest correlation strength between SFR and energy of the gas near $\rvir$ out of the three halos examined here. It is also the most quiescent of the three halos, with few very strong bursts of SFR that would be expected to drive the halo away from virial equilibrium, and this lack of strong peaks in both the SFR and the energy (see bottom panel of Fig.~\ref{fig:virial_time}) is likely what produces the weakest correlation in Figure~\ref{fig:energy_xcorr}. However, there is still a general trend of positive correlation between VE and $\KEnt$ and SFR, with a weak peak at $\sim100$ Myr, that declines over 1 Gyr. Interestingly, there is an anti-correlation between SFR and $\KEth$ in Maelstrom roughly constant with time-delay, which is not seen in either Squall or Tempest, but a strong positive correlation with $\KEnt$. This indicates that despite Maelstrom's general quiescence compared to the other two halos where it appears bursts of star formation do not heat the gas much, the non-thermal kinetic motions triggered by star formation feedback are still important to the overall energy balance (or lack thereof) of the gas at $\rvir$.

\begin{figure}
    \centering
    \includegraphics[width=\linewidth]{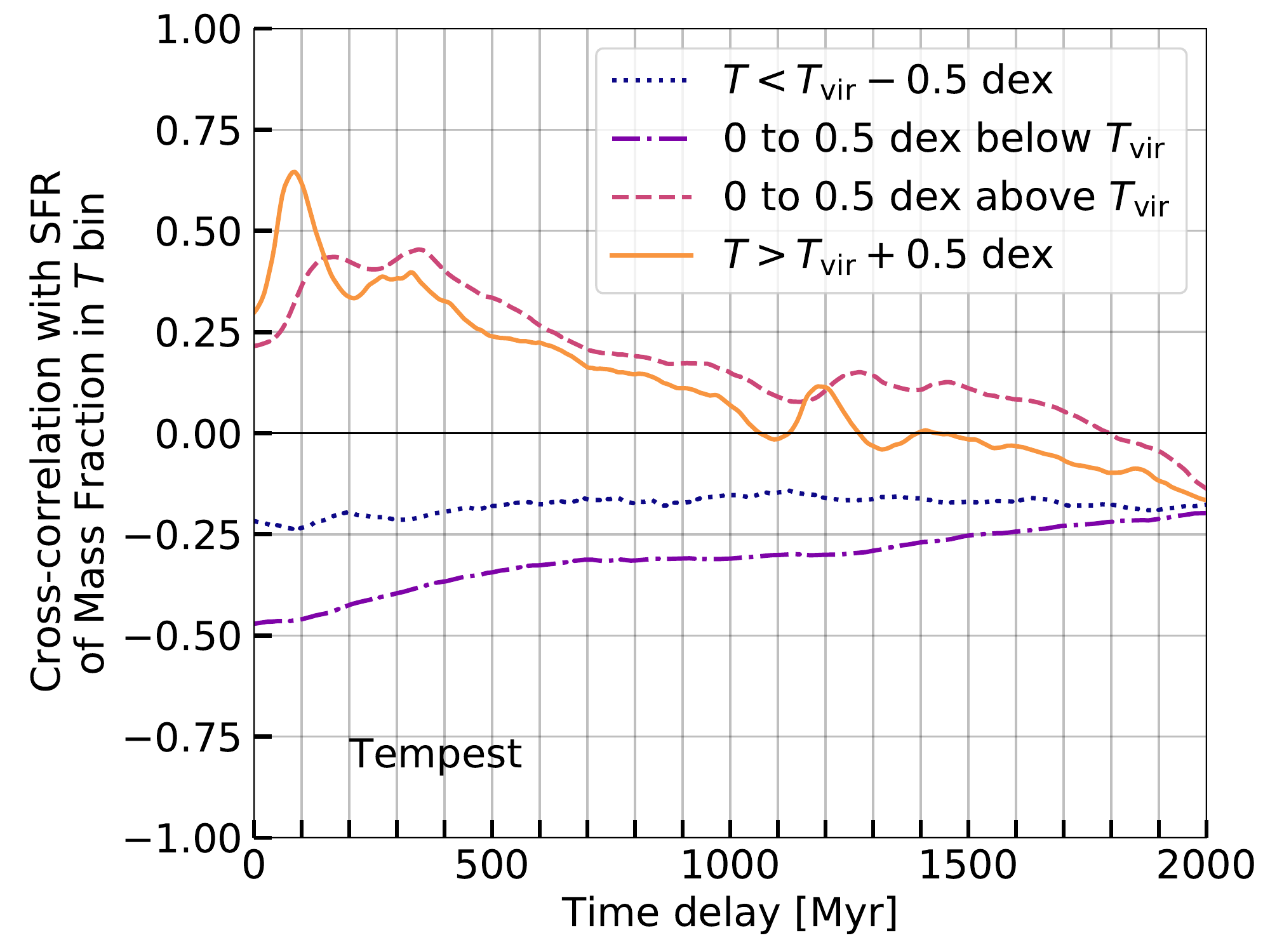}
    \includegraphics[width=\linewidth]{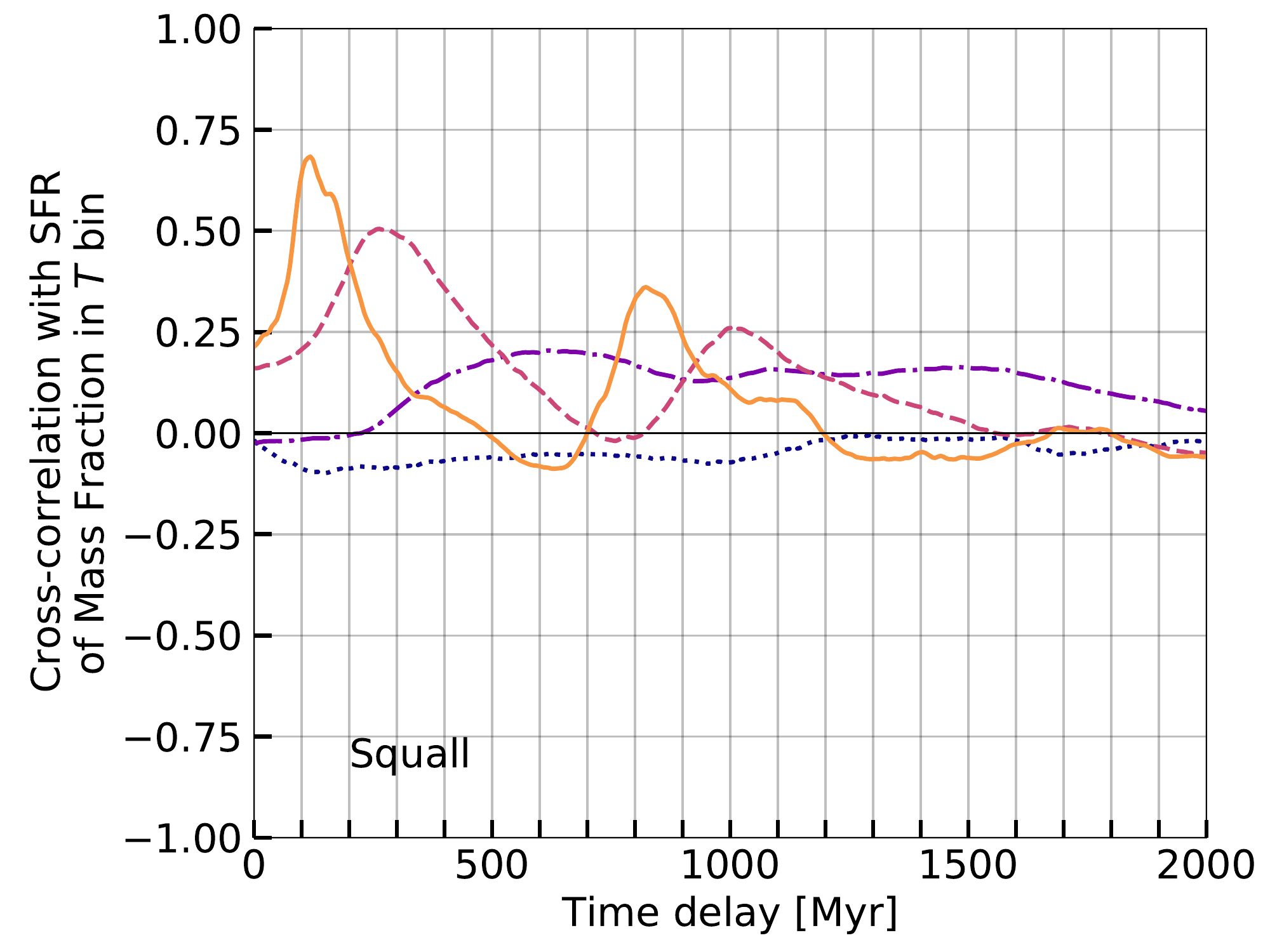}
    \includegraphics[width=\linewidth]{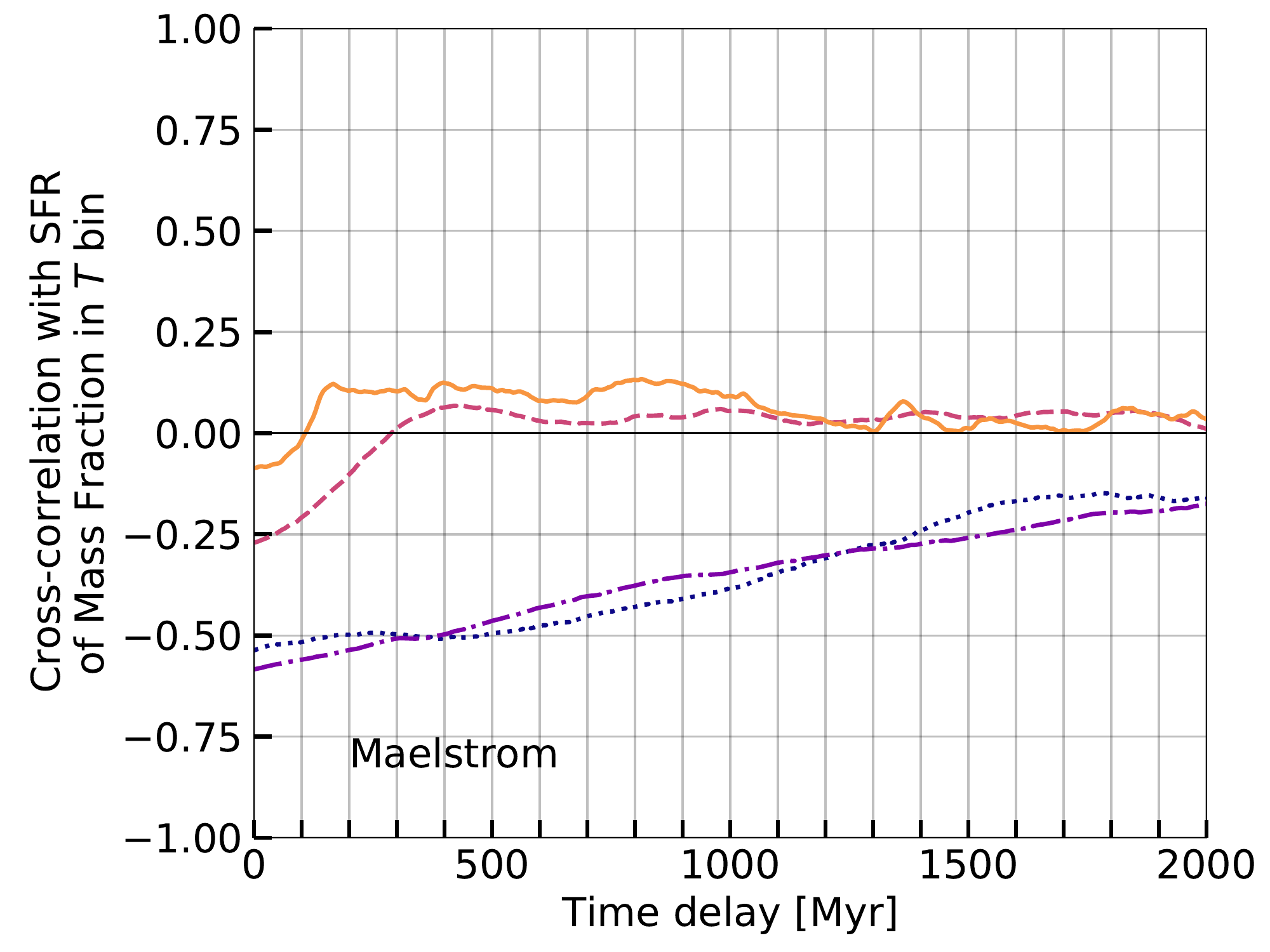}
    \caption{The cross-correlation between SFR and mass in different temperature bins relative to the standard $\tvir$ as indicated in the figure, as a function of time delay $\tau$ (Equation~\ref{eq:xcorr}). The mass in the warmer temperature bins is positively correlated with SFR while the mass in the cooler bins is anticorrelated, quantifying what is easy to see by eye in Figure~\ref{fig:temp_time}, that temperatures $\gtrsim\tvir$ are only achievable following a burst of SFR and \emph{not} when the halo is in relaxed virial equilibrium.}
    \label{fig:temp_xcorr}
\end{figure}

Figure~\ref{fig:temp_xcorr} shows a similar time-delay cross-correlation, this time correlating the SFR with the mass of gas at $\rvir$ within different temperature bins relative to the standard $\tvir$ as marked on the figure (calculated with Equation~\ref{eq:xcorr}, but replacing E$(t_i)$ and $\overline{\mathrm{E}}$ with the mass in a temperature bin at $t_i$ and averaged over time, respectively). The two hottest temperature bins ($\gtrsim\tvir$) in Tempest and Squall are positively correlated with bursts of SFR, indicating that the presence of $\sim\tvir$ gas at $\rvir$ is due to star formation feedback, \emph{not} that the gas at $\rvir$ is naturally at $\tvir$ when the halo is fully relaxed. The two coolest temperature bins ($\lesssim\tvir$) are anticorrelated with the SFR in Tempest and Maelstrom, indicating that bursts of star formation remove cool gas mass from $\rvir$. This trend can be seen by eye in Figure~\ref{fig:temp_time}, where the only time the gas at $\rvir$ is close to or greater than the standard $\tvir$ (orange dashed line in that figure) is shortly following a burst of feedback, after which the temperature drops again as the halo relaxes. Like the energy cross-correlation, the temperature cross-correlation takes $\sim1-1.5$ Gyr to relax to the state it was prior to the burst of SFR. The highest temperature bin (solid orange) peaks sooner than the next-highest (dashed pink) temperature bin, a signature of the general declining temperature trend after a strong burst of star formation seen in Figure~\ref{fig:temp_time}.

In Squall, while the two highest temperature bins are positively correlated with SFR, the two lowest temperature bins are not particularly correlated with SFR in either a positive or a negative direction. This may indicate that Squall has significant cool gas mass in the halo regardless of feedback, and that bursts of feedback simply add more hot gas without removing cool gas near $\rvir$. Just like above with the energy-SFR cross-correlation, the secondary peak in the hot gas mass in Squall is driven by just two of the strongest starburst events and is greatly reduced if we cap the SFR at $20 M_\odot$ yr$^{-1}$ (not shown). Maelstrom shows the opposite trend, where the two hottest temperature bins are not particularly correlated with SFR while the two coolest temperature bins are strongly anti-correlated with SFR. The anti-correlation of cool gas mass with SFR is expected if feedback heats the gas near $\rvir$, but lack of correlation with hot gas is unexpected and may just suggest that there are not enough significant peaks in the SFR to drive the gas temperature near $\rvir$ significantly away from its equilibrium value. This seems to be corroborated by Figure~\ref{fig:temp_time} (bottom panel), where the temperature of the gas near $\rvir$ does not exhibit as many short deviations to high temperatures as in the other halos.

\section{Implications of a Cooler CGM}
\label{sec:implications}

In Sections~\ref{sec:virial} through~\ref{sec:xcorr}, we found that across cosmic time and throughout the outskirts of a galactic halo, non-thermal gas motions are critical to understanding the energy partition and the temperature of the halo gas. For the simulated halos studied here, virial equilibrium holds near $\rvir$ only if non-thermal kinetic energies are included in the energy balance and only when the halo is not being perturbed by strong feedback, proving that the halo gas can be virialized without being fully thermalized. The consequence of this finding is that the standard virial temperature $\tvir$ overestimates the peak of the gas temperature distribution by a factor of $\sim2$ when the halo is relaxed. Instead, the modified virial temperature, which is calculated taking these non-thermal gas motions into account, is a better descriptor of the halo gas near $\rvir$. The end result is a somewhat cooler galactic halo than expected from standard virial arguments.

While we carefully calculate the contribution to the energy budget of bulk flows here, this may not be possible in many cases, such as in interpreting observations or in analytic models or idealized simulations of $L^*$ galaxy halos. In these cases, we suggest using a halo temperature roughly a factor of 2 lower than the standard virial temperature. The factor of two arises due to the roughly equal contributions of thermal and non-thermal energies to the halo's energy budget throughout most of its evolution (see Figure~\ref{fig:virial_time}), which we suggest as a general rule of thumb in cases where emergent non-thermal kinetic energies cannot be explicitly calculated.

Observations of the diffuse gas making up the CGM, especially in the outskirts of galaxy halos, are typically done in absorption line studies against the light from a bright background source. This generally restricts the derived information to at most a handful of lines of sight through any given galaxy's halo \citep[an important exception is M31, for which multiple sightlines through the same galaxy's CGM can be obtained,][]{Lehner2015,Lehner2020}, and restricts the dimensionality of that information to only line-of-sight velocities. Galaxy formation simulations or cosmological simulations can provide more information than these pencil-beam observations, but they may be under-resolving the small-scale structure of the CGM, especially on scales far from the galaxy. In addition, the implementation of star formation and feedback in the central galaxy varies across different simulations and with resolution. It is necessary to develop and analyze analytic models for the CGM to tie the differing information from observations and simulations back to the gas physics that govern the CGM, and it is in the context of these analytic models where a cooler-than-expected CGM has significant implications that we outline below.

\subsection{Thermal Pressure of the CGM}
\label{subsec:th_pres}

Absorption-line surveys of the CGM routinely discover both hot ($T\sim10^6$ K) and cool ($T\sim10^4$ K) gas, frequently in the same line of sight and at the same line-of-sight velocity \citep{Tumlinson2013,Bordoloi2014,Borthakur2015,Werk16, Keeney2017,Berg2018,Chen2018,Muzahid2018}. Ionization modeling of low-ionization state absorption lines produces estimates of the gas density that tend to show that the cool gas density is similar to the hot gas density \citep{Werk2014,Stern2016} despite large temperature differences, and thus the two gas phases are out of pressure equilibrium. However, multiphase CGM models that pose the cold phase is found in small clouds embedded within the volume-filling hot phase generally expect that the cool and hot gas are in pressure equilibrium at the pressure of the hot phase \citep{Mo1996,Maller2004}. An overall cooler CGM can help alleviate this discrepancy somewhat, as a lower temperature for the hot phase reduces its thermal pressure and thus reduces the thermal pressure needed in the cold phase to match it, allowing the cold phase to be more diffuse. A factor of two decrease in hot-phase temperature leads to a factor of two decrease in the expected cold-phase density if the phases are in pressure equilibrium. The detailed multiphase ionization modeling of \citet{Haislmaier2021} finds the warm and cool gas phases may actually be in pressure equilibrium in some cases, but out-of-equilibrium solutions are not entirely ruled out, and those authors also find that the thermal gas pressure of all phases (regardless of whether they are in pressure equilibrium with each other) is lower than typically expected for $\sim L^\star$ halos. A lower thermal pressure may be explained by the cooler-than-expected volume-filling phase of the halo we present in this paper, although it may not be enough to fully rectify thermal pressure differences between phases and a shallower pressure gradient \citep{Voit2019a,Voit2019b} or non-thermal pressure from turbulence or cosmic rays is likely needed \citep{Salem2016,Oppenheimer2018,Butsky2020,Ji2020,Lochhaas2020}.

\subsection{Cooling Time of CGM Gas}
\label{subsec:tcool}

The efficiency of radiative cooling is strongly dependent on temperature and peaks around $T\sim10^5$ K for metal-enriched gas. At these \ temperatures, CGM gas produces many intermediate ions such as \ion{C}{4}, \ion{Si}{4}, and potentially \ion{O}{6}, all species that are frequently found in absorption in the CGM surrounding Milky Way-like galaxies. This intermediate-temperature gas may live in radiatively-cooling interface regions between hot and cold phases, which grow as hot and cold gas turbulently mix together \citep[e.g.,][]{Begelman1990,Slavin1993,Wakker2012,Kwak2015,Ji2019,Fielding2020a,Tan2020}. A cooler hot phase means less mixing is required for gas to reach intermediate temperatures and cool efficiently, perhaps allowing cooling to proceed more quickly than would be expected in a hotter medium. This could lead to more cool gas forming by entraining mass from the hot phase, explaining observations of the cool and intermediate phases. While we find that cooling does not dominate the overall energy balance of the halo at more than a $\sim10\%$ level (see Appendix~\ref{sec:cooling}), the lower gas temperature may seed cooling on small scales that can explain observations without upsetting the global balance.

\subsection{CGM Mass Estimates from X-Ray Observations}
\label{subsec:xray}

X-ray studies observe the hot component of the CGM of both the Milky Way and external galaxies. X-ray emission is strongly dependent on the gas density, so typically the CGM is only observed in X-ray emission in the densest regions closest to massive galaxies. A popular strategy for characterizing the hot CGM gas that emits in X-ray is to fit its density profile with a $\beta$ model, which is a power-law profile where the parameter $\beta$ describes the power. To estimate the total mass of hot gas in a galaxy's CGM, the $\beta$ model is extrapolated out to the virial radius and integrated \citep{Anderson2010,Gupta2012,Das2019}, finding $\sim10^{9-10} M_\odot$ of hot gas within the halo \citep{Anderson2010}. This method assumes that the gas maintains its hot temperature out to the virial radius and that it is only the decline in density that leads to the decrease in X-ray surface brightness below detection limits in the outskirts of galaxy halos. In addition, fits to X-ray spectra may infer gas temperatures higher than the peak of the temperature distribution \citep{Vijayan2021}.

Figure~\ref{fig:temp_radius} shows that the gas temperature decreases with increasing radius approaching the virial radius, and is a factor of $\sim2$ lower than the commonly-assumed virial temperature, so that mass estimates of the hot gas from $\beta$ models may not be accurate. If the decrease of gas temperature with increasing radius is what drives the low X-ray surface brightness in the outskirts of galactic halos, rather than lack of gas mass, the gas mass in the halo's volume-filling phase may be higher than estimated. However, the volume-filling gas is not ``hot" ($T > 10^6$ K) but ``warm" ($T = 10^5 - 10^6 K$, see Figure~\ref{fig:temp_radius}), so the \emph{hot gas} contribution to the mass of the halo may be lower than estimated even if the \emph{volume-filling} gas phase contribution to the halo mass is higher than estimated. A detailed analysis of the relative contribution of each gas phase to the mass budget of the CGM is beyond the scope of this paper.

\subsection{The Origin of \ion{O}{6}}
\label{subsec:OVI}

The UV doublet of \ion{O}{6} is the highest-ionization tracer of warm gas that is readily accessible outside the X-ray. \citet{Tumlinson2011} presented correlations of CGM \ion{O}{6} abundance with galaxy properties, finding a bimodality in presence of \ion{O}{6} that depends on SFR of the central galaxy: star-forming galaxies have more \ion{O}{6} in their halos than quiescent galaxies. \citet{Oppenheimer2016} proposed a different source for gas traced by \ion{O}{6}: the ionization fraction of \ion{O}{6} peaks near the virial temperature of roughly Milky Way-mass galaxies such that the \ion{O}{6} bimodality is actually a bimodality in halo mass (and thus virial temperature), rather than relating primarily to SFR. Galaxies living in massive halos have virial temperatures too high for \ion{O}{6} to be prevalent (oxygen ions are instead ionized further to \ion{O}{7} or \ion{O}{8}), and these galaxies are also typically quenched, thus explaining the \ion{O}{6} bimodality with SFR as well.

However, \citet{McQuinn2018} showed that there is a slight offset between the halo mass where virial-temperature-tracing \ion{O}{6} is expected to be prevalent and the halo mass where \ion{O}{6} is most frequently observed, such that \ion{O}{6} is detected surrounding more massive halos whose virial temperatures are too large, seemingly, for \ion{O}{6} to survive (see their Figure 3). \citet{McQuinn2018} proposed a spread in gas temperature around the virial temperature as one possible solution to this dilemma, such that some of the gas exists at lower temperatures where \ion{O}{6} can be found. 

In this paper, we showed the temperature of CGM gas is lower than expected from the classical virial analysis. While we do not yet have enough halos simulated at high resolution to examine this trend with halo mass, the prevalence of the hydrostatic mass bias in galaxy clusters (see \S\ref{subsec:turb}) shows that even very high-mass halos have significant non-thermal gas motions, which should also affect the virial balance and reduce the virial temperature. If the actual temperature of halo gas is lower than expected for halos of all masses, then the halo mass at which the \ion{O}{6} ionization fraction peaks at the virial temperature is larger than expected, potentially explaining why \ion{O}{6} is seen with such abundance surrounding more massive galaxies than expected. The highest-mass halo explored in this paper, Maelstrom, does still show a temperature lower than the standard virial temperature, so it seems likely that halos of all masses will have lower temperatures than expected. We note, however, that without a more rigorous study of what drives the turbulence and other non-thermal gas motions and how those processes may change with halo mass, we cannot derive a halo mass scaling for the modified virial temperature to confirm this scenario.

A substantial amount of the \ion{O}{6} in a galaxy halo may arise from a cooler, photoionized phase rather than the warm, volume-filling virialized phase \citep{Stern2018,Strawn2020}. The scenario outlined in this subsection assumes most of the observed \ion{O}{6} arises from the warm phase in collisional ionization equilibrium, rather than from a cool phase in photoionization equilibrium. If \ion{O}{6}-hosting gas is primarily cool, then a trend of \ion{O}{6} column density with halo mass would not be tracing the virial temperature of the halo but rather the amount of cool gas in a halo as a function of halo mass. Reality is likely to be a mixture of both scenarios, and distinguishing between them is beyond the scope of this paper.

\subsection{The Importance of Turbulence}
\label{subsec:turb}

The main result of this study is that non-thermal gas motions, such as turbulence, are important to the energy partition of a virialized halo, and this has important consequences for the temperature of the CGM. Significant turbulent motions also have effects that go beyond just modifying the virial temperature: turbulence can provide pressure support to the CGM \citep{Oppenheimer2018,Lochhaas2020} and it can affect how cool gas condenses out of the hot medium \citep{Voit2018} or mixes with the hot medium to efficiently create more cool gas \citep[e.g.,][]{Fielding2020a}.

In particular, turbulent pressure drives halo gas away from purely hydrostatic solutions where it is assumed that thermal pressure exactly balances the gravitational potential. The importance of non-thermal pressure support has been known for some time in galaxy clusters, where the idea of a ``hydrostatic mass bias" is well known \citep{Nagai2007,Piffaretti2008,Lau2009,Lau2013,Shi2015,Shi2016,Biffi2016,Shi2018,Simionescu2019,Gianfagna2020}. The hydrostatic mass bias is the difference between inferred cluster mass from a hydrostatic assumption for cluster gas and the true cluster mass, and most studies find differences on the order of $\sim10-20\%$, driven by a non-thermal pressure contribution on the order of $\sim20-30\%$ of the total pressure \citep{Vazza2011,Nelson2014,He2020}. Clearly, non-thermal gas motions are important in galaxy clusters, and there is no reason to suspect that galaxy-scale halos lack significant non-thermal pressure or energy. Indeed, we have shown in this paper that non-thermal kinetic energy is a significant contribution to the energy balance of galaxy-scale halos, and that this has consequences for the temperature of the halo gas. The consequences of significant non-thermal energy on the pressure of the halo gas will be explored in a forthcoming paper.

\citet{Rudie2019} observed the CGM of star-forming galaxies at $z\sim2$ and found that the non-thermal broadening of most ions' absorption lines was small, indicating that thermal motions dominate over turbulence. However, the broadening of \ion{O}{6}, the highest ionization state ion probed in that study, was larger, roughly a few tens of km s$^{-1}$. If \ion{O}{6} traces the warmest phase of gas, this would indicate turbulence roughly on the scale of the simulated turbulence in the warmest gas phases in the FOGGIE halos. \citet{Lehner2014} found a similar result for \ion{O}{6}, but also found roughly half of \ion{C}{4} and \ion{Si}{4} absorption lines at high redshift were broader than would be expected from pure thermal broadening, indicating that there may be turbulence in the warm gas phase probed by these mid-ions as well. \citet{Rudie2019} found lower ionization state ions had narrower absorption lines, indicating less turbulent broadening. If lower ions are found in cool clouds embedded within a warmer halo, each individual cloud may not have significant internal turbulence, leading to narrow individual absorption components, but the collection of cool clouds may trace the turbulence of the hot phase in which they are embedded. If that is the case, it is the velocity dispersion between individual cool-phase components that traces the hot-phase turbulence, which \citet{Rudie2019} find to be $\sim100-200$ km s$^{-1}$. Some of these components may be tracing fast-moving coherent structures like outflows or accretion filaments (and indeed they find a subset of absorbers with velocities above the escape velocity of their host halo). \citet{Zahedy2019} carried out a similar analysis at lower redshift ($z\sim0.4$), and found the low-ion absorption lines had a modest amount of non-thermal broadening. Turbulence clearly plays some role in the CGM, but it is unclear as yet how much, and in what gas phases.

Turbulence drives motions that cascade down to smaller scales, all the way to the single-cell resolution scale in simulations. If the resolution in a simulation is poor, meaning that the turbulent cascade is cut off, then the turbulent energy in the smaller scales will not be captured. If the small-scale energy in turbulence is significant, a simulation with poor spatial resolution will underestimate the amount of energy in non-thermal, turbulent motions. A deeper analysis of the driving and structure of turbulence in the FOGGIE simulations is beyond the scope of this paper, but we note that FOGGIE's high spatial resolution in the CGM may be required to capture the consequences of substantial non-thermal motions. For example, \citet{Bennett2020} showed an increase of $\sim80\%$ in turbulent energy near the virial radius with increasing resolution (see their Figure 13), which was balanced by a decrease in thermal energy and thus likely temperature, although they do not discuss temperature explicitly. Assuming a turbulent cascade from large to small scales, the majority of the turbulent energy is located in the large scales on which turbulence is first driven, so it may be that this driving scale is all that needs to be resolved in order to capture the majority of the turbulent energy. \citet{Li2020b} found the driving scale for turbulence in galaxy clusters to be on the order of the scale of feedback, so an analysis of the impact of feedback at different scales in CGM simulations may specify the driving scale and thus enlighten the resolution needed to resolve the bulk of the turbulent energy in the CGM.

\section{Summary and Conclusions}
\label{sec:summary}

In this paper, we derived a modified virial temperature by explicitly including the kinetic energy of non-thermal gas motions in the virial equation for a galaxy halo (Equation~\ref{eq:Tvirmod}). We made two estimates for the non-thermal kinetic energy: one that includes only turbulence (Equation~\ref{eq:KEnt_turb}) and one that includes both turbulence and bulk outflows (Equation~\ref{eq:KEnt_turb_out}). We used the Figuring Out Gas \& Galaxies In Enzo simulations to show how non-thermal kinetic energy contributes to $\sim L^*$ galaxy halos roughly equally to thermal kinetic energy, motivating the need for non-thermal kinetic energy in considerations of virial equilibrium. Even when all forms of energy are accounted for, the gas in galaxy halos is generally not in virial equilibrium throughout much of the halos' evolution (Figure~\ref{fig:virial_time}) and only approaches equilibrium at low redshifts when the halo mass surpasses $\mathrm{few}\times10^{11}M_\odot$ and only when strong bursts of stellar feedback are not perturbing the halo gas (Figure~\ref{fig:energy_xcorr}). Finally, we showed that the modified virial temperature is a closer description of the gas temperature for most of the gas mass in the outskirts of a galaxy halo than the standard virial temperature, which is $\sim2\times$ too large, \emph{even when the halo gas is virialized} (Figures~\ref{fig:temp_time} and~\ref{fig:temp_radius}), suggesting that even the gas in \emph{virialized} halos is not fully \emph{thermalized}. The only times when the standard $\tvir$ is a good descriptor of the gas near $\rvir$ is for a short time following a strong burst of feedback (Figure~\ref{fig:temp_xcorr}), which may be difficult to ``catch" in observations and only occurs when the halo gas is not in virial equilibrium --- giving the expected temperature for the wrong reason.

A lower-than-expected gas temperature in galaxy halos has important implications for analytic CGM models and the initial conditions of idealized CGM simulations. If gas is cooler, thermal pressures are lower, radiative cooling is more efficient, expected X-ray surface brightnesses are lower, and galaxy halos may be able to maintain higher \ion{O}{6} column densities at larger halo masses than expected (\S\ref{sec:implications}). These consequences of lower temperatures may affect analytic models that derive gas physics processes starting from initial assumptions of virial temperature. They also affect idealized simulations of isolated galaxies, where the initial hot halo is frequently put in by hand at the standard virial temperature at the start of the simulation. \textbf{We suggest that analytic models and idealized simulations adopt the modified virial temperature at a factor of $\sim2$ lower than the standard virial temperature for the initial conditions of any model or simulation.}

A lower-than-expected halo temperature due to energy contributions from non-thermal motions is not a unique feature in FOGGIE. This phenomenon should be measurable in any self-consistent cosmological simulation where gaseous halos are built up along with galaxies. However, other cosmological simulations with lower spatial resolution than FOGGIE may not be capturing enough of the energy contained in the small scales of the turbulent cascade in order to make a considerable difference to the overall energy of the halo gas. Indeed, it is possible that at the resolution of FOGGIE, there is still some turbulent energy below the resolution scale that we do not capture, so the magnitude of the difference between standard and modified virial temperatures may be even larger than what we find here. A full analysis of the structure of turbulence in the CGM in FOGGIE is forthcoming.

\acknowledgments{
This study was primarily funded by NASA via an Astrophysics Theory Program grant 80NSSC18K1105.
We are especially grateful to Mark Voit for comments that helped clarify the key concepts of the paper. We also thank Nicolas Lehner, John O'Meara, Philipp Grete, David H.\ Weinberg, and Claire Kopenhafer for useful discussion. We thank the referee for suggestions that ultimately improved the clarity of the paper. CL and RA were additionally supported by HST GO \#15075. BWO acknowledges support from NSF grants no. AST-1517908, OAC-1835213, and AST-1908109, by NASA ATP grants NNX15AP39G and 80NSSC18K1105, and by HST AR \#14315. 
RCS appreciates support from a Giacconi Fellowship at the Space Telescope Science Institute, which is operated by the Association of Universities for
Research in Astronomy, Inc., under NASA contract NAS 5-26555. This research was supported in part by the National Science Foundation under Grant No. NSF PHY-1748958.

Resources supporting this work were provided by the NASA High-End Computing (HEC) Program through the NASA Advanced Supercomputing (NAS) Division at Ames Research Center and were sponsored by NASA's Science Mission Directorate; we are grateful for the superb user-support provided by NAS. Resources were also provided by the Blue Waters sustained-petascale computing project, which is supported by the NSF (award number ACI-1238993 and ACI-1514580) and the state of Illinois. Blue Waters is a joint effort of the University of Illinois at Urbana-Champaign and its NCSA. 

\texttt{Enzo} \citep{Bryan2014,BrummelSmith2019} and \texttt{yt} \citep{Turk2011} are developed by a large number of independent researchers from numerous institutions around the world.  This research made use of Astropy (http://www.astropy.org), a community-developed core Python package for Astronomy \citep{Astropy1, Astropy2}. Their commitment to open science has helped make this work possible.

This work benefited greatly from copious pictures of the FOGGIE team's cats and dog posted to our Slack channel.

\facilities{NASA Pleiades}

\software{astropy \citep{Astropy1,Astropy2},
          Cloudy \citep{Ferland2017}, 
          Enzo \citep{Bryan2014,BrummelSmith2019},
          grackle \citep{Smith2017},
          yt \citep{Turk2011}
          }
}

\appendix{
\section{Definitions of Virial Radius}
\label{sec:Rvir_def}

Throughout this work, we use the radius enclosing 200 times the critical density of the universe, frequently referred to as $\rvir$, as the ``virial" radius. However, the true ``virial radius" is more strictly defined from the collapse of a top-hat dark matter structure to include all matter that is bound to the halo. The virial radius defined in this way is \emph{not} simply $\rvir$ because the overdensity factor evolves with redshift in a universe with dark energy \citep{Bryan1998}. Instead, the overdensity factor at $z=0$ is actually closer to 100, not 200, for the \citet{Planck2016} cosmological parameters $\Omega_M=0.3075$ and $\Omega_\Lambda=0.6925$. We chose to use a non-evolving overdensity factor of 200 in this work because this is what is frequently done in observational surveys, and because $\rvir$ falls at least partially within the zoomed refine box for each of the FOGGIE halos at all redshifts (see \S\ref{sec:FOGGIE}), while the true virial radius, $\realrvir$, does not. At $z=0$, the difference between these different definitions of ``virial radius" for the Tempest halo is $\sim50$ kpc and this difference is smaller at higher redshift. At $z=0$, $\rvir\approx170$ kpc while $\realrvir\approx220$ kpc, for the Tempest halo. We show in this appendix that the precise definition of the virial radius does not matter for our qualitative conclusions.

The left panel of Figure~\ref{fig:virial_radius_def} compares the energies within the halo as functions of radius at $z=0.15$ in the Tempest halo. We choose $z=0.15$ because at lower redshift, $\realrvir$ entirely exits the forced refinement region of the Tempest halo. It is clear there is no strong difference in the virial energy between the two radii, as the virial energy becomes fairly flat with radius past $\sim100$ kpc in both cases. The difference between the virial energy including non-thermal gas motions and the thermal-only virial energy is far larger than the differences in virial energy for different halo radius definitions.

\begin{figure}
    \centering
    \includegraphics[width=0.49\linewidth]{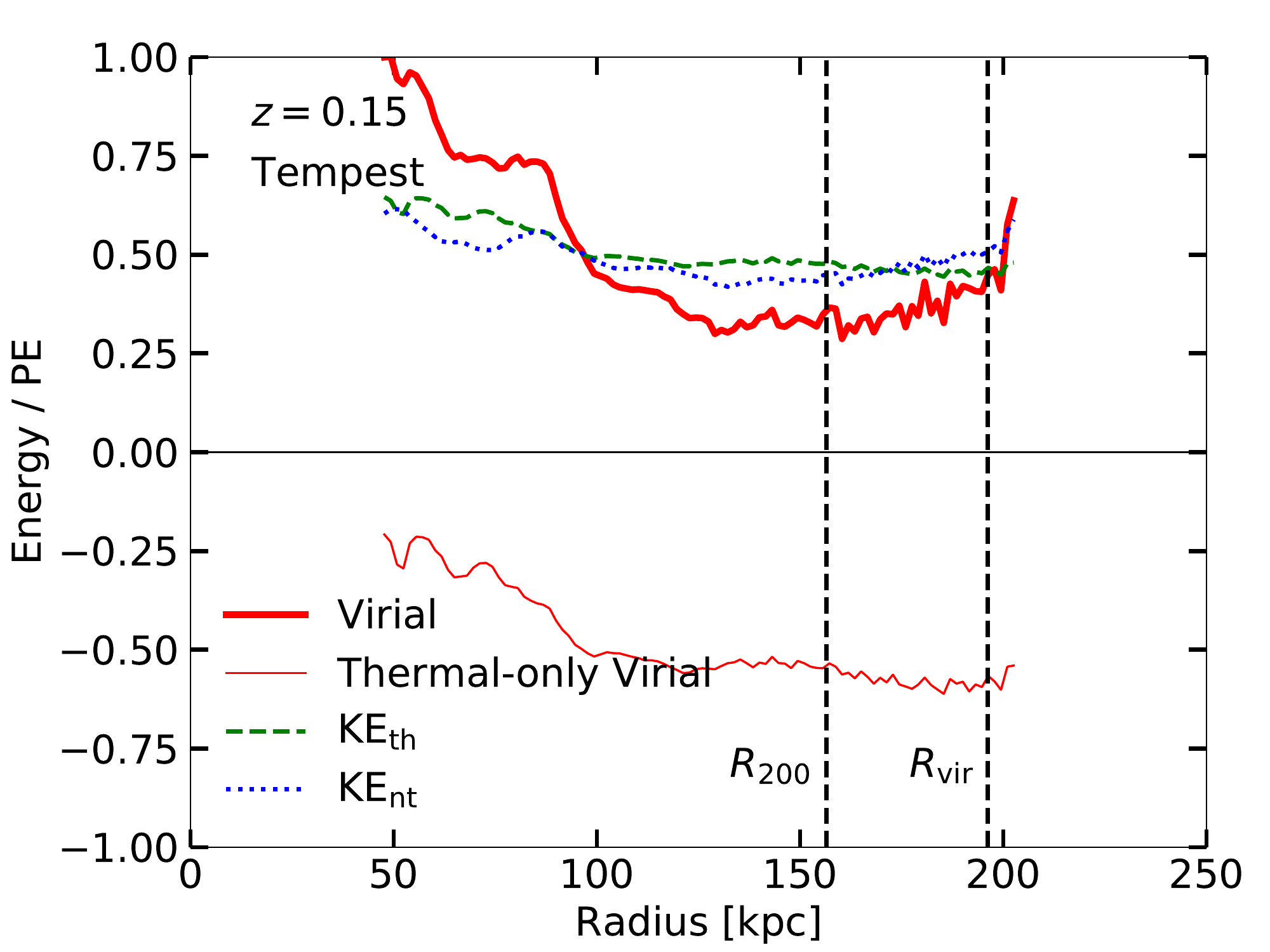}
    \includegraphics[width=0.49\linewidth]{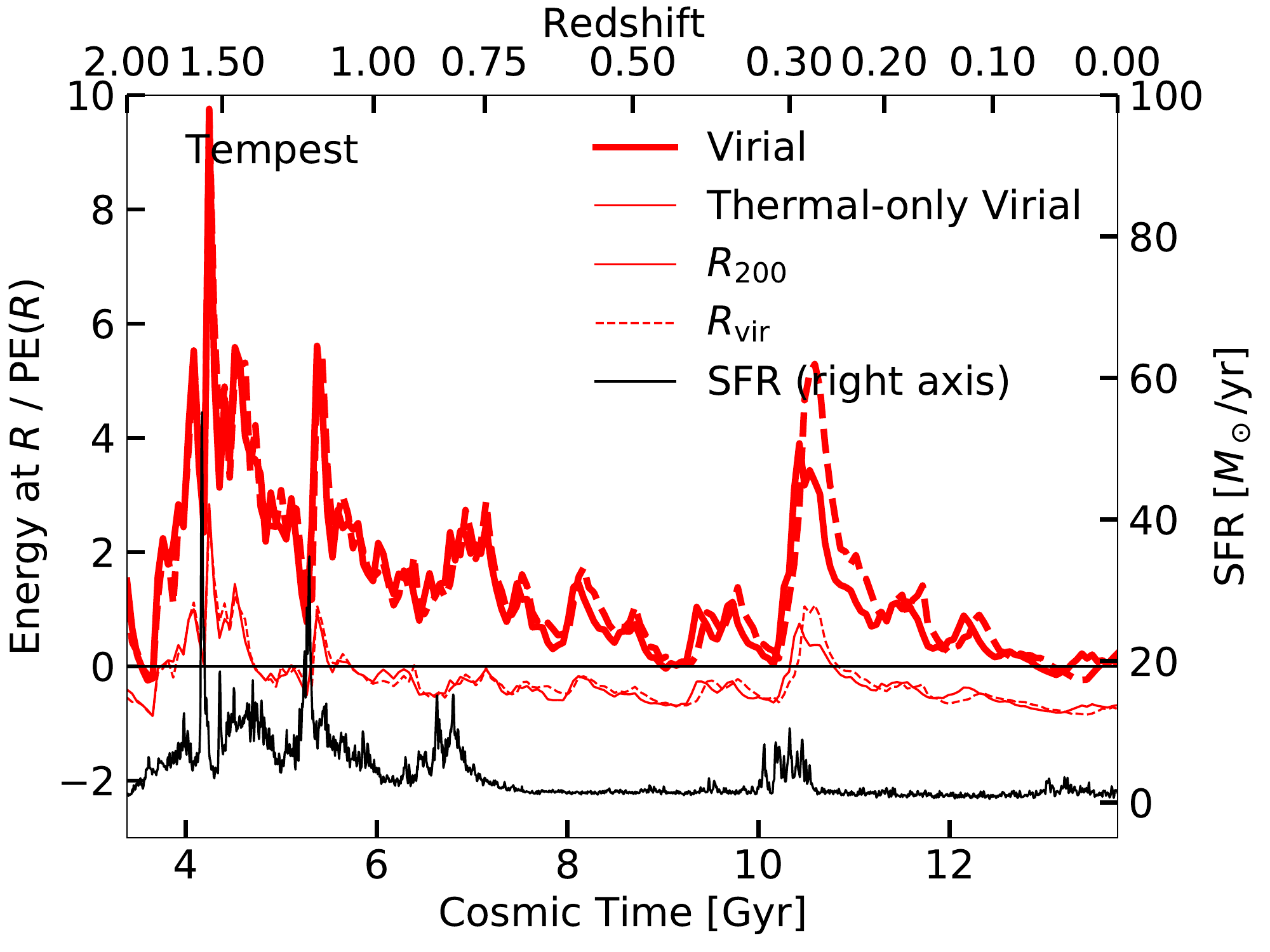}
    \caption{{\it Left:} The various energies of the gas in the Tempest halo at $z=0.15$, similar to Figure~\ref{fig:virial_radius}. Also marked as vertical dashed lines are two definitions of the halo radius, the $\rvir$ used throughout this paper and the virial radius $\realrvir$. {\it Right:} The virial energies of the gas in the Tempest halo at $\rvir$ (solid lines) or $\realrvir$ (dashed lines), similar to Figure~\ref{fig:virial_time}. There is little substantial difference in the virial energy (thick red line), between the definitions of halo radius.}
    \label{fig:virial_radius_def}
\end{figure}

The right panel of Figure~\ref{fig:virial_radius_def} shows the virial energy (Equation~\ref{eq:VE}) of gas within $0.99\rvir < r< \rvir$ (solid) or within $0.99\realrvir<r<\realrvir$ (dashed) as functions of cosmic time, similar to Figure~\ref{fig:virial_time}. We show the virial energies in this case only down to $z=0.15$, because at smaller $z$, $\realrvir$ entirely exits the zoom-in, high-resolution box and is instead located only in the low-resolution simulation domain, and we lose the valuable information about small-scale non-thermal kinetic energy such as turbulence that is so crucial to the virial energy and temperature of the halo gas. The difference between the virial energy (thick lines) and the thermal-only virial energy (thin lines) is once again larger than the difference in virial energy (or thermal-only virial energy) between different radii (solid vs dashed) so our conclusions are clearly not strongly dependent on the exact definition of virial radius.

\section{Testing the Singular Isothermal Sphere Assumption}
\label{sec:SIS}

The virial theorem as presented in Equation~(\ref{eq:virial}) is originally defined for the sum of all matter in the system, not in thin radial shells as we do in this paper. However, in a galaxy simulation where star formation feedback and radiative cooling provide sources and sinks of energy to the halo gas, a sum of energy of all halo gas within $\rvir$ or $\realrvir$ is not expected to satisfy the virial theorem (and $\realrvir$ is not a hard ``edge" to the system in any case). To reduce the impact of energy sources and sinks, we neglected the inner parts of the halo near the stellar component of the galaxy and calculated the virial energy in radial shells. In order for this method to be valid, the halo gas must adhere to a singular isothermal sphere (SIS) profile in the outskirts of the halo where we focus. In a SIS profile, the potential energy and gas density are both proportional to $r^{-2}$ so the energy budget of a shell of gas at $r$ will independently satisfy the virial theorem if the whole SIS system does. This is not necessarily the case for other potential energy or gas density profiles, so here we test the SIS assumption compared to an explicit calculation of energies over the outskirts of the system.

Figure~\ref{fig:den_profile} shows a mass-weighted histogram of the filament-removed halo gas density as a function of radius. Overplotted as the black solid line is a density profile with the $r^{-2}$ dependence of the SIS profile, normalized to the average gas density at $\rvir$. The SIS profile cuts through the middle of the gas density distribution at each radius, signifying it is a decent assumption for the structure of the warm halo gas.

\begin{figure}
    \centering
    \includegraphics[width=0.5\textwidth]{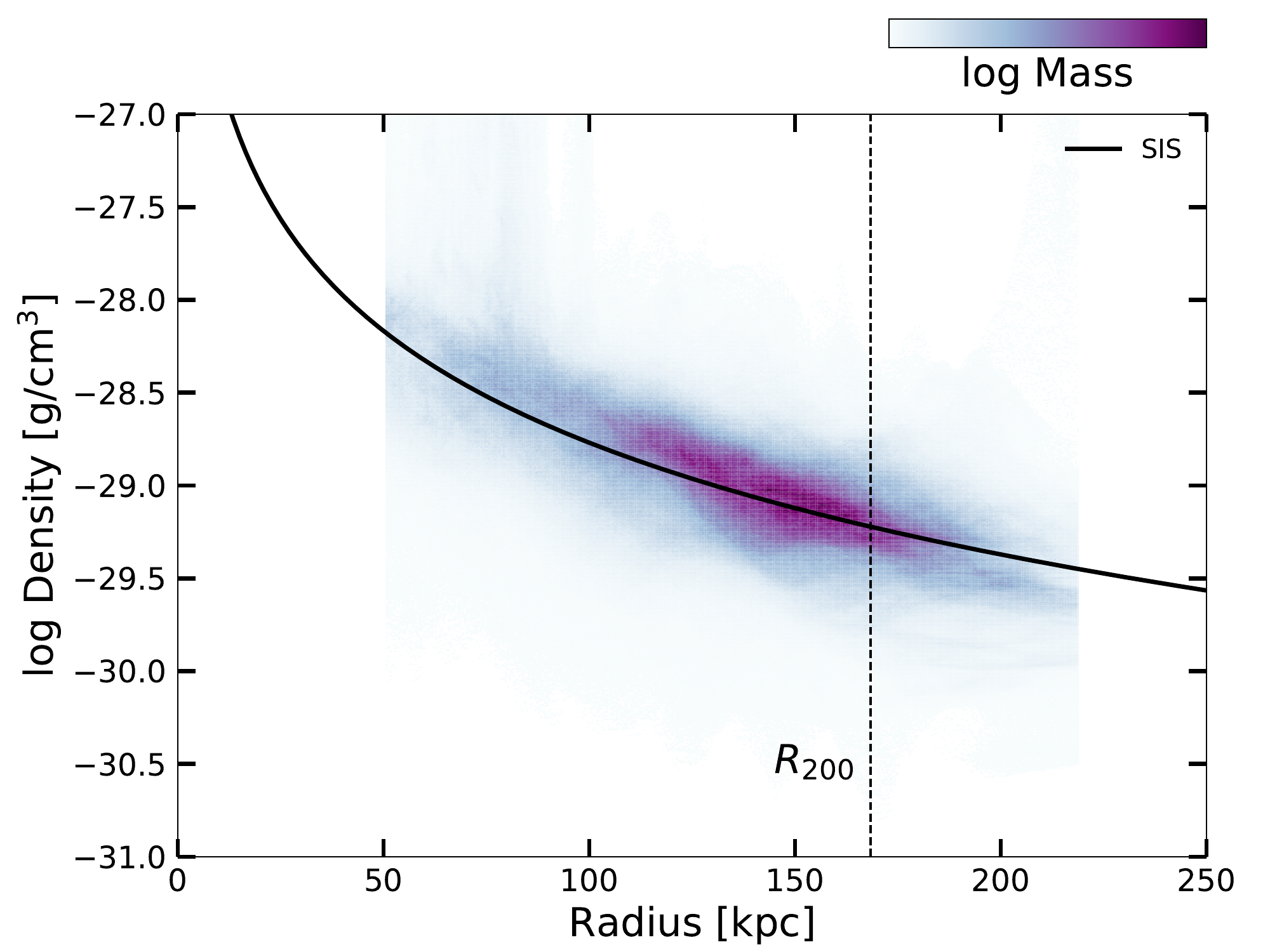}
    \caption{A mass-weighted histogram of warm halo gas density (filaments removed) as a function of radius. The black solid line shows a SIS density profile that is normalized to the gas density at $\rvir$. The SIS profile describes the gas profile reasonably well.}
    \label{fig:den_profile}
\end{figure}

The SIS assumption for the virial theorem requires not just the gas density profile to follow a SIS, but also the gravitational potential energy profile. The left panels of Figure~\ref{fig:SIS_compare} show the specific energies of the warm halo gas as a function of radius for the Tempest halo at $z=0$, similar to Figure~\ref{fig:virial_radius}. Unlike Figure~\ref{fig:virial_radius}, these energies are not normalized by $GM(<r)/r$, and instead there are curves shown for the potential energy (solid black) and boundary pressure (magenta dot-dashed) terms. The thick curves show these energies in thin shells, as is done throughout the paper, using the SIS assumption. The thin curves show a cumulative sum of these energies, summing from an inner radius $r_i$ outward. We pick different inner radii to examine the impact of summing over different parts of the outer halo, and each row shows the calculation using a different value of $r_i=0.3\rvir,\ 0.5\rvir,\ 0.7\rvir$, increasing from top to bottom. The thermal and non-thermal kinetic energies are calculated by summing over these energies in each cell of the simulation, so the only difference between the thick and thin curves for these energies are how much of the halo is entering into the sum. There is clearly little difference between the shells and the cumulative sum in the kinetic energies.

The main difference between the explicit sum and the SIS assumption in shells is in the potential energy and boundary pressure terms. For the shell calculation, the potential energy is given by
\begin{equation}
    \mathrm{PE}=\int_{r_{n}}^{r_{n+1}} \frac{GM(<r)}{r}\ \mathrm{d}r
\end{equation}
where $r_n$ and $r_{n+1}$ are the radii of the edges of radial shell $n$. For the exact, cumulative calculation, the potential energy of the gas within a boundary radius $r_b$ is given by
\begin{equation}
    \mathrm{PE}=\int_{r_i}^{r_b} \frac{GM(<r)}{r}\ \mathrm{d}r
\end{equation}
where the inner boundary $r_i$ is chosen as an intermediate value such that the very inner regions of the halo near the galaxy are excluded. Note that although we integrate over a range in radius near the outskirts of the system, the potential energy of the gas depends on the total mass \emph{within} the radius $r$. The exact calculation of the potential energy deviates from the SIS assumption only when smaller values of $r_i$ are used, indicating that it is only in the inner regions of the halo where the gas cannot be approximated as following a SIS profile. When a larger $r_i$ is used, the exact calculation and the SIS assumption lead to potential energies that are very similar.

The specific boundary pressure term $\Sigma$ is given by
\begin{equation}
    \Sigma=\frac{4\pi r_b^3 P_b}{M_\mathrm{gas}(r_i<r<r_b)}
\end{equation}
where
\begin{equation}
    P_b = P_\mathrm{th}(r_b) + \frac{1}{2}\rho(r_b)\sigma_r(r_b)^2
\end{equation}
describes the sum of thermal and non-thermal pressures in the radial direction at the boundary radius $r_b$, where $\sigma_r$ is the radial component of the gas velocity dispersion. $\Sigma$ describes the energy of a pressure that is exerted on the virialized system, so to find its specific energy, it must be normalized by the gas for which we are applying the virial theorem, which is gas between the inner radius $r_i$ and the boundary radius $r_b$. In the exact calculation shown by the thin curves in Figure~\ref{fig:SIS_compare}, $\Sigma$ is computed directly from the gas density, pressure, and radial velocity dispersion at each radius $r_b$, where the boundary radius $r_b$ is given by the value of the radius on the horizontal axis. In the SIS assumption (thick curves), $\Sigma$ is assumed to simply be $\frac{1}{2}\frac{GM(<r_b)}{r_b}$ (equation~\ref{eq:Sigma}, not shown).

Finally, the solid red curves in Figure~\ref{fig:SIS_compare} show the virial energy VE (equation~\ref{eq:VE}) calculated in both the SIS assumption in shells (as done throughout this paper, thick curves) and in the exact calculation as a cumulative sum using the terms defined in the preceding paragraphs (thin curves). We perform the sum with different inner radii $r_i$ to determine the inner radius at which the SIS assumption in shells deviates from the exact, cumulative calculation. At the time snapshot shown in the left panels of Figure~\ref{fig:SIS_compare}, which was chosen as a time the halo is fairly quiescent and perhaps closest to virial equilibrium, the virial energies computed in the two different ways become more in agreement as the inner radius value increases. This suggests that the SIS assumption for the potential energy and the boundary pressure terms of the virial equation approximately holds in the outer regions of these halos, where we focus in this paper. The exact calculation deviates from the SIS assumption in the inner parts of the halos closer to the galaxies, as expected.

\begin{figure}
    \centering
    \includegraphics[width=0.49\linewidth]{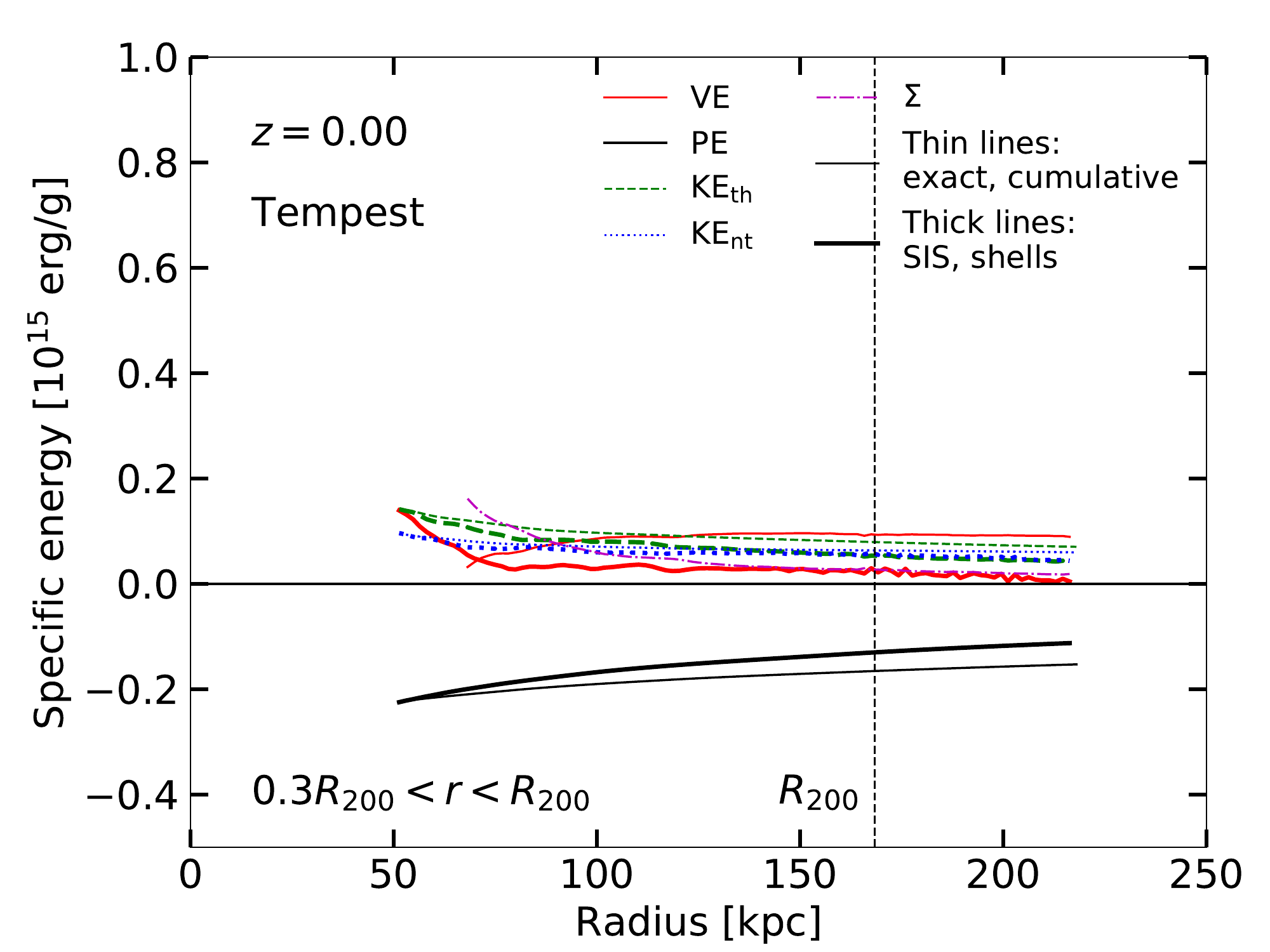}
    \includegraphics[width=0.49\linewidth]{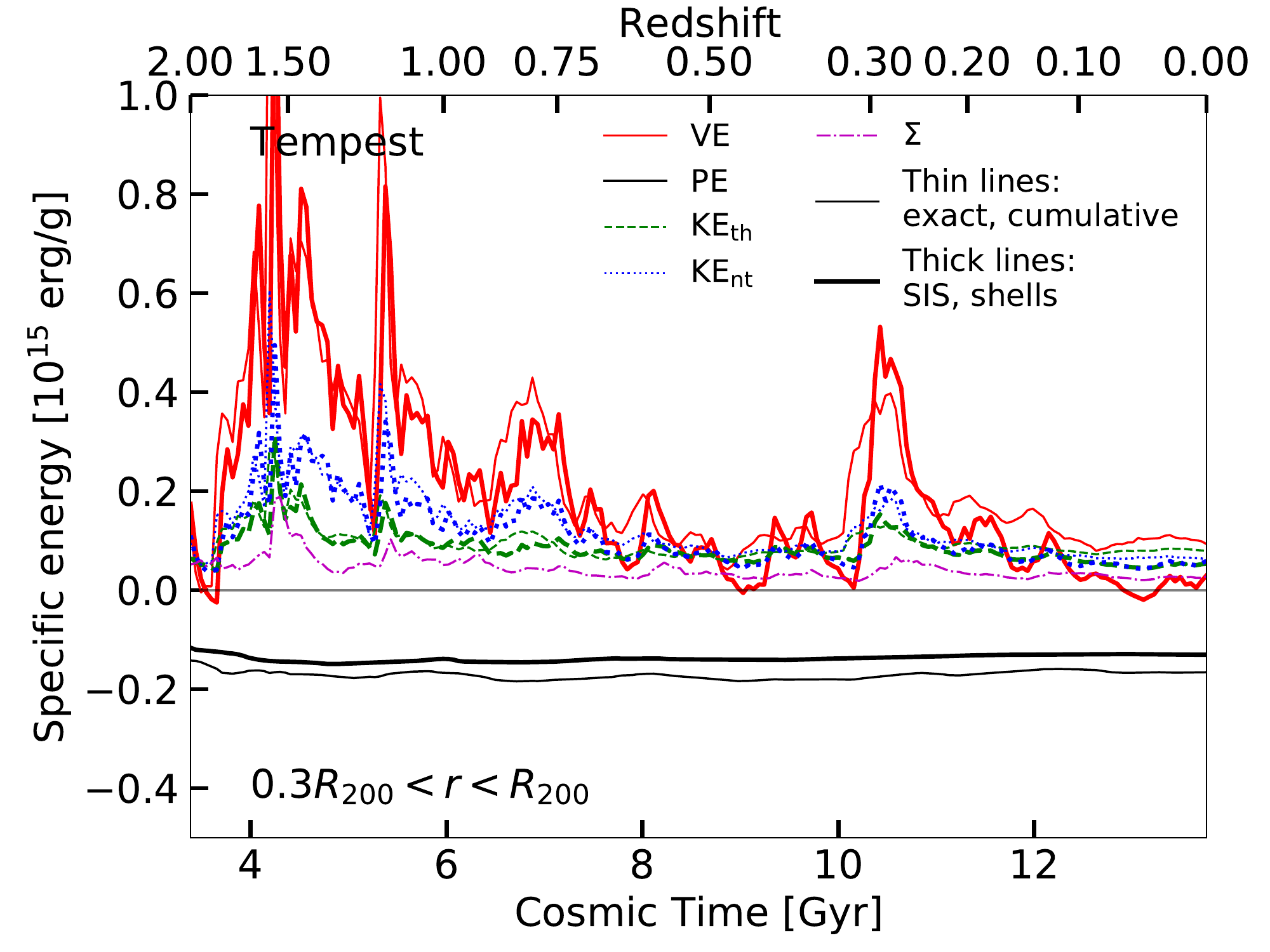}
    \includegraphics[width=0.49\linewidth]{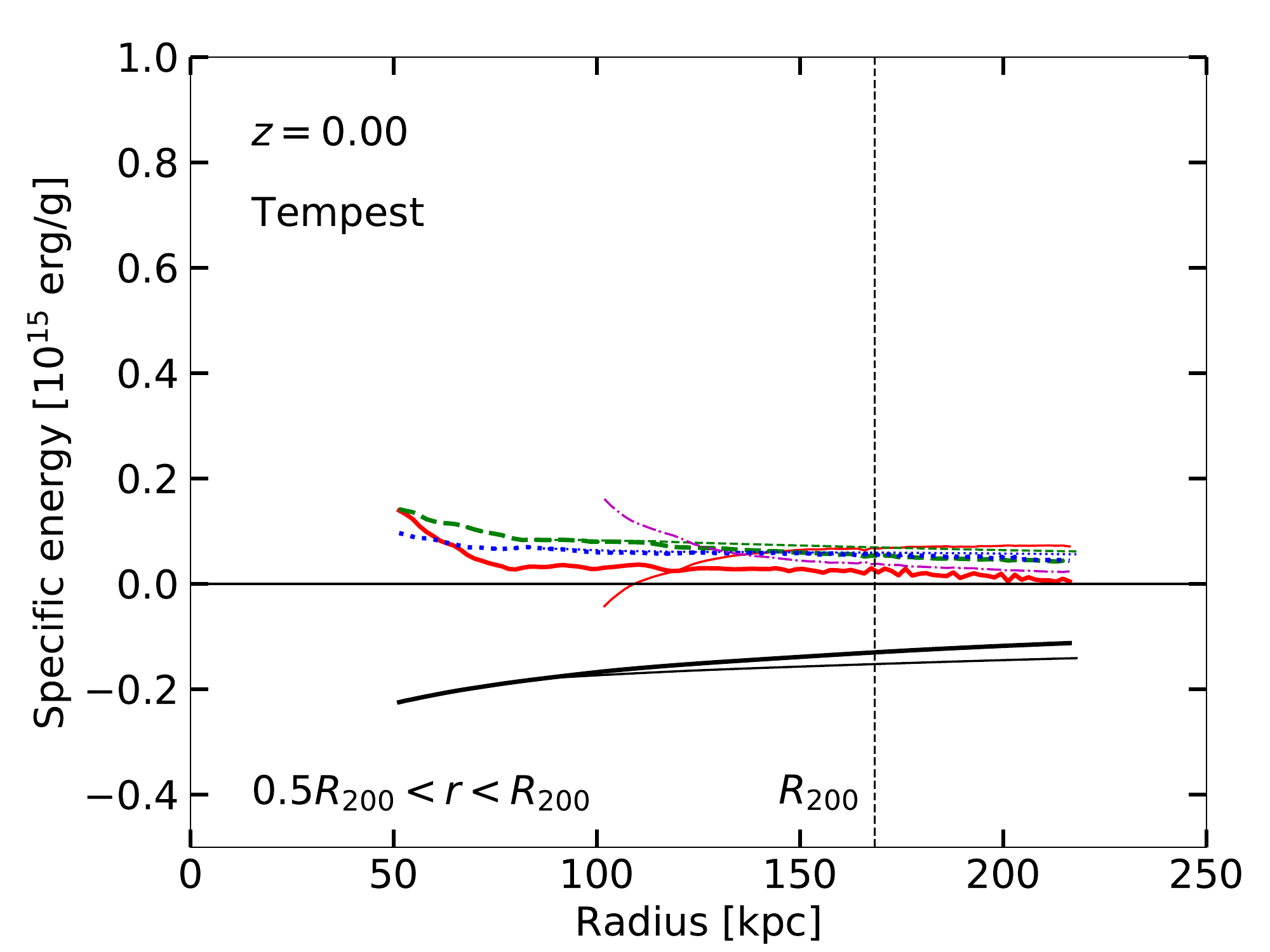}
    \includegraphics[width=0.49\linewidth]{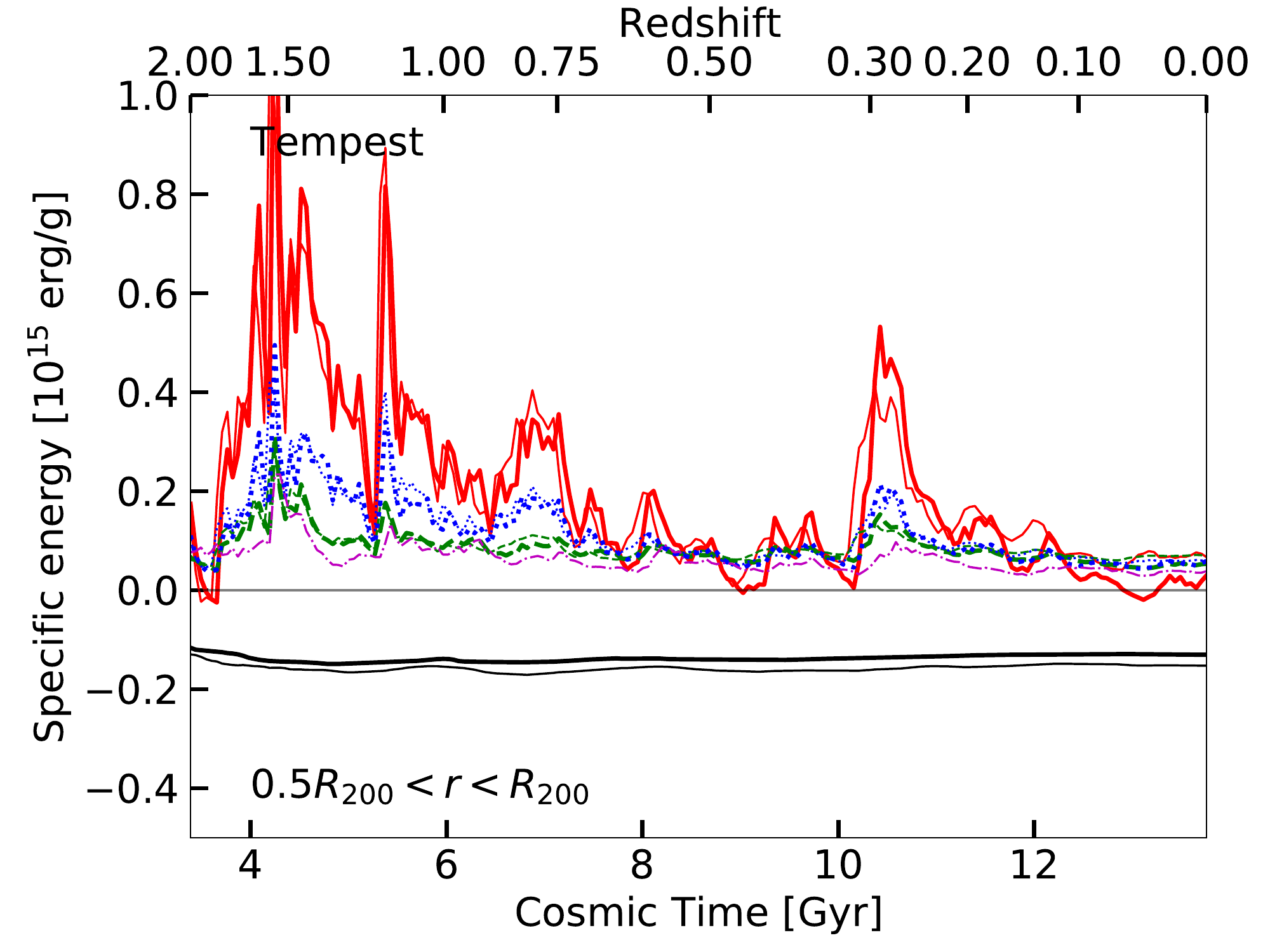}
    \includegraphics[width=0.49\linewidth]{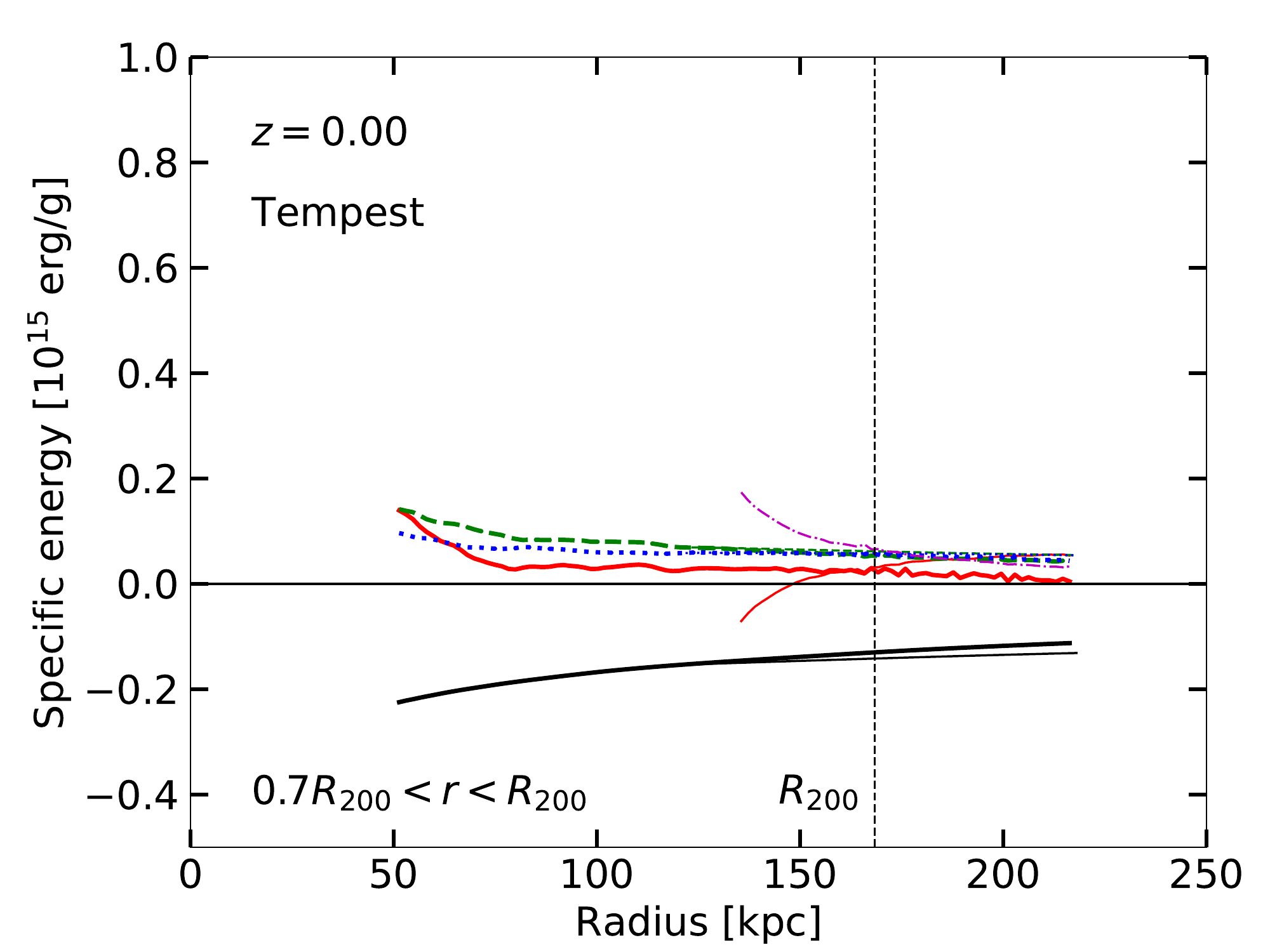}
    \includegraphics[width=0.49\linewidth]{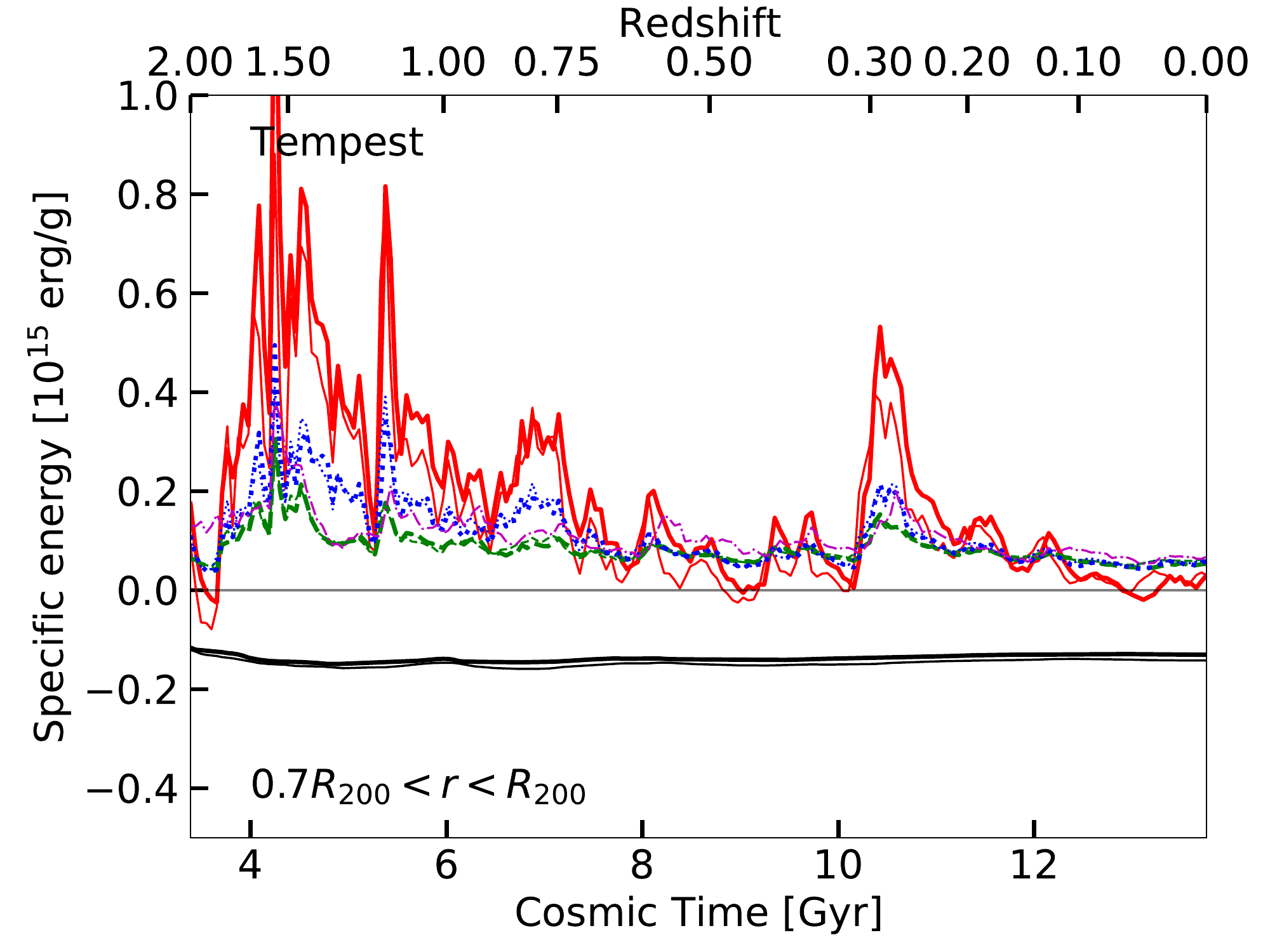}
    \caption{Various specific energies at $z=0$ for the Tempest halo. The thin curves show exact, cumulative calculations of energies between an inner radius $r_i$ and the boundary radius, while the thick curves show calculations of energies in shells assuming SIS profiles. Left panels show energies as functions of radius (cumulative between $r_i$ and $r$ for the thin curves or within a thin shell at $r$ for the thick curves), while right panels show energies over time (cumulative between $r_i$ and $\rvir$ for thin curves or within a shell at $\rvir$ for thick curves). Different rows show different choices for the inner radius $r_i$, which only affects the cumulative energy sum, increasing from top to bottom. The two types of calculations for the energies produce similar values of VE (solid red curves) as $r_i$ approaches large values in the outskirts of the halo, indicating that the SIS assumption is valid in the regions near $\rvir$ but not in the inner regions of the halo.}
    \label{fig:SIS_compare}
\end{figure}
}

Finally, in the right panels of Figure~\ref{fig:SIS_compare}, we show the time evolution of the various energies in the Tempest halo, as in Figure~\ref{fig:virial_time}. The thick lines show the energy calculations assuming SIS in thin shells at $\rvir$ while the thin lines show the energy calculations without any assumptions, cumulatively between an inner radius $r_i$ and $\rvir$. The rows show increasing values of the inner radius, from $0.3\rvir$ (top) to $0.5\rvir$ (middle) to $0.7\rvir$ (bottom). As the inner radius increases and the calculation is done over only those parts of the halo approaching $\rvir$, the difference between the SIS assumption (thick lines) and the exact calculation (thin lines) decreases. This is further proof that the SIS assumption holds only in the outskirts of the halo, near $\rvir$. We do not show the energy evolution over time for Squall and Maelstrom for brevity, but we see the same trends and results there as we do with Tempest.

Thus, we conclude that while the SIS assumption for the virial theorem does not hold generically throughout the halo, it does hold well enough in the outer regions of the halo where we are most interested in measuring the temperature of the warm halo gas. This allows us to use the SIS assumption throughout this paper and in our definition of $\tmod$, which is beneficial because it allows us to estimate a temperature without requiring knowledge of the sum of gas energies throughout the entire halo. We also note that the exact calculation tends to produce a VE $>0$, especially when $r_i$ is small, indicating an excess of energy over that expected from virial equilibrium in the inner regions of the halo. This energy excess is likely due to stellar feedback heating and accelerating the gas, which is not a virial process and thus is not expected to adhere to the virial theorem.

\section{The Role of Radiative Cooling}
\label{sec:cooling}

Throughout this paper, we have considered the balance of energies in the halo gas neglecting the possible energy sink of radiative cooling under the assumption that there is little radiative cooling occurring at large distances from the galaxy, near the virial radius. Here, we test this assumption.

\begin{figure}
    \centering
    \includegraphics[width=0.5\linewidth]{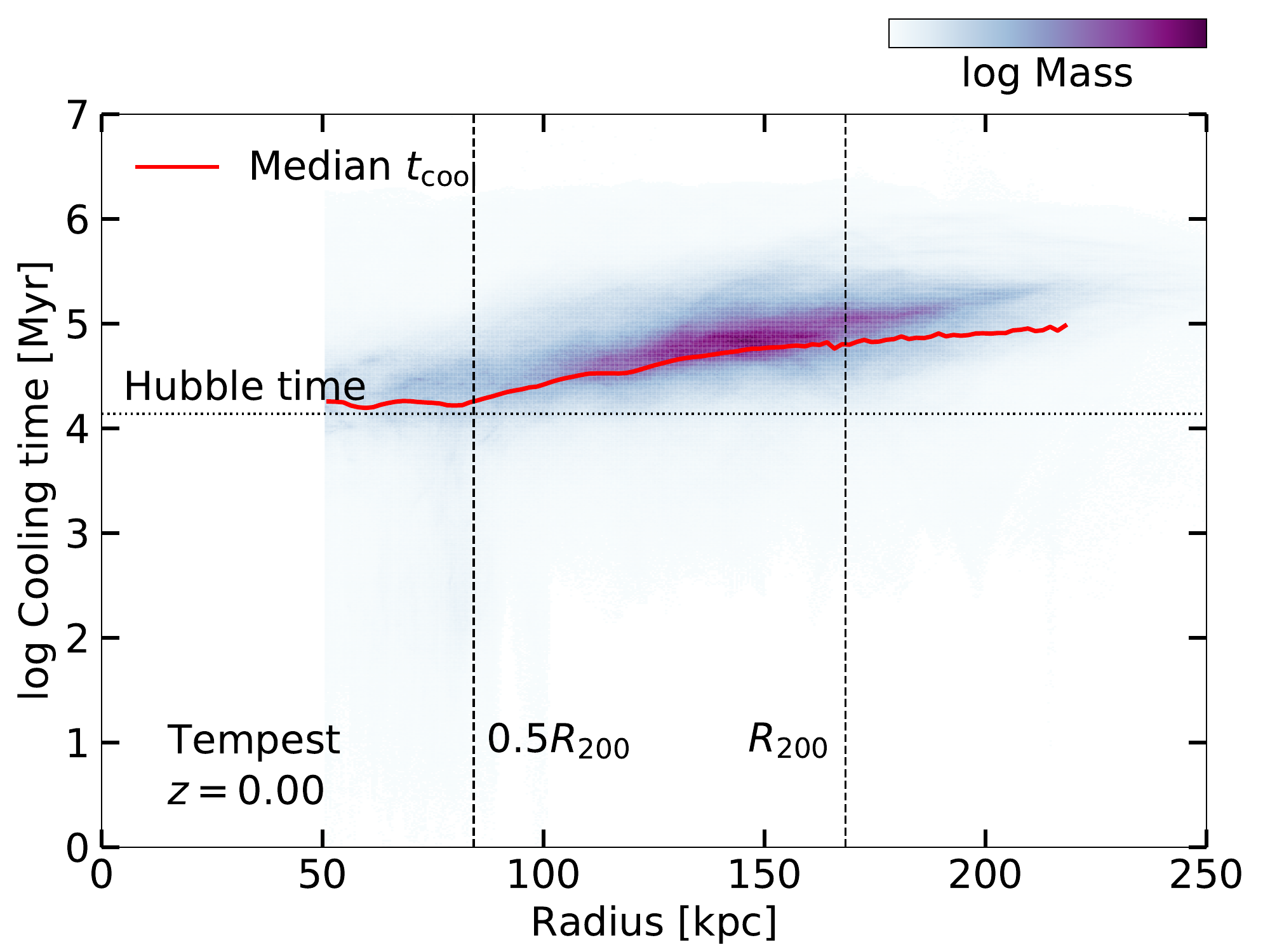}
    \includegraphics[width=0.5\linewidth]{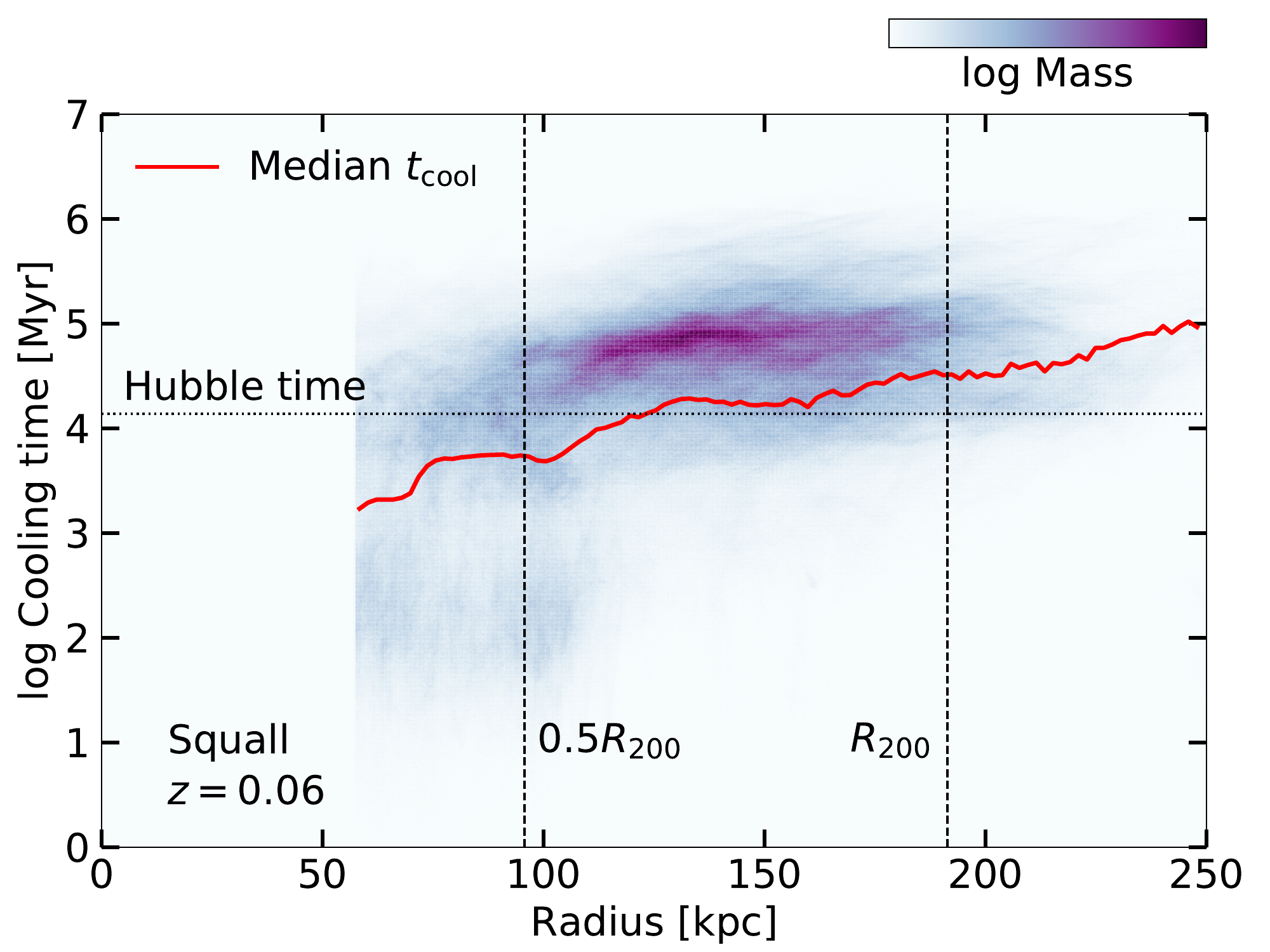}
    \includegraphics[width=0.5\linewidth]{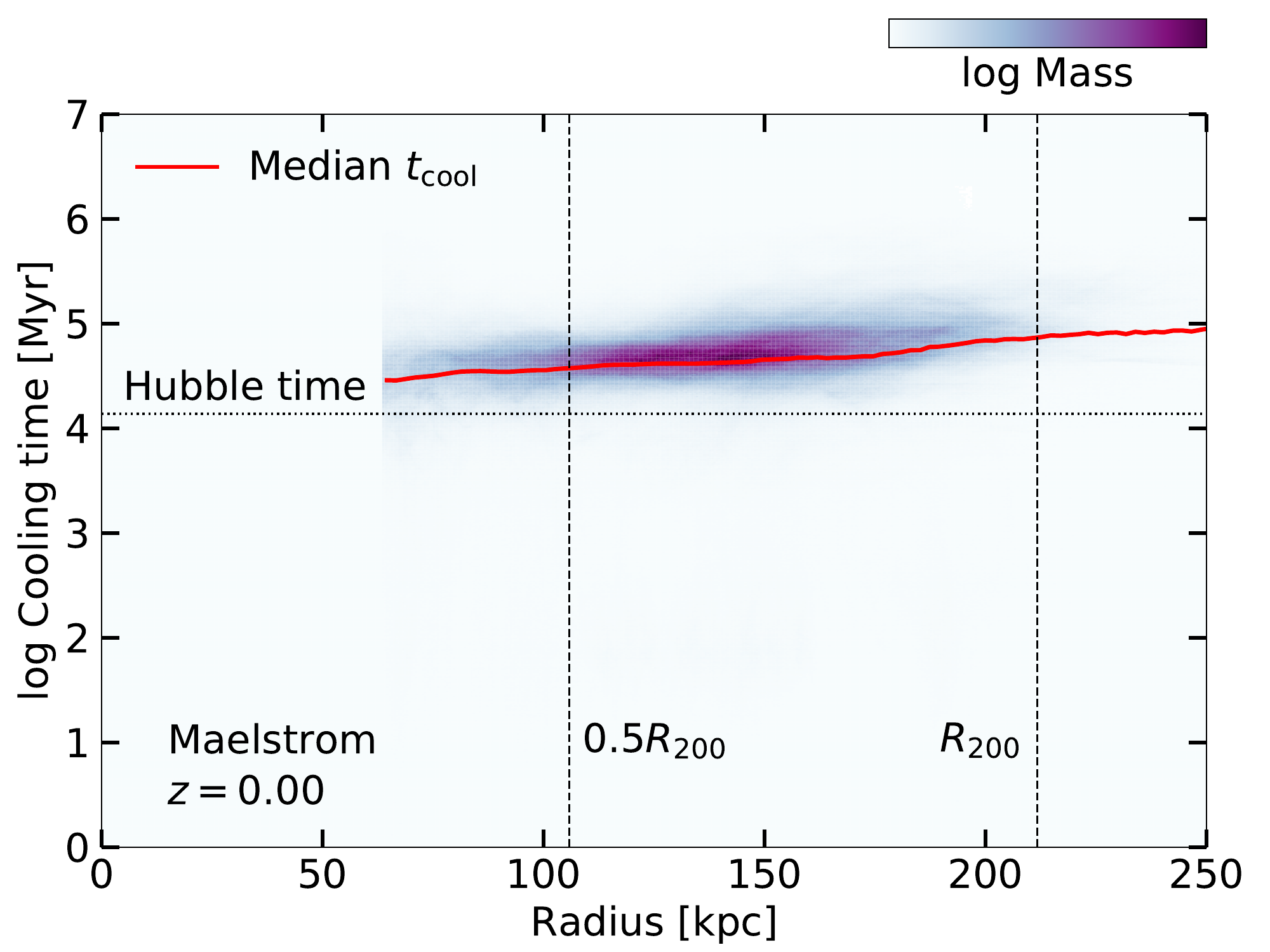}
    \caption{Histograms of gas cell cooling times (white-purple color shading) as functions of radius for each of Tempest, Squall, and Maelstrom at times when the halo gas is approximately virialized. The median cooling time is plotted as a red solid line. The bulk of the gas mass (dark purple colors) between $0.5\rvir$ and $\rvir$ (dashed vertical lines) has cooling times a factor of $\sim2-8\times$ the Hubble time (horizontal dashed line), but there is a tail of cells with short cooling times that pull the median downward, especially in Squall, which is the furthest out of virial equilibrium.}
    \label{fig:cooling_hist}
\end{figure}

Figure~\ref{fig:cooling_hist} shows mass-weighted histograms of simulation cell cooling times as a function of radius for the three halos as the white-purple colors, and the median cell cooling time as a solid red line. A horizontal dashed line marks the Hubble time. The bulk of the gas mass, indicated by the darkest colors in the histogram, hasa typical cooling times a factor of $2-10\times$ the Hubble time in the outskirts of the halo, between $0.5\rvir$ and $\rvir$ (marked with vertical dashed lines). This indicates that the bulk of the cells in the CGM are not rapidly cooling.

\begin{figure}
    \centering
    \includegraphics[width=0.5\linewidth]{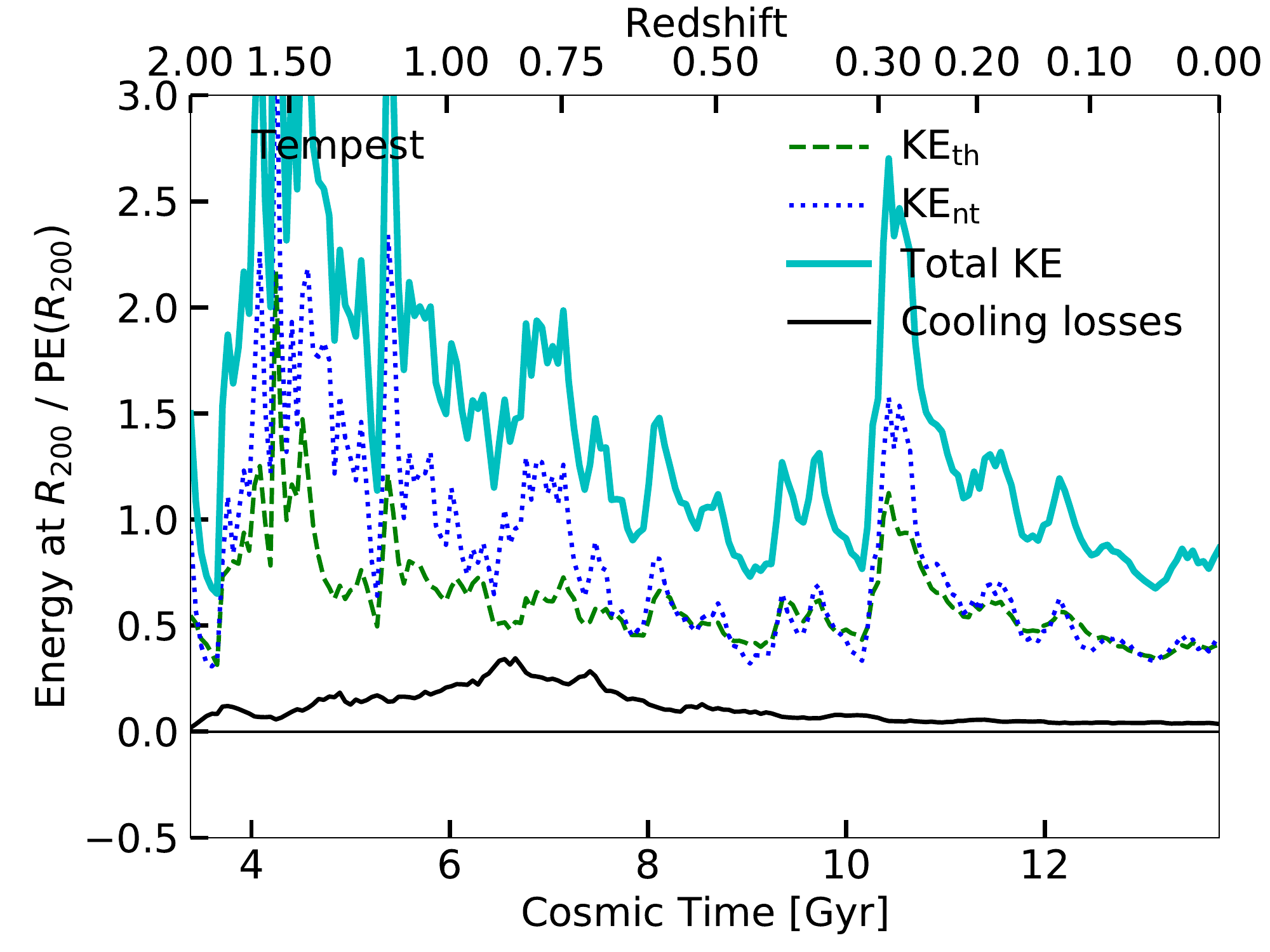}
    \includegraphics[width=0.5\linewidth]{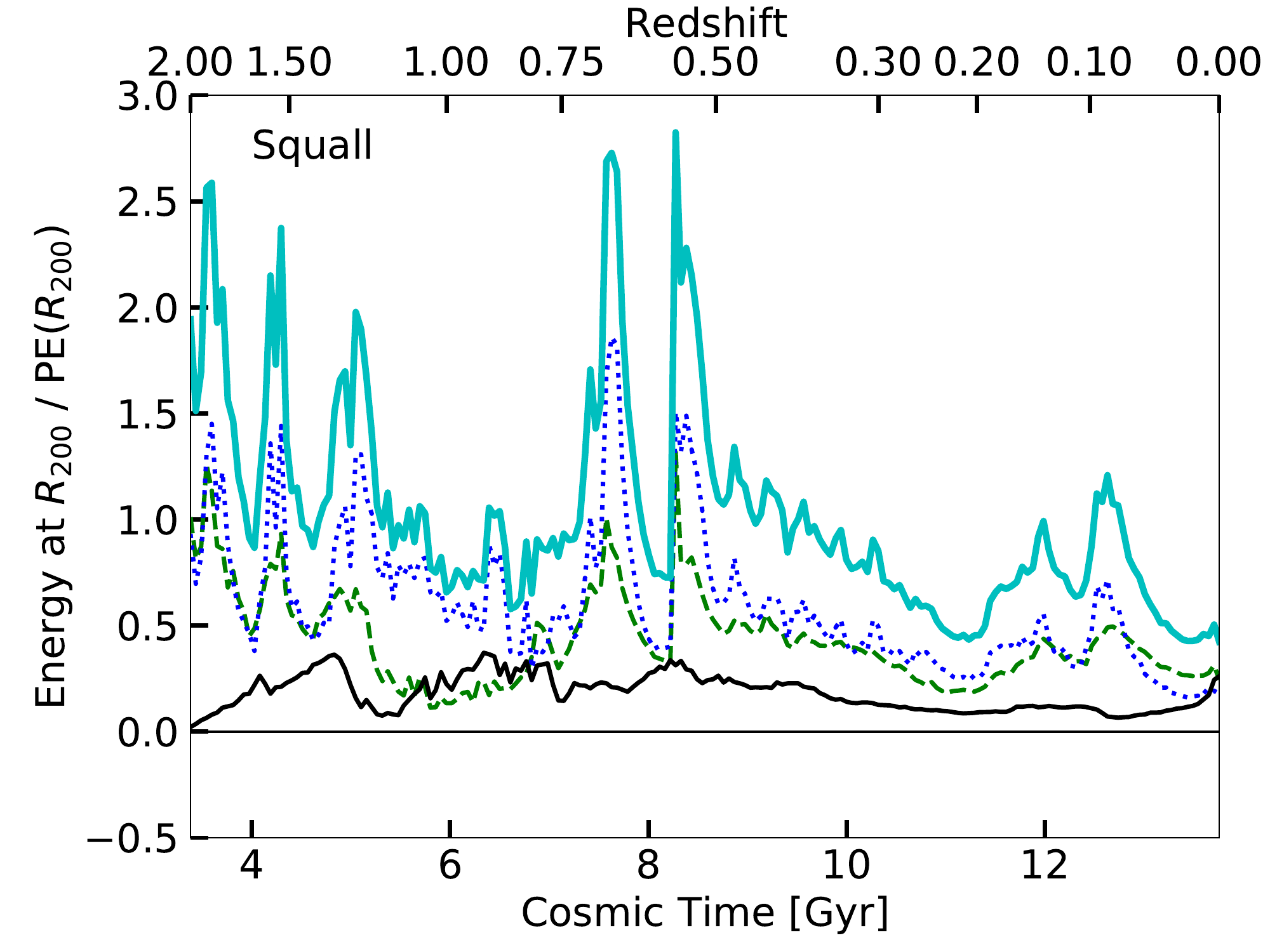}
    \includegraphics[width=0.5\linewidth]{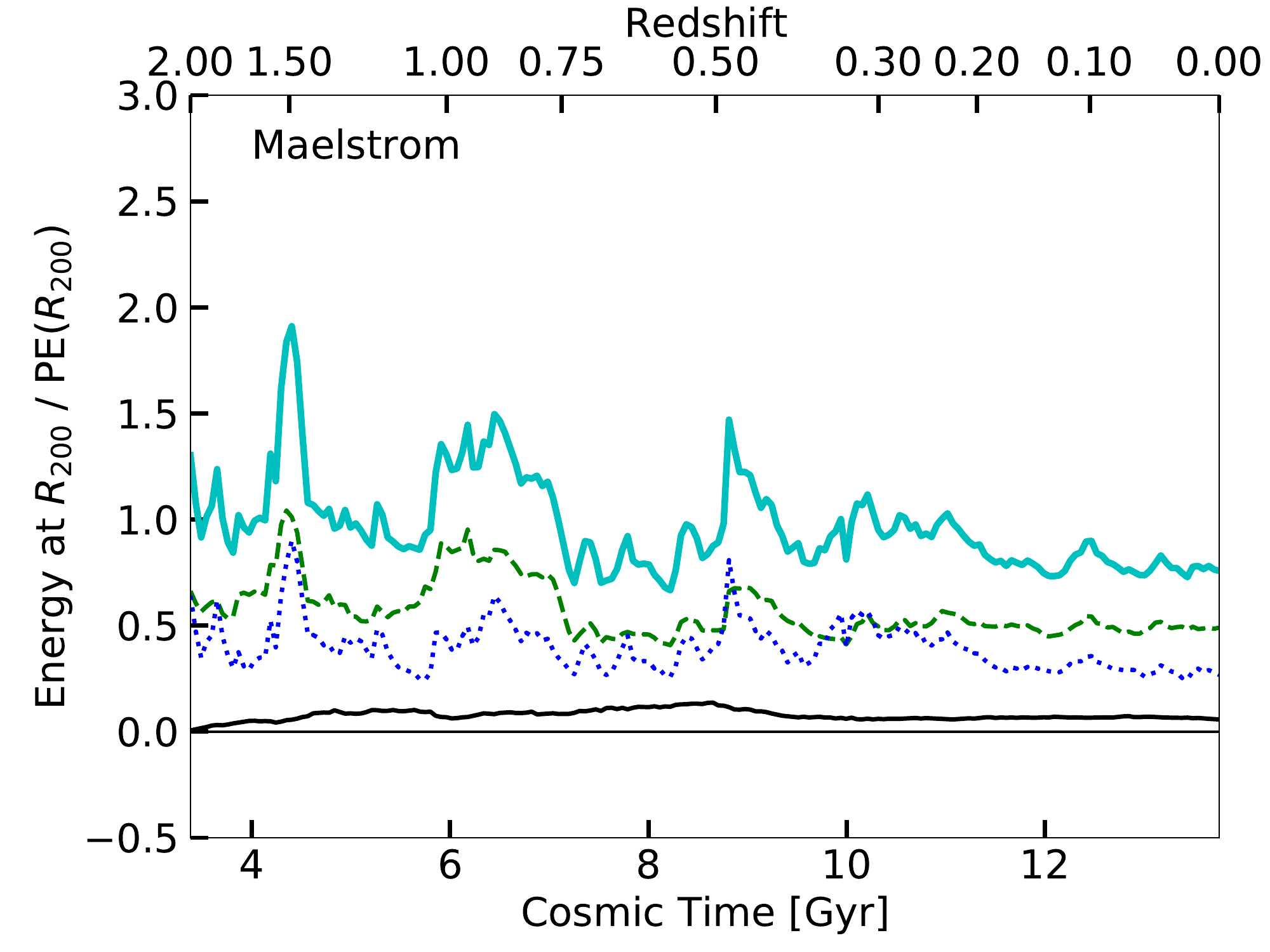}
    \caption{Energies of the halo gas in a shell at $\rvir$ over time, as in Figure~\ref{fig:virial_time}, compared to the radiative cooling losses integrated over the previous 3 Gyr. The thermal kinetic energy (green dashed), non-thermal kinetic energy (blue dotted), and their sum (light blue solid) are all significantly larger than the integrated cooling losses (black solid) at the times when the halo gas near $\rvir$ is close to virial equilibrium.}
    \label{fig:cooling_energy_vs_time}
\end{figure}

However, there is a tail down to short cooling times, most noticeable in Squall (middle panel). The median cooling time indicated by the red line lies below the darkest part of the histogram, pulled downward by the gas cells with short cooling time. While those cells with short cooling times are not dominant in the halo, it is possible they have strong enough cooling to affect the overall energy balance of the entire system. To check the effect of cooling losses on the halo energy balance, Figure~\ref{fig:cooling_energy_vs_time} shows the thermal and non-thermal kinetic energies of the gas in a shell at $\rvir$, and their sum, compared to the radiative cooling losses in the shell integrated over the previous 3 Gyr. At low redshift, $z\sim0.5-0$, the sound crossing time to $\rvir$ is $2-3$ Gyr, and shorter at higher redshifts, so the integrating of cooling losses is performed conservatively over one sound crossing time or longer. When Tempest and Maelstrom are closest to virial equilibrium, at low redshift, the integrated cooling losses are less than 10\% of the total kinetic energy in the halo. The gas in Squall's halo exhibits significantly more radiative cooling losses compared to its total kinetic energy, but the cooling loss is lowest at the time Squall is closest to virial equilibrium at $z=0.1-0.05$.

Figure~\ref{fig:cooling_energy_vs_time} suggests that while there are some cooling losses in the overall energy of the halo gas, they are smallest during the times when halo gas near $\rvir$ is closest to virial equilibrium. Cooling does not dominate the energy balance in virialized halos, and thus likely has a small impact on the factor of two decrease in temperature in the outskirts of the halo \emph{when the halo gas is virialized}. Following bursts of star formation, or during minor mergers, the loss of halo gas energy to radiation is much stronger and may be more important to the overall energy balance of the halo.

\bibliography{modified_virial_temp}{}
\bibliographystyle{aasjournal}

\end{document}